\documentclass{aa}
\usepackage{graphicx}
\usepackage{mwe}
\usepackage{siunitx}
\usepackage[table]{xcolor}
\usepackage{caption}
\DeclareCaptionType{equ}[][]
\usepackage{chemist}
\usepackage{multirow}
\usepackage[flushleft]{threeparttable}
\usepackage{subcaption}
\usepackage{float}
\usepackage{txfonts}
\titlerunning{}

\usepackage{natbib,twoopt}
\usepackage[colorlinks=true, pdfstartview=FitV, linkcolor=blue, citecolor=blue, urlcolor=blue, breaklinks=true]{hyperref} 
\bibpunct{(}{)}{;}{a}{}{,}             
\makeatletter
  \newcommandtwoopt{\citeads}[3][][]{\href{http://adsabs.harvard.edu/abs/#3}%
    {\def\hyper@linkstart##1##2{}%
     \let\hyper@linkend\@empty\citealp[#1][#2]{#3}}}
  \newcommandtwoopt{\citepads}[3][][]{\href{http://adsabs.harvard.edu/abs/#3}%
    {\def\hyper@linkstart##1##2{}%
     \let\hyper@linkend\@empty\citep[#1][#2]{#3}}}
  \newcommandtwoopt{\citetads}[3][][]{\href{http://adsabs.harvard.edu/abs/#3}%
    {\def\hyper@linkstart##1##2{}%
     \let\hyper@linkend\@empty\citet[#1][#2]{#3}}}
  \newcommandtwoopt{\citeyearads}[3][][]%
    {\href{http://adsabs.harvard.edu/abs/#3}
    {\def\hyper@linkstart##1##2{}%
     \let\hyper@linkend\@empty\citeyear[#1][#2]{#3}}}
\makeatother

\usepackage{color}

\def\Mo{M_{\odot}}

\usepackage{color}

\usepackage{cleveref}
\usepackage{multirow}
 
\newcommand\mH{m_\mathrm{H}}
\newcommand\Tstar{T_\ast}
\newcommand\Rstar{R_\ast}
\newcommand\rhos{\rho_\mathrm{s}}

\newcommand\gff{g_\mathrm{ff}}
\newcommand\gbf{g_\mathrm{bf}}
\newcommand\Ts{T_\mathrm{s}}
\newcommand\Rs{R_\mathrm{s}}
\newcommand\Ms{M_\mathrm{s}}
\newcommand\e{\mathrm{e}}

\newcommand{\Msolar}{\mbox{$\mathrm{M}_{\odot}\,$}}

\usepackage{chemist}

\begin{document}
\title{A thin shell of ionized gas explaining the IR excess of classical Cepheids}
\titlerunning{A thin shell of ionized gas explaining the IR excess of Cepheids}
\authorrunning{Hocd\'e et al. }
\author{V. Hocd\'e \inst{1} 
\and N. Nardetto \inst{1}
\and E. Lagadec \inst{1}
\and G. Niccolini\inst{1}
\and A. Domiciano de Souza\inst{1}
\and A. M\'erand \inst{2}
\and P. Kervella \inst{3}
\and A.~Gallenne\inst{1,4}
\and M.~Marengo\inst{5}
\and B.~Trahin\inst{3}
\and W.~Gieren\inst{7} 
\and G.~Pietrzy\'nski\inst{6}
\and S. Borgniet\inst{3}
\and L.~Breuval\inst{3}
\and B.~Javanmardi\inst{3}
}

\institute{Universit\'e Co\^te d'Azur, Observatoire de la C\^ote d'Azur, CNRS, Laboratoire Lagrange, France,\\
email : \texttt{vincent.hocde@oca.eu}
\and European Southern Observatory, Karl-Schwarzschild-Str. 2, 85748 Garching, Germany
\and LESIA (UMR 8109), Observatoire de Paris, PSL, CNRS, UPMC, Univ. Paris-Diderot, 5 place Jules Janssen, 92195 Meudon, France  
\and European Southern Observatory, Alonso de C\'ordova 3107, Casilla 19001, Santiago 19, Chile 
\and Department of Physics and Astronomy, Iowa State University, Ames, IA 50011, USA
\and Nicolaus Copernicus Astronomical Centre, Polish Academy of Sciences, Bartycka 18, PL-00-716 Warszawa, Poland
\and Universidad de Concepción, Departamento de Astronomía, Casilla 160-C, Concepción, Chile}

\date{Received ... ; accepted ...}
\abstract{Despite observational evidences, InfraRed (IR) excess of classical Cepheids are seldom studied and poorly understood, but probably induces systematics on the Period-Luminosity (PL) relation used in the calibration of the extragalactic distance scale.
} {This study aims to understand the physical origin of the IR excess found in the spectral energy distribution (SED) of 5 Cepheids~: RS~Pup ($P=41.46$d), $\zeta$~Gem ($P=10.15$d), $\eta$~Aql ($P=7.18$d), V~Cen ($P=5.49$d) and SU~Cyg ($P=3.85$d).}
 {A time series of atmospheric models along the pulsation cycle are fitted to a compilation of data, including  optical and near-IR photometry, \textit{Spitzer} spectra (secured at a specific phase), interferometric angular diameters, effective temperature estimates, and radial velocity measurements. \textit{Herschel} images in two bands are also analyzed qualitatively. In this fitting process, based on the SPIPS algorithm, a residual is found in the SED, whatever the pulsation phase, and for wavelengths larger than about $1.2\mu$m, which corresponds to the so-determined infrared excess of Cepheids. This IR excess is then corrected from interstellar medium absorption in order to infer or not the presence of dust shells, and is finally used in order to fit a model of a shell of ionized gas.} {For all Cepheids, we find a continuum IR excess increasing up to $\approx$-0.1 magnitudes at 30$\mu$m, which cannot be explained by a hot or cold dust model of CircumStellar Environment (CSE). 
However, a weak but significant dust emission at 9.7 $\mu$m is found for $\zeta$~Gem, $\eta$~Aql and RS~Pup, while clear interstellar clouds are seen in the \textit{Herschel} images for V~Cen and RS~Pup. We show, for the first time, that the IR excess of Cepheids can be explained by free-free emission from a thin shell of ionized gas, with a thickness of  $\simeq$15\% of the star radius, a mass of $10^{-9}-10^{-7}$\Msolar and a temperature ranging from 3500 to 4500K.}{The presence of a thin shell of ionized gas around Cepheids has to be tested with interferometers operating in visible, in the mid-IR or in the radio domain. The impact of such CSEs of ionized gas on the PL relation of Cepheids needs also more investigations.}

\maketitle

\section{Introduction}\label{s_Introduction}


Cepheids have been the keystone of distance scales determination in the Universe for a century, because their pulsation period correlates directly with their luminosity, through the Leavitt law \citep{leavitt08,1912HarCi.173....1L} also called the period-luminosity relation (hereafter PL relation). 

The recent one percent precision on the Large Magellanic Cloud distance \citep{LMC2019} has led to a new determination of the Hubble constant H$_0$ \citep{Riess2019}. Moreover, upcoming space and ground-based telescopes such as the  \textit{James Webb Space Telescope} (JWST) and the \textit{Extremely Large Telescope} (ELT) will make it possible to obtain light curves of extragalactic Cepheids up to one hundred  megaparsecs.
However, this distance ladder is still largely based on Cepheids PL relation whose uncertainties on both zero point and slope are today one of the largest contributors to the error on H$_0$ \citep{Riess2019}. One possible bias could be due to IR excesses from CSEs such as the ones discovered using near- and mid-infrared interferometry around nearby Cepheids \citep{kervella06a,merand06}. Indeed, if the brightness of CSEs is found to be significantly different in the Milky Way, the SMC, the LMC and in galaxies hosting SNIa due to metallicity effects for instance, then the use of an universal PL relation could introduce a bias in the distance scale calibration.

Envelopes around Cepheids have been discovered by long-baseline interferometry in the K-Band with VLTI and CHARA \citep{kervella06a,merand06}, and four Cepheids CSEs have been observed in the N band with VISIR and MIDI \citep{Kervella2009,gallenne13b} and one with NACO in the near-IR \citep{gallenne11}. The presence of a motionless H$\alpha$ absorption component using high-resolution spectroscopy around l~Car \citep{nardetto08a} has also been attributed to a CSE, and, recently, a CSE was detected in the visible domain with the VEGA/CHARA facility around $\delta$ Cep \citep{nardetto16}. These various studies determined a  CSE radius of around 3 stellar radii and a flux contribution in the K band, ranging from 2\% to 10\% of the continuum, for medium- and long-period Cepheids respectively, while it is around 10\% or more in the N band. However, we still do not know how these CSEs are produced, neither their nature, nor their characteristics (density and temperature profiles, chemical composition...).

This paper aims at building a phase-dependent Spectral Energy Distribution (SED) of a sample of Cepheids from visible to mid-IR wavelengths and compare it with dedicated atmospheric models in order to quantify and study their IR excess. We present the IR excess of the stars in the sample in Sect.~\ref{observation} using photometric and \textit{Spitzer}   observations in various bands, and we study qualitatively far-infrared images from \textit{Herschel}. In Sect. \ref{deredden} we correct the spectra from interstellar foreground absorption along the line-of-sight and seek for residuals at 9.7$\mu$m that could be due to a dusty CSE. In Sect. \ref{dust}, we use the radiative transfer code \texttt{DUSTY} \citep{DUSTY} to model the IR excess continuum, and show that such continuum cannot be due to a dusty CSE. In Sect. \ref{ionized}, we instead propose a model of CSE composed of ionized gas in order to reproduce the observed IR excess. Our results are discussed in Sect. \ref{Discussion} and we conclude in Sect.~\ref{Conclusion}

\section{Observations and data reduction of RS~Pup, $\zeta$~Gem, $\eta$~Aql, V~Cen, and SU~Cyg}\label{observation}

\subsection{Star sample}
In order to study the IR excess of Cepheids we selected a sample of Galactic Cepheids with \textit{Spitzer} observations. The combination of its sensitivity and wide IR wavelength coverage from 5$\mu m$ to 38$\mu m$, makes it the best tool to study dust features and IR excess continuum emission. We selected the five Cepheids having \textit{Spitzer} low resolution spectra with the highest signal-to-noise ratio 
:  RS~Pup ($P=41.46$d), $\zeta$~Gem ($P=10.15$d), $\eta$~Aql ($P=7.18$d) , V~Cen ($P=5.49$d) and SU~Cyg ($P=3.85$d). In addition, we retrieved \textit{Herschel} data at 70$\mu m$ and 160$\mu m$ to study larger and cold environment of dust. These space based observations have the advantage of not being biased by the Earth atmosphere, e.g. the entire flux is preserved and it is not hidden by ozone band which is essential to study dust spectral features. 

\subsection{Building the infrared excess using the SPIPS algorithm}\label{spips.photometry}
SpectroPhoto-Interferometric modeling of Pulsating Stars (SPIPS) is a model-based parallax-of-pulsation code which includes photometric, interferometric, effective temperature and radial velocity measurements in a robust model fit \citep{2015A&A...584A..80M}. To compute synthetic photometry to match those from the dataset SPIPS uses a grid of ATLAS9 atmospheric models\footnote{\url{http://wwwuser.oats.inaf.it/castelli/grids.html}} \citep{castelli2003} with solar metallicity and a standard turbulent velocity of 2km/s.

SPIPS models are available for $\eta$~Aql \citep{Merand2015}, $\zeta$~Gem \citep{breitfelder16} and RS~Pup \citep{kervella17}. We updated them with new datasets including Gaia photometry \citep{GaiaDR2} and we also applied the SPIPS algorithm to SU Cyg and V Cen. Fig. 1 to 5 in Appendix \ref{app:spips} show the results of the fitting process. For each star in the sample, SPIPS provides a fit of the photometry along the pulsation cycle, which is in agreement with the observations. While the fit is satisfactory from the visible domain to the near infrared, a disagreement is obtained for wavelengths larger than about 1.2$\mu$m. For $\lambda > 1.2\mu$m, the observed photometries ($m_\mathrm{obs}$) are indeed significantly brighter than the synthetic ones ($m_\mathrm{kurucz}$), which corresponds to an infrared excess. SPIPS does not physically model this IR excess, but instead takes it into account using an {\it ad-hoc} analytic power law $\mathrm{IR}_\mathrm{ex}$, which is defined as:

   
\begin{figure*}[htbp]
\centering
\begin{tabular}{cc}
\includegraphics[width=0.45\textwidth]{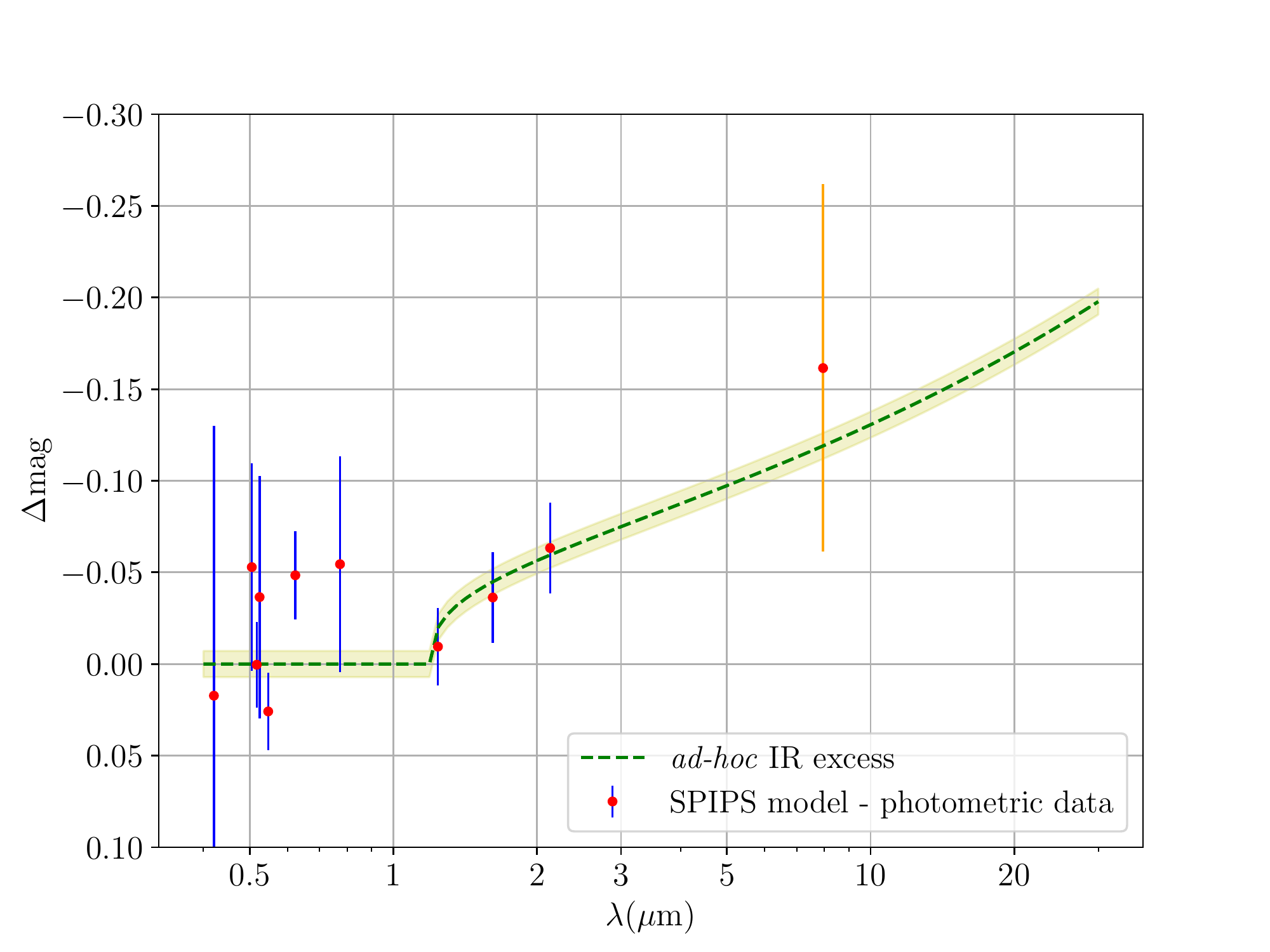}\\
(a) RS Pup, $\mathrm{IR}_\mathrm{ex}=-0.061(\lambda-1.2)^{0.350}\mathrm{mag}$
\end{tabular}
\begin{tabular}{cc}
\includegraphics[width=0.45\textwidth]{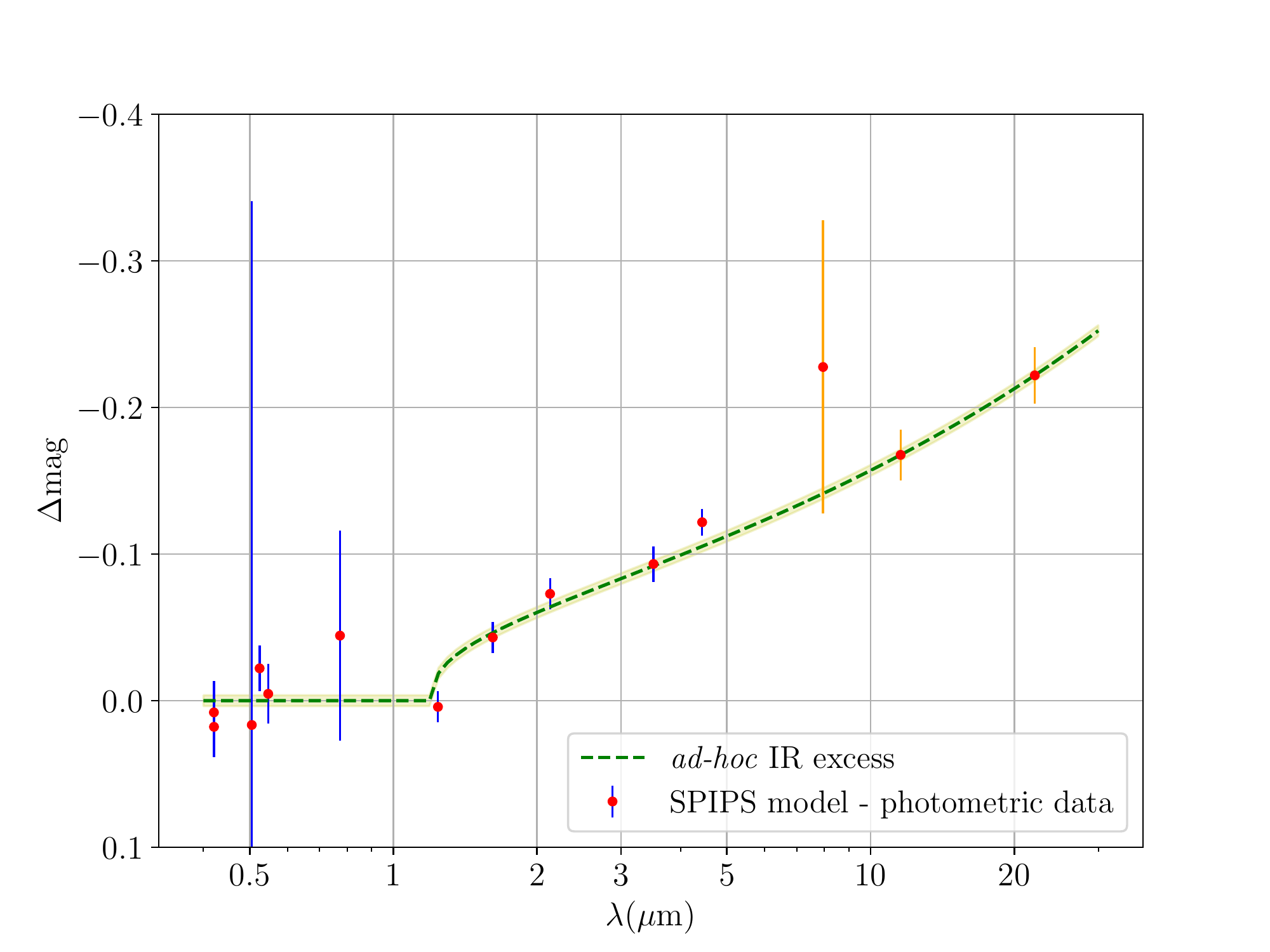}&
\includegraphics[width=0.45\textwidth]{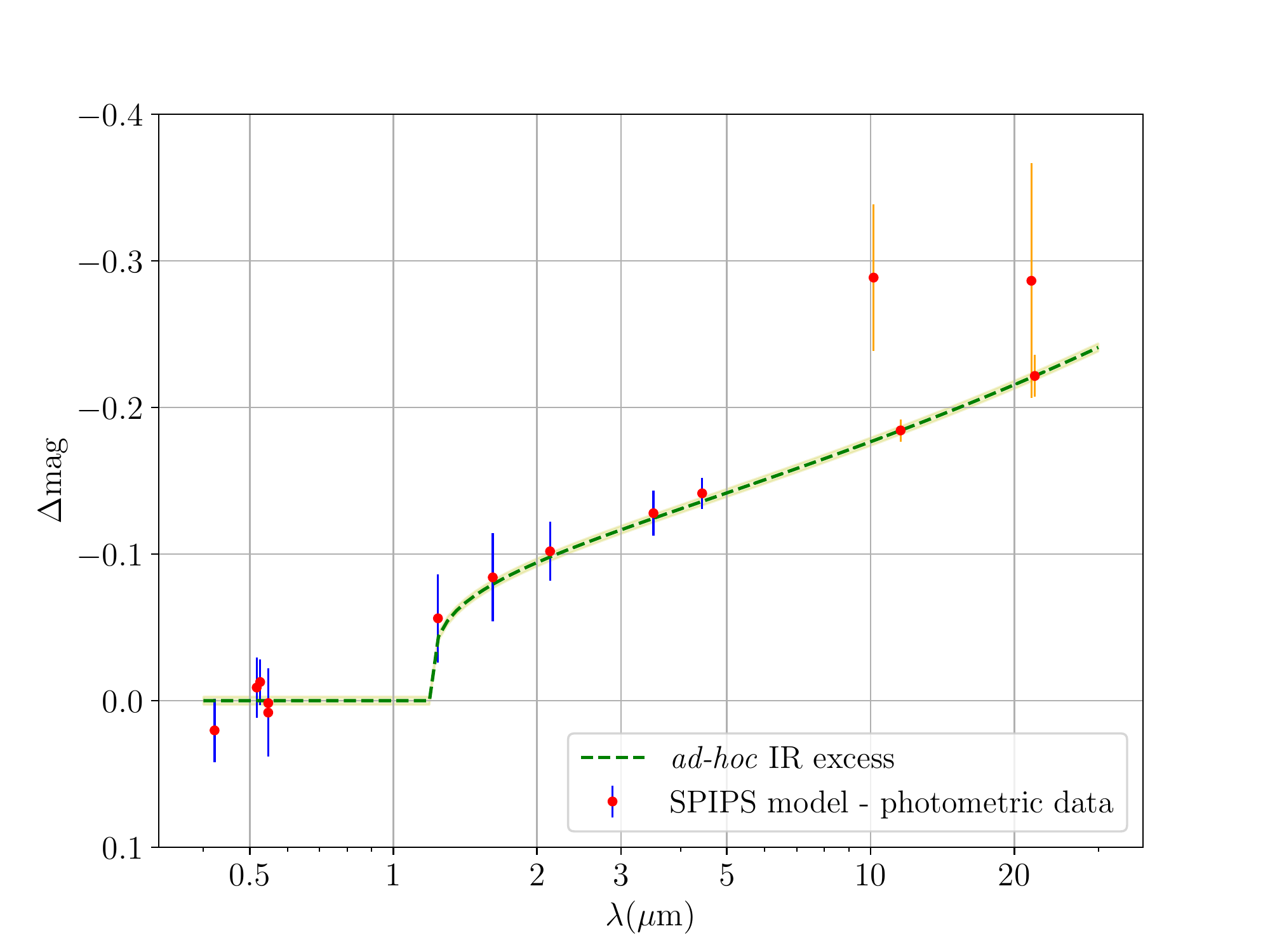}
\\
(b) $\zeta$ Gem, $\mathrm{IR}_\mathrm{ex}=-0.066(\lambda-1.2)^{0.400}\mathrm{mag}$& (c)  $\eta$ Aql, $\mathrm{IR}_\mathrm{ex}=-0.100(\lambda-1.2)^{0.262}\mathrm{mag}$ 
\end{tabular}

\begin{tabular}{cc}
\includegraphics[width=0.45\textwidth]{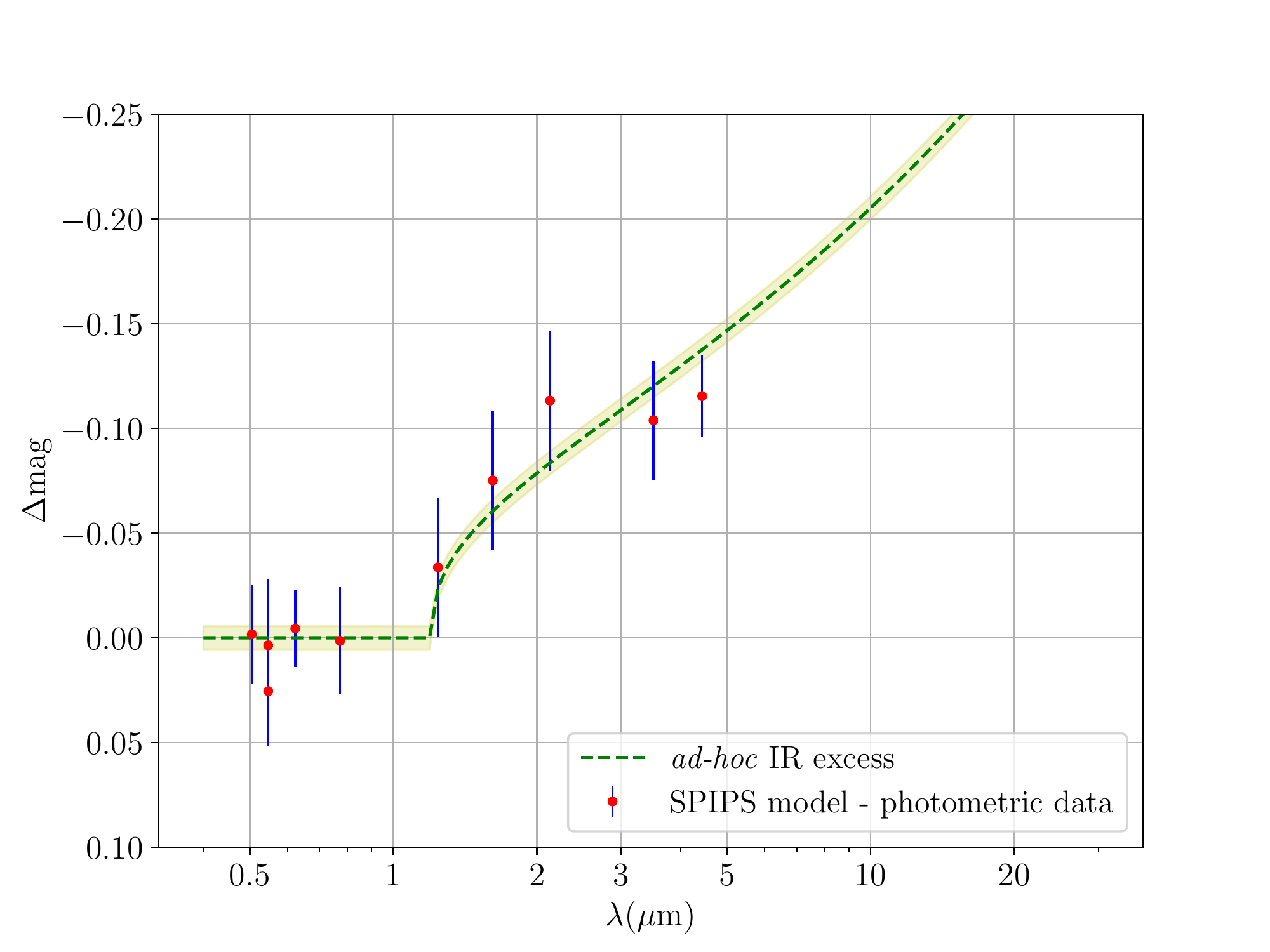} &
\includegraphics[width=0.45\textwidth]{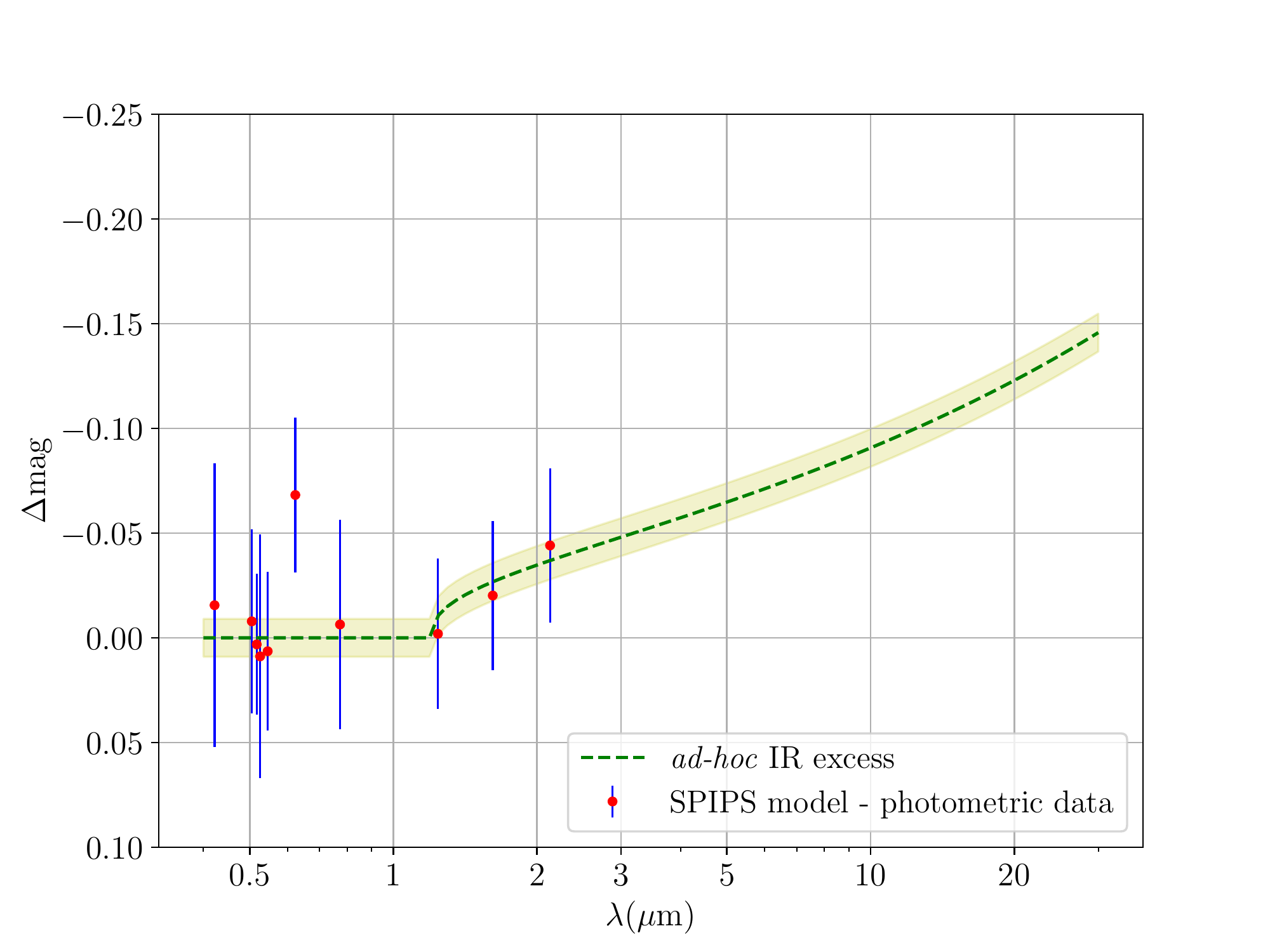}
\\
(d) V Cen, $\mathrm{IR}_\mathrm{ex}=-0.086(\lambda-1.2)^{0.400}\mathrm{mag}$& (e) SU Cyg, $\mathrm{IR}_\mathrm{ex}=-0.038(\lambda-1.2)^{0.400}\mathrm{mag}$ 
\end{tabular}

\caption{\small \label{spips.adhoc} IR excess analytic law for all star sample together with the measurements. IR excess as derived from the SPIPS algorithm and represented by {\it ad-hoc} analytic laws (Eq.~\ref{eq:mag}) for each star in the sample. For each photometric band, red dots with error bars are the mean excess value over the cycle of the Cepheid and the corresponding standard deviation. Red point with orange bars are photometric band discarded in this work (see Sect. \ref{spips.photometry}). The green zone is the error on the magnitude obtained using the covariance matrix of SPIPS fitting result.}
\end{figure*}

\begin{equation} \label{eq:mag}
\mathrm{IR}_\mathrm{ex}=\Delta \mathrm{mag} = m_\mathrm{obs} - m_\mathrm{kurucz}=\begin{cases} 0, & \mbox{for } \lambda < 1.2\mu \mathrm{m}  \\\alpha (\lambda - 1.2)^\beta, & \mbox{for } \lambda \geqslant 1.2 \mu \mathrm{m}  \end{cases}
\end{equation} 
with two parameters, $\alpha$ and $\beta$. 
These \textit{ad-hoc} laws take the visible domain as a reference, and thus assume that there is no excess nor deficit in this wavelength domain due to the CSE ($\Delta \mathrm{\textbf{m}}=0$ for $\lambda < 1.2 \mu m$). The choice of the \textit{ad-hoc} analytic power law in Eq.~(\ref{eq:mag}) was adopted originally in SPIPS modeling because it provides a reasonable description of the IR excess without changing the results of the fit.
 
In  Fig.~\ref{spips.adhoc}, we show the IR excess analytic law obtained for the star sample together with the measurements. These measurements correspond to the cycle-averaged magnitude difference $\Delta \mathrm{mag} = m_\mathrm{obs} - m_\mathrm{kurucz}$ (in a specific band), while the uncertainties are the corresponding standard deviations. Since these deviations over the cycle are quite small ($\approx 0.05$mag) from Fig.~\ref{spips.adhoc}, we conclude that the IR excess of Cepheids is not time dependent or, at most, slightly time dependent. 

In the fitting process described above, the photometric band corresponding to the carbon-monoxyde (CO) band-head at 4.6$\mu m$ has to be considered with a special care. Indeed,   CO can form at low temperature \citep{scowcroft2016} and this is indeed probably seen in IRAC I2 Spitzer bands for $\zeta$~Gem and V~Cen (see Figs. \ref{spips_fit_v_cen} and \ref{spips_fit_zet_gem}, respectively). In these bands, the agreement between the ATLAS9 model and the observations is in disagreement at certain phases. Even if ATLAS9 includes the modeling of CO molecules, these models are static and do not reproduce the dynamical structure of the atmosphere of Cepheids satisfactorily. However, the cycle-averaged infrared excess we obtain at this specific wavelength of 4.6$\mu m$ is consistent with the general trend of the IR excess law (see Fig.~\ref{spips.adhoc}), and is considered in the following of the analysis. 

Besides, the photometric bands longward 5 microns are problematic for two reasons. First their cycle coverage is poor i.e. one data point in each band with undetermined phase, this is the case for $\eta$~Aql, $\zeta$~Gem and RS Pup at $\lambda$>5$\mu$m (see for instance the A MSX band of RS~Pup in Fig.~\ref{rs_pup_spips}). Hence the reliability of the fitted ad-hoc IR excess model depending on pulsation phase is questionable. Secondly these bands are more sensitive to interstellar dust emission around the star than \textit{Spitzer} because of their lower spatial resolution. Thus we have decided to discard these photometric bands from the analysis. The measurements which are not considered are indicated by orange bars in Fig. \ref{spips.adhoc}.




\subsection{Spitzer data}\label{spitzer.data}
The spectroscopic observations were made with the InfraRed Spectrograph IRS \citep{2004ApJS..154...18H} onboard the \textit{Spitzer} telescope \citep{2004ApJS..154....1W} and the full spectra were retrieved from the CASSIS atlas \citep{CASSIS}. CASSIS identified the five stars as point-like sources therefore we retrieved the best flux calibrated spectrum obtained from optimal extraction method \citep{Lebouteiller2010}.
The \textit{Spitzer} dataset of this paper is presented in Table~\ref{Tab.cals}. Short-Low (SL) and Long-Low (LL) are IRS modules placed in the focal plane instrument providing low spectroscopic resolution (R=60 - 128) from 5.2 $\mu m$ to 38 $\mu m$. 
In this study we use low-resolution IRS spectra obtained with the SL and LL modules, for which aperture sizes together with covered spectral ranges are described in Table~\ref{Tab.modules}. 
\begin{table}[tbp]
\caption{\label{Tab.cals} \small The \textit{Spitzer} data set is presented with the Astronomical Observation Requests (AORs) used in this paper, the date of observation, the corresponding Modified Julian Date (MJD), and the pulsation phase ($\phi$). RS~Pup is from observation program number 50346 and others stars are from program number 485 \citep{ardila2010}.}
\begin{center}
\begin{tabular}{ccccc}
\hline
\hline
Object	&	AORs	&	Date & MJD & $\phi$	\\
\hline									
RS Pup	&	26021120	&		2009.05.04&	54955.160 & 0.938\\
$\zeta$ Gem	&	27579392	&			2008.11.11& 54781.833& 0.410\\	
$\eta$ Aql	&	27579136	&		2008.06.07&	54624.819 &0.408\\
V Cen	&	27566336	&				2008.10.01&	54740.445&0.960\\
SU Cyg	&	27592960	&				2008.11.06&	54776.575 &0.013\\
\hline																		\end{tabular}
\normalsize
\end{center}
\end{table}
\begin{table}[]
\caption{\label{Tab.modules} Properties of the \textit{Spitzer}/IRS AORs Short-Low (SL) and Long-Low (LL) modules used.}
\begin{center}
\begin{tabular}{ccccc}
\hline
\hline
Module	&	Aperture size ('')	&	Orders	&	$\lambda_{min}$-$\lambda_{max}$	\\
\hline											
SL	&		$3.7\times57$	&	1	&		7.4-14.5	\\
	&			&	2	&	5.2-7.7	\\
\hline											
LL	&		$10.7\times168$	&	1	&		19.5-38.0	\\
	&			&	2&	14.0-21.3	\\
\hline									
\end{tabular}
\normalsize
\end{center}
\end{table}

\begin{figure*}[]
\centering
\begin{tabular}{cc}
\includegraphics[width=0.45\textwidth]{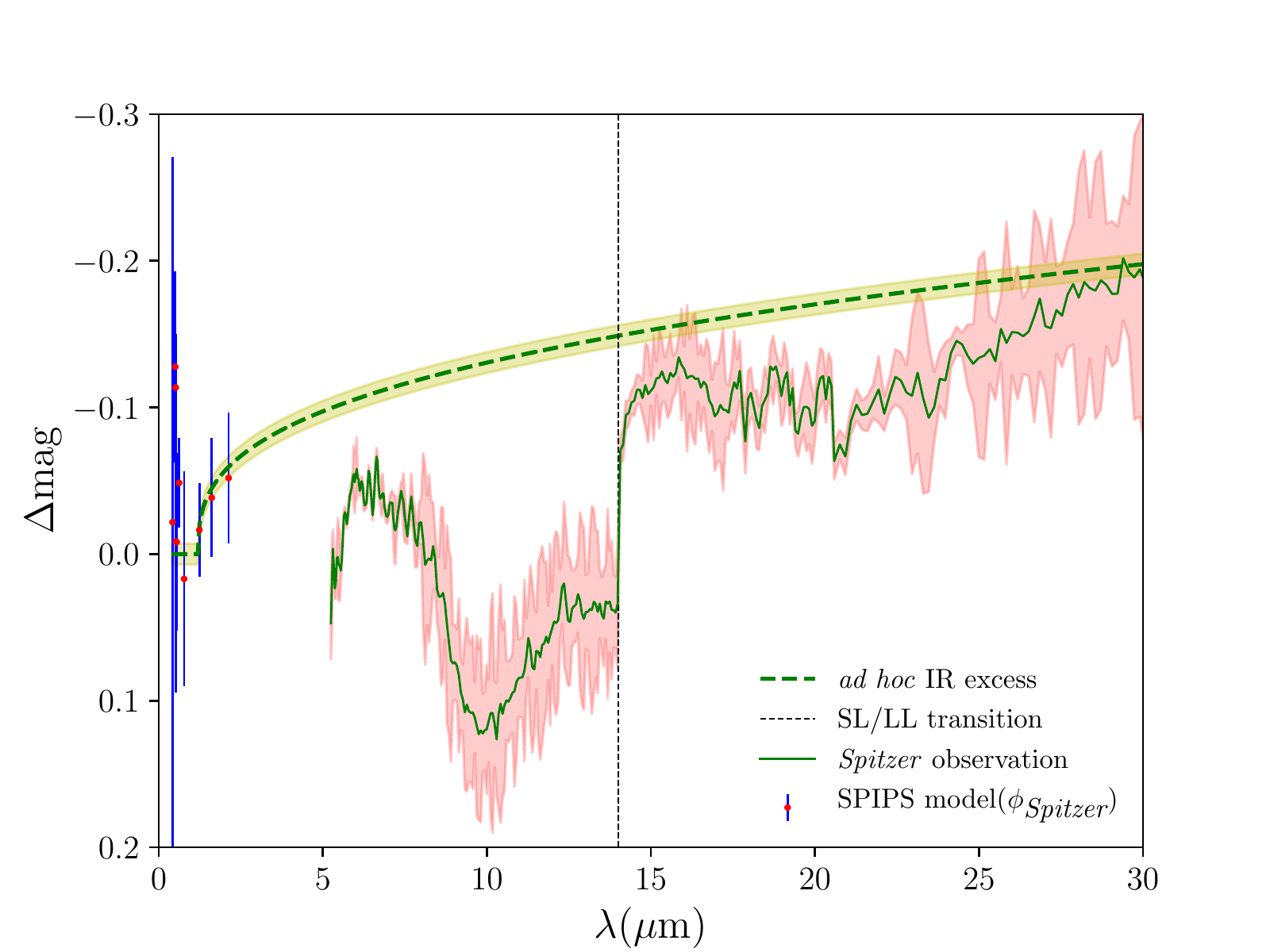}\\
(a) RS Pup, $\phi=0.938$
\end{tabular}
\begin{tabular}{cc}
\includegraphics[width=0.45\textwidth]{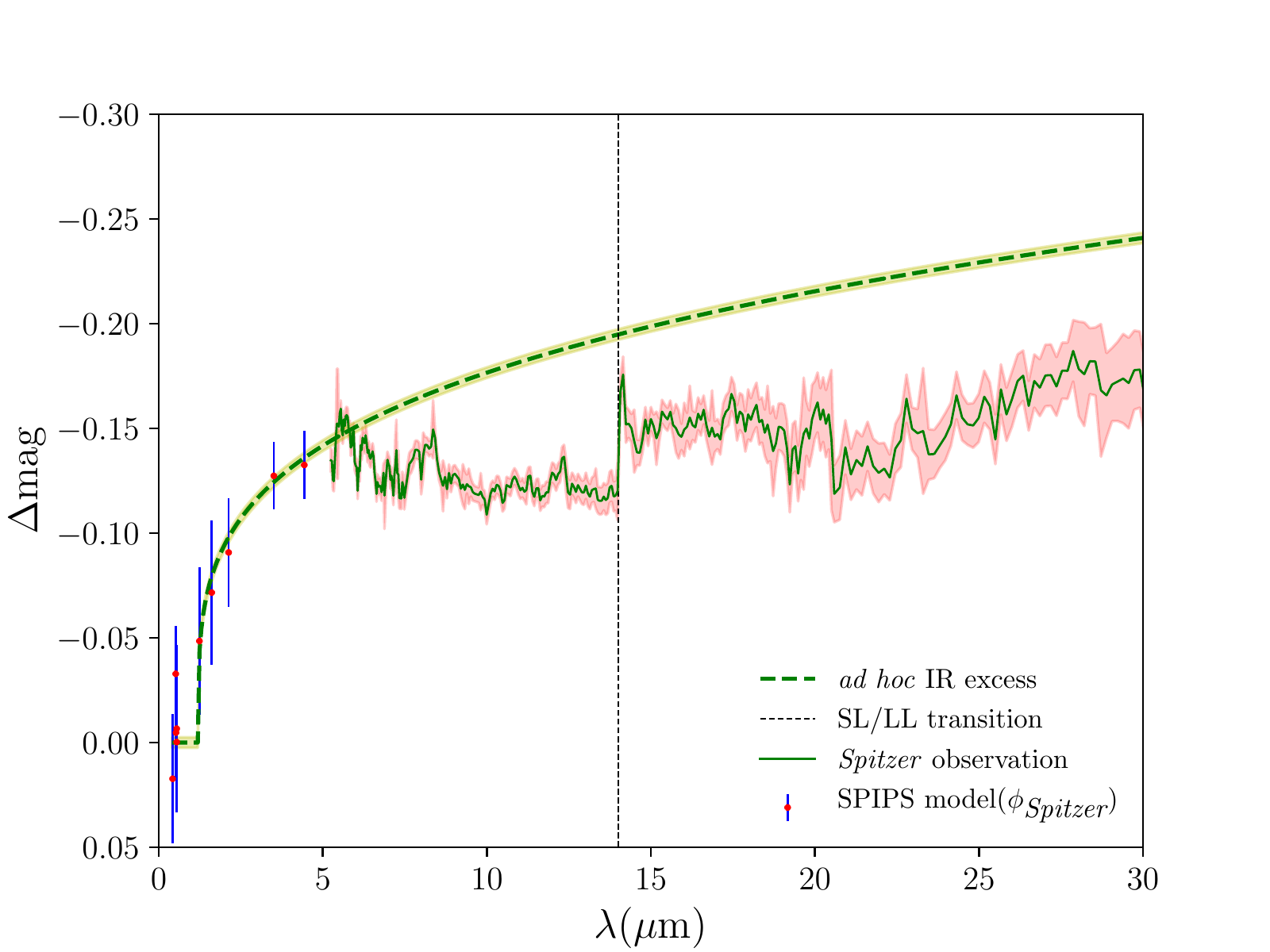} &
\includegraphics[width=0.45\textwidth]{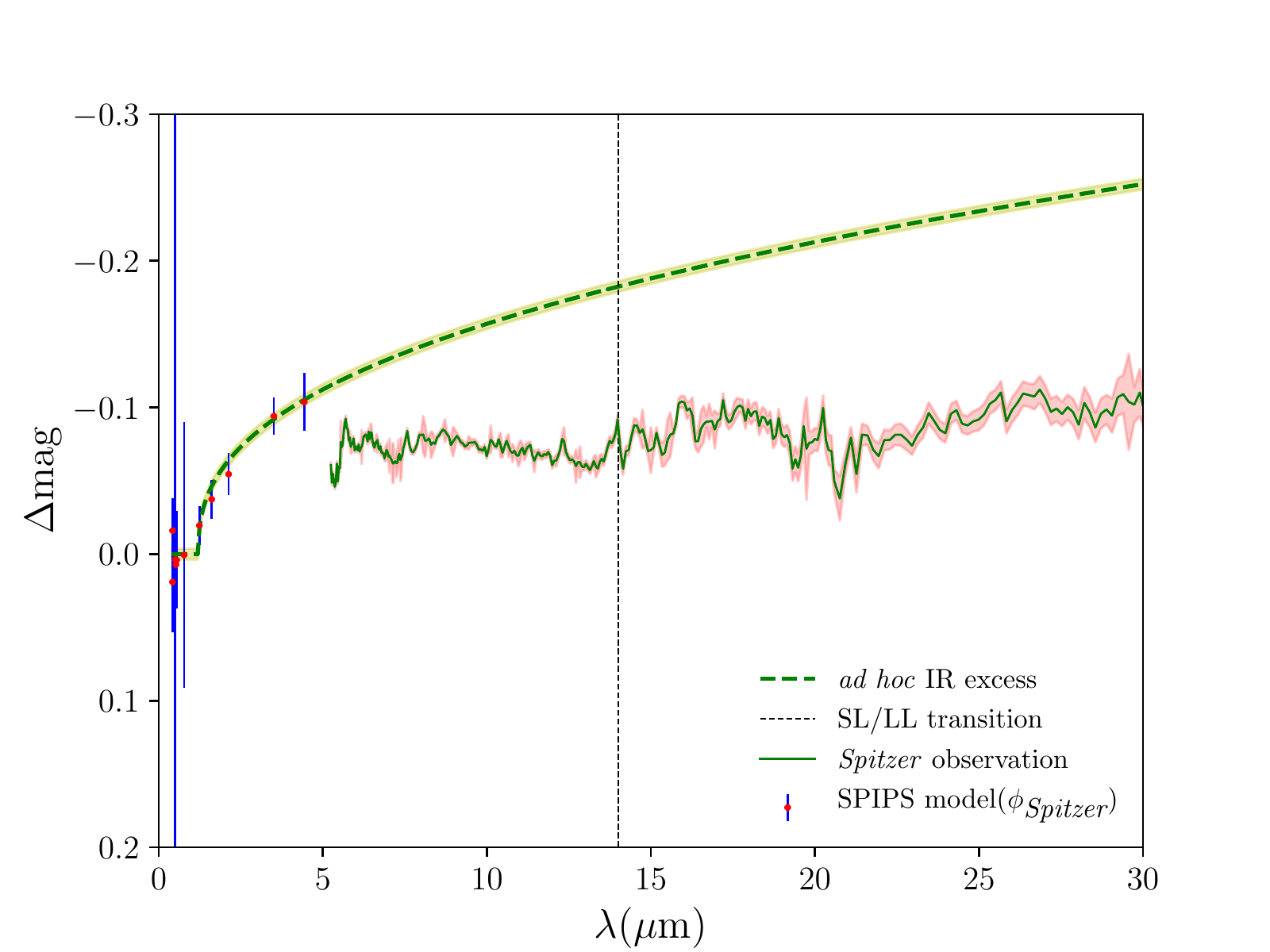} \\
 (b) $\zeta$ Gem, $\phi=0.410$ & (c) $\eta$ Aql, $\phi=0.408$
\end{tabular}
\begin{tabular}{cc}
\includegraphics[width=0.45\textwidth]{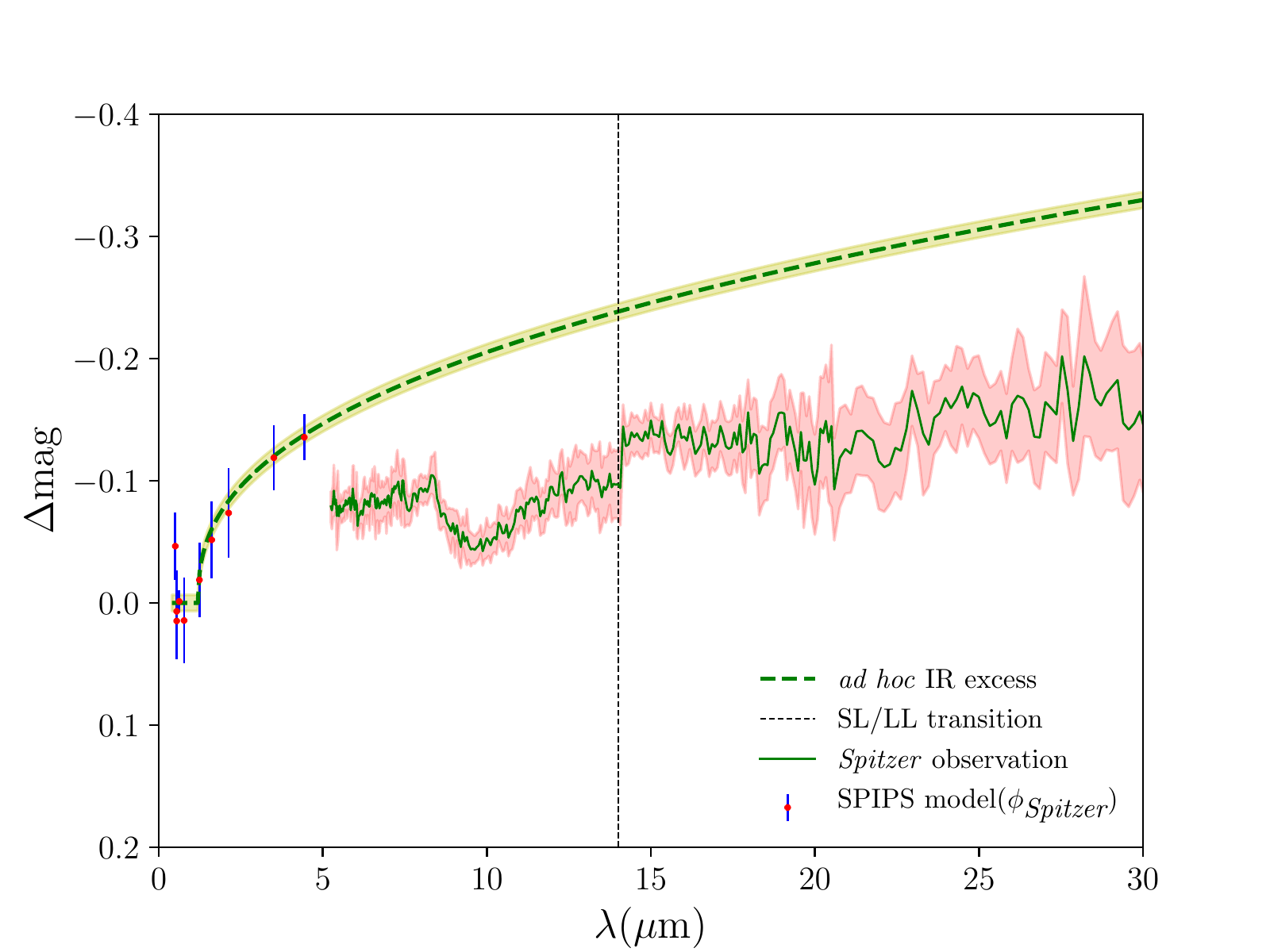} &
\includegraphics[width=0.45\textwidth]{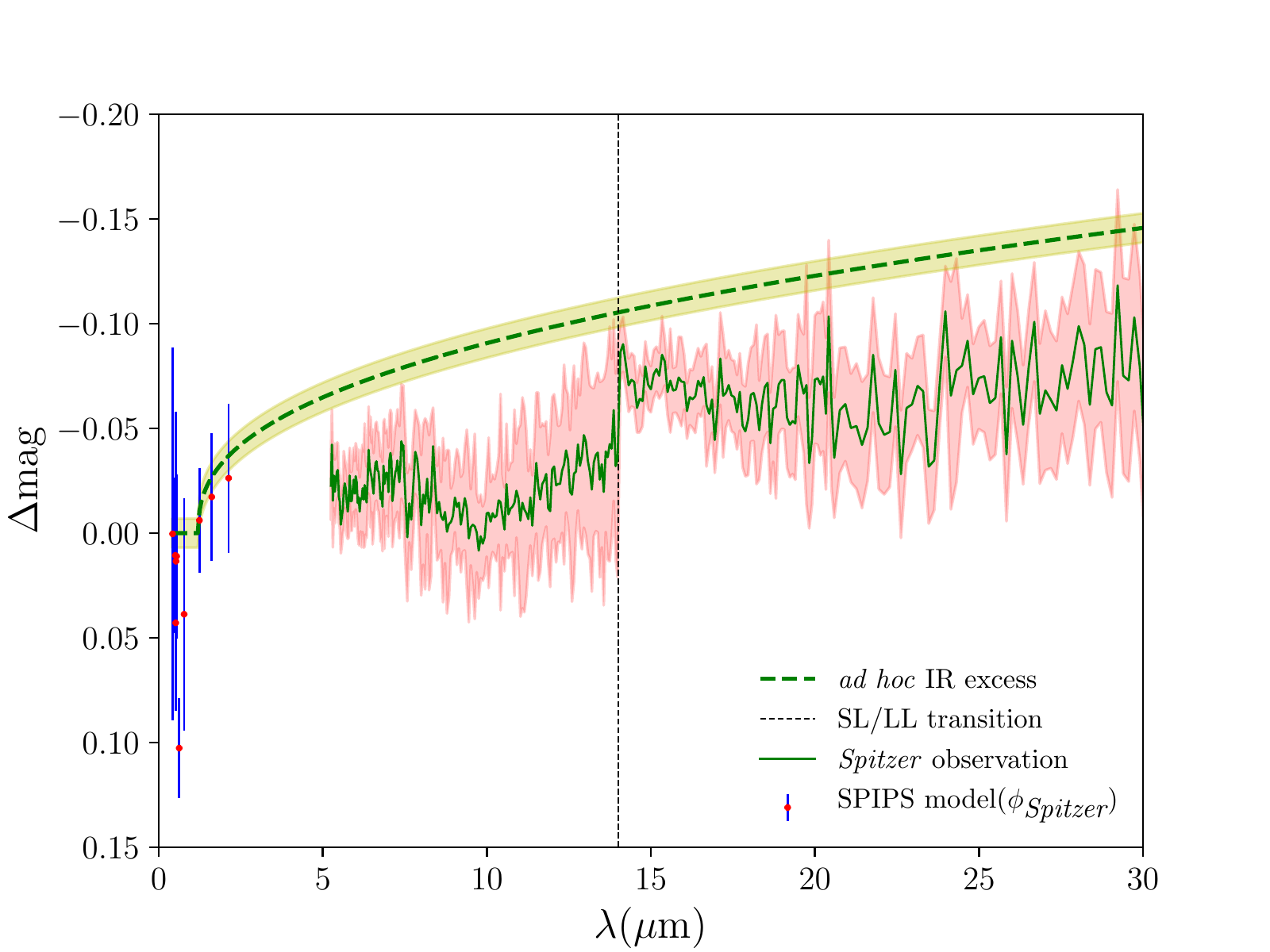}\\
(d) V Cen, $\phi=0.960$& (e) SU Cyg, $\phi=0.013$
\end{tabular}

\caption{\small \label{fig:data1} Photometric observations interpolated at the specific phases of \textit{Spitzer} data using the SPIPS algorithm. The transition between SL and LL detectors is indicated by a dashed line at $\approx 14 \mu m$. The cycle-averaged {\it ad-hoc} analytic laws from SPIPS are represented by a dashed green for comparison only. Pulsation phase $\phi$ of the \textit{Spitzer} observations is indicated in the caption of each panel.}
\end{figure*}

We derive the IR excess of each star in the sample at the specific phase of \textit{Spitzer} using:
\begin{equation}\label{eq:mag_spitzer}
\Delta \mathrm{mag} = m_\textit{Spitzer} - m_\mathrm{kurucz}[\phi_\textit{Spitzer}]
\end{equation}
where  $m_\textit{Spitzer}$ is the magnitude of the \textit{Spitzer} observation and $m_\mathrm{kurucz}[\phi_\textit{Spitzer}]$ is the magnitude of the ATLAS9 atmospheric model interpolated at the phase of \textit{Spitzer} observations ($\phi_\textit{Spitzer}$). The $T_{\mathrm{eff}}$ and log $g$ values of the star at the phase of \textit{Spitzer} are provided by the SPIPS algorithm, while the interpolation is then done in a ATLAS9 grid of models with steps of  250K in effective temperature and 0.5 in log $g$, respectively. The angular diameter derived by SPIPS is then used to calculate $m_\mathrm{kurucz}[\phi_\textit{Spitzer}]$. The corresponding stellar parameters are summarized in Table~\ref{Tab.param} and the $\Delta \mathrm{mag}$ values we finally obtain are shown by green dots and red uncertainties in Fig.~\ref{fig:data1}.   The same is done for all the photometric bands of observations, i.e. we do not use the average IR excess values shown in Fig. 1, but instead we recalculate the actual values corresponding to the phase of \textit{Spitzer} observations (red dot with blue uncertainties in Fig.~\ref{fig:data1}). Note that in Fig.~\ref{fig:data1}, the cycle averaged {\it ad-hoc} analytic IR excess law is shown only for comparison.
\begin{table}[tbp]
\caption{\label{Tab.param} \small Physical parameters of the stars derived by SPIPS at the phase of \textit{Spitzer} observations. $T_\mathrm{eff}$ and log $g$ are used to interpolate a ATLAS9 atmosphere model.} The limb-darkened angular diameter $\theta$ is then used to derived the observed flux. Temperature and angular diameter are in Kelvin and milliarcsecond respectively. Uncertainties on $T_\mathrm{eff}$ and $\theta$ are provided by SPIPS whereas uncertainty on log $g$ is arbitrarily set to 10\%.
\begin{center}
\begin{tabular}{ccccc}
\hline
\hline
	& $T_\mathrm{eff}(\phi_\textit{Spitzer})$ & log $g(\phi_\textit{Spitzer})$& $\theta(\phi_\textit{Spitzer}) $	\\
\hline									
RS Pup		& $5860^{+15}_{-15}$& $1.03^{+0.10}_{-0.10}$&$0.792^{+0.002}_{-0.002}$	 \\[0.15cm]
$\zeta$ Gem		&			$5178^{+7}_{-7}$& $1.52^{+0.15}_{-0.15}$& $1.699^{+0.002}_{-0.002}$\\[0.15cm]	
$\eta$ Aql		&$5543^{+6}_{-6}$&$1.68^{+0.17}_{-0.17}$&$1.784^{+0.002}_{-0.002}$	\\[0.15cm]
V Cen			&				$6368^{+15}_{-15}$&$1.89^{+0.19}_{-0.19}$&$0.510^{+0.001}_{-0.001}$	\\[0.15cm]
SU Cyg		&				$6781^{+17}_{-17}$&$2.03^{+0.20}_{-0.20}$&$0.353^{+0.001}_{-0.001}$	\\
\hline																		\end{tabular}
\normalsize
\end{center}
\end{table}

First, we correct several discontinuities in the \textit{Spitzer} spectra. Even if absolute uncertainties \citep{CASSIS,Sloan2015} are not excluded, we find a discontinuity at 14$\mu m$ between SL and LL spectra, indicated by a dashed line in Fig.~\ref{fig:data1}. According to CASSIS atlas a residual offset appears in the case of slightly extended sources \citep{CASSIS}, because more flux enters the larger LL aperture. Indeed the two \textit{Spitzer} detectors have indeed different Fields Of View (FOV, see Table~\ref{Tab.modules}) with different orientations on the sky (see Fig.~\ref{fig:herschel}), which explains the different level of flux measured. As an indication, the FOV of both detectors are overlaid on \textit{Herschel} images (see Fig.~\ref{fig:herschel} in Sect. 2.4). 

For RS~Pup, we obtain a jump of $7\%$ in flux between the SL and LL detectors, while for other stars we find a difference of around $2\%$ ($1.3\%$ in the case of $\zeta$~Gem). Extended dust emission has been discovered around RS~Pup \citep{Westerlund1961,mcalary1986}, so that this jump is not surprising. 
We correct these discontinuities at 14$\mu m$ in the spectra by scaling up SL flux (with the latter flux ratio) so that it corresponds to LL flux at 14 $\mu m$. This correction assumes that the LL flux calibrated by CASSIS pipeline is reliable and the emission of the environment is homogeneous between SL and LL apertures.

Also, orders mismatches can appear because of telescope pointing accuracy and causing discontinuity in the spectra between 2 adjacent orders. We corrected this feature in SU~Cyg and $\zeta$~Gem spectra at 7.5 $\mu m$ between SL orders 1 and 2 by scaling up the lower part. Note also that the \textit{Spitzer} data at larger wavelengths than 30$\mu$m were not considered due to their extremely large uncertainties. 

Second, for the five stars in the sample, we find an agreement between the level of IR excess at $5\mu$m in all photometric bands  ($\lambda < 5\mu$m) and in the \textit{Spitzer} spectroscopic observation ($\lambda > 5\mu$m). This is verified at the 0.05 magnitude level, and before the correction of different \textit{Spitzer} discontinuities (see Fig. \ref{fig:dereddened} for a detailled analysis of corrected IR excess). 
The agreement between IR excesses from SPIPS models and \textit{Spitzer} data shows the consistency of this approach. It  reveals that Cepheids exhibit a continuum IR excess from about 2$\mu$m up to 30$\mu$m in the star sample. This result is different from \cite{Schmidt2015} who found no IR excess for short-period classical Cepheids (see Discussion \ref{sect:6.1}). We find however that the \textit{ad-hoc} analytic laws of SPIPS overestimate the IR excess in the Spitzer wavelength domain (see Fig. \ref{fig:data1}).

Third, RS~Pup, SU~Cyg and V~Cen present unambiguous silicates absorption bands at $9.7\mu m$ and $20\mu m$. The latter absorption is known to peak usually at 18$\mu m$ however it can be shifted to longer wavelength depending on various conditions such as temperature or minerals proportions into silicates \citep{koike2006,henning1997}. These silicate absorptions are likely due to the presence of interstellar clouds on the line of sight.

\subsection{\textit{Herschel} images \label{sect.herschel}}
We retrieved \textit{Herschel} images from the PACS instrument at $70$ and $160\mu m$ to study the presence of extended environment around cepheids. No data have been found for SU~Cyg. \textit{Herschel} observation products are ordered according to the level of the data processing ranging from raw data (level 0) to highly processed scientific data (level 3). All the images in the sample are highly processed data with a level of 2.5. The FOV of \textit{Herschel} is of 8' and its spatial resolution is of ~7'' at 100$\mu m$. As an indication, we performed aperture photometry measurements of the flux within 1' in order to include most of the emission due to the environment of the cepheids. Then we derived the IR excess at $70$ and $160\mu m$ by comparing with atmospheric model in the same way than Eq. \ref{eq:mag_spitzer}. 

\begin{figure}[]
    \centering
    \begin{subfigure}[b]{0.48\textwidth}
        \centering
        \includegraphics[width=\textwidth]{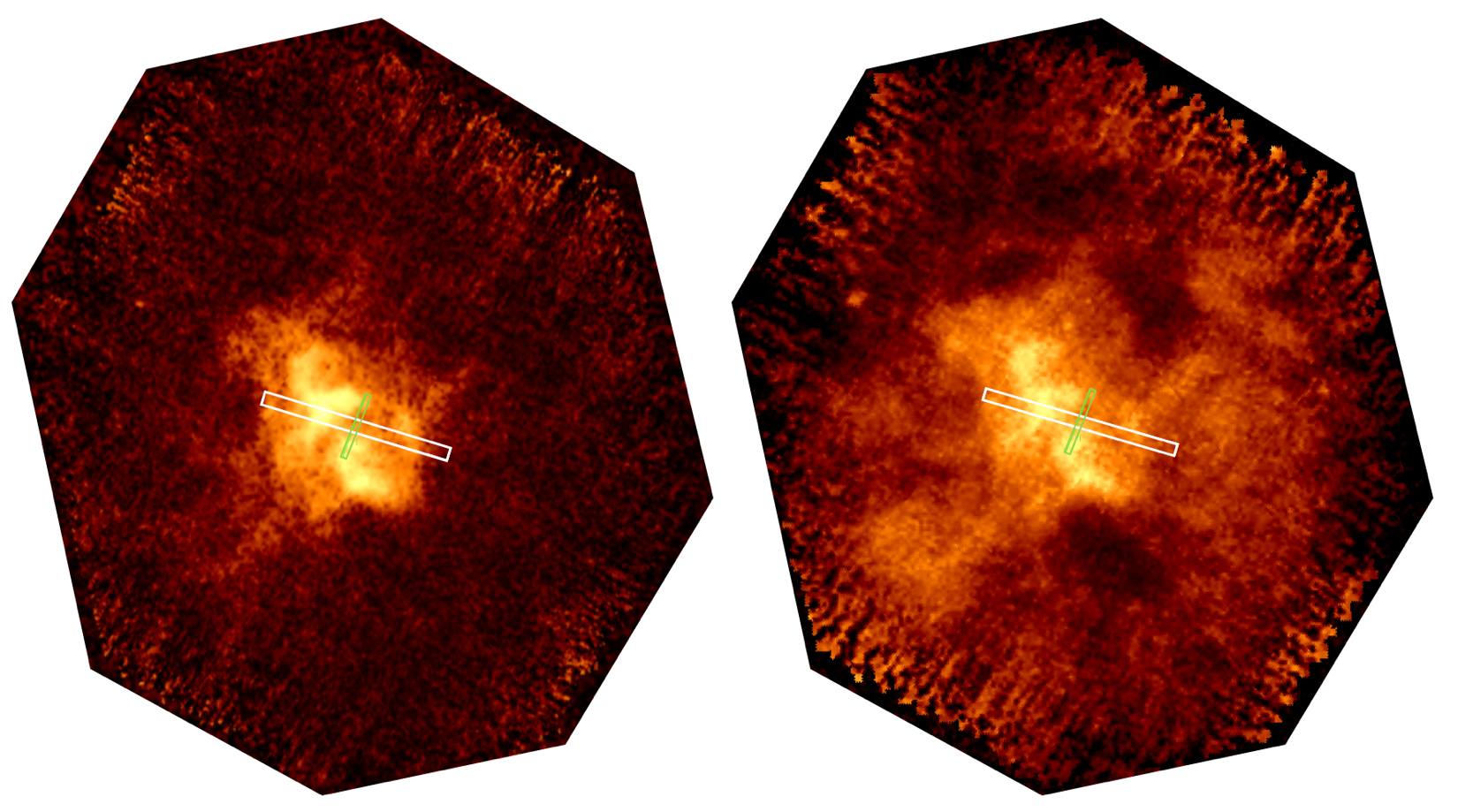}
        \caption[Network2]%
            {{\small RS~Pup}}  
            \vspace{-0.3cm}
            \label{fig:rs_pup_herschel}
        \end{subfigure}
        \vskip\baselineskip
        \begin{subfigure}[b]{0.49\textwidth}   
            \centering 
            \includegraphics[width=\textwidth]{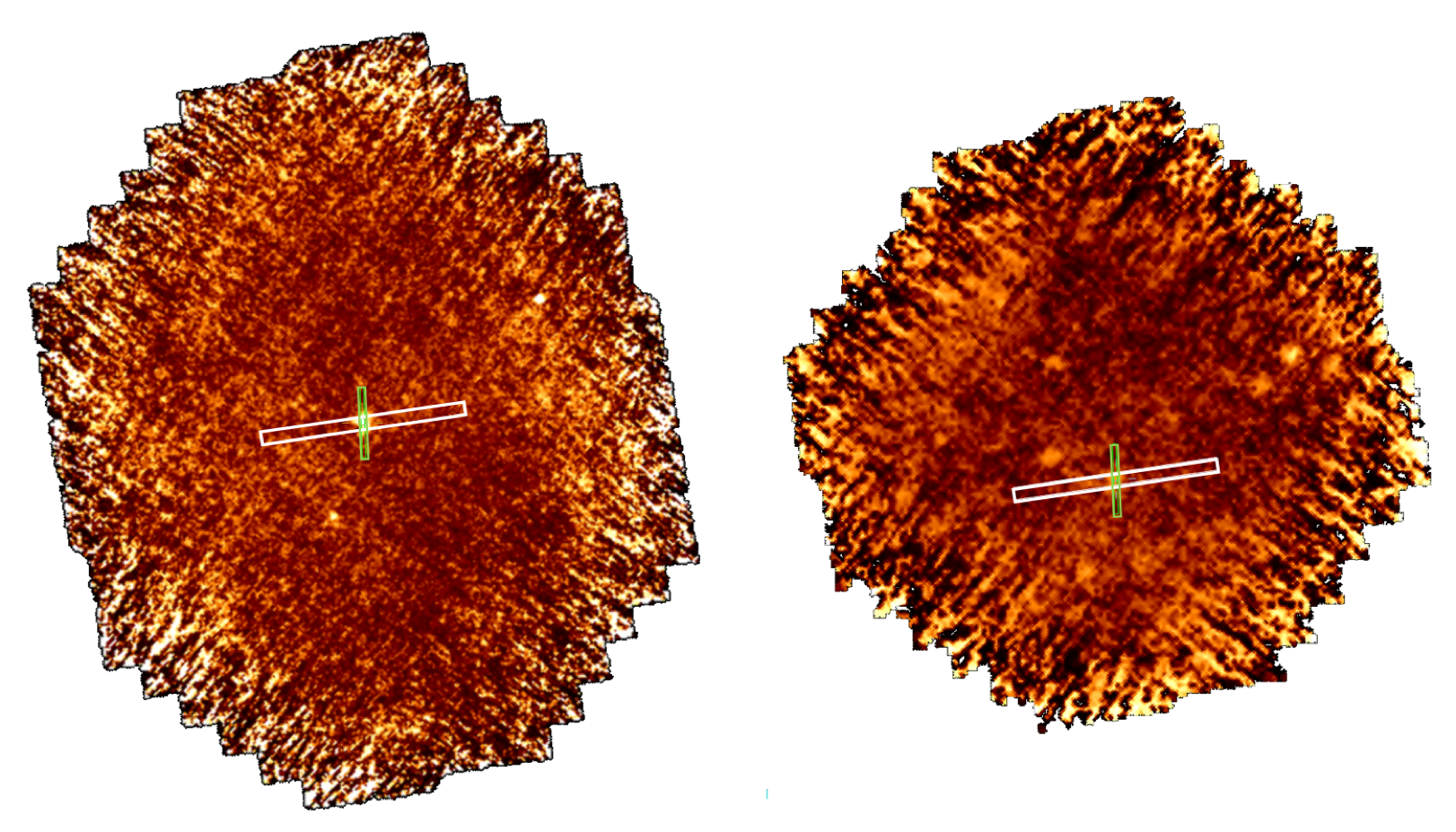}
            \caption[]%
            {{\small $\zeta$~Gem}}  
                     \vspace{-0.3cm}
            \label{fig:zet_gem_herschel}
        \end{subfigure}
        \vskip\baselineskip
        \begin{subfigure}[b]{0.49\textwidth}   
            \centering 
            \includegraphics[width=\textwidth]{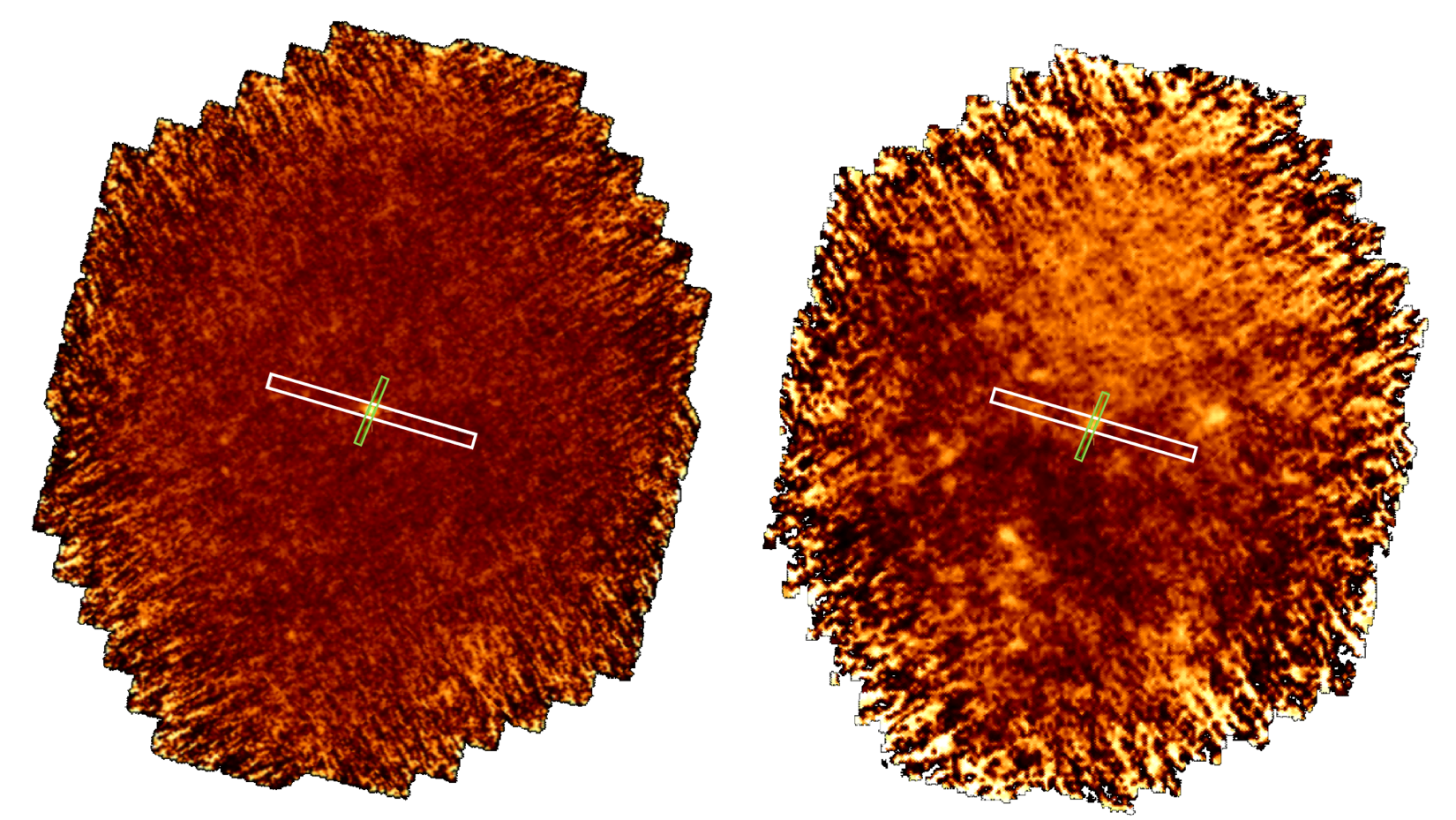}
            \caption[]%
            {{\small $\eta$~Aql}}
                     \vspace{-0.3cm}
            \label{fig:eta_aql_herschel}
        \end{subfigure}
         \vskip\baselineskip
        \begin{subfigure}[b]{0.49\textwidth}  
            \centering 
            \includegraphics[width=\textwidth]{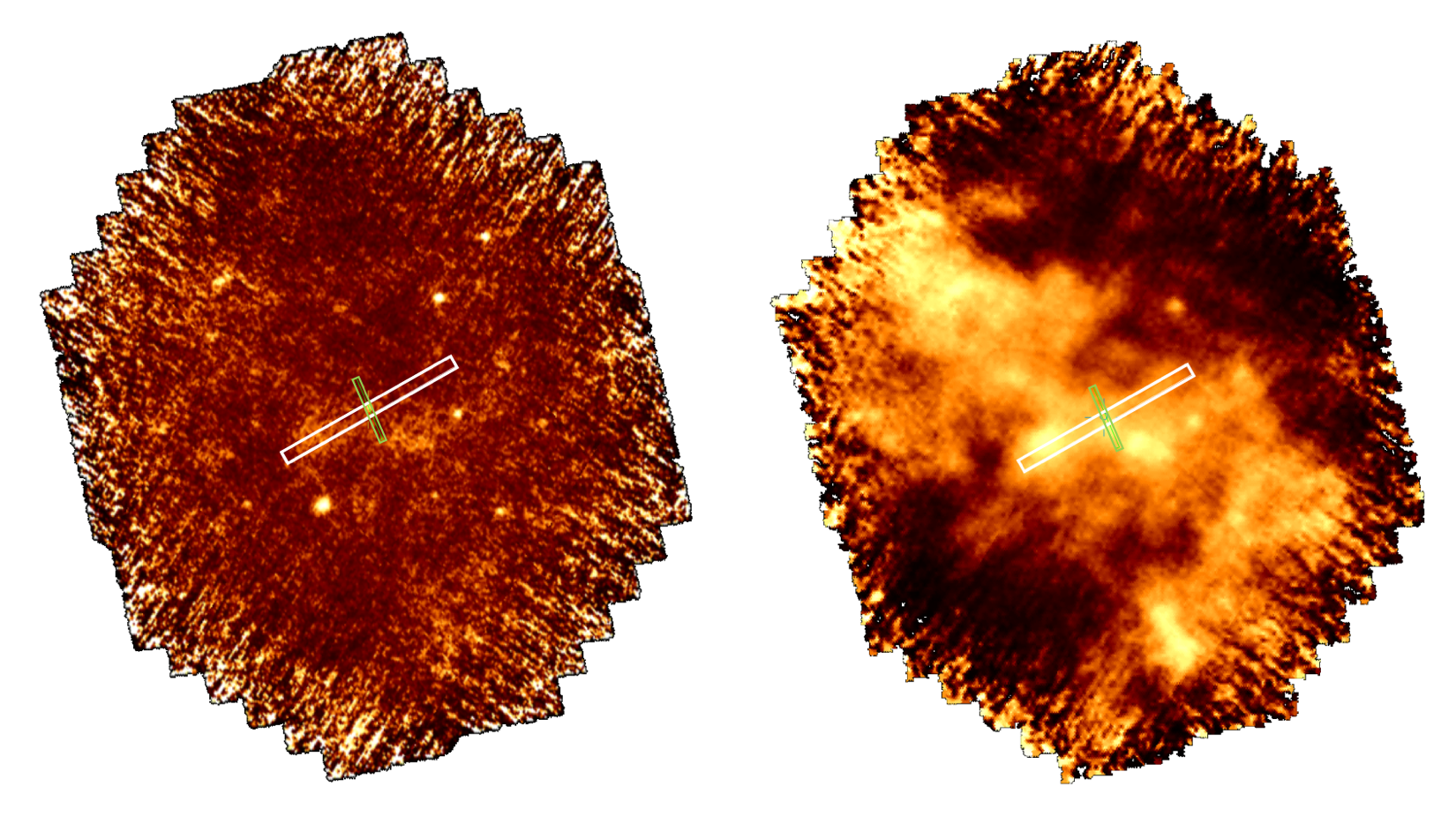}
            \caption[]%
            {{\small V~Cen}}  
                     \vspace{-0.3cm}
            \label{fig:v_cen_herschel}
        \end{subfigure}
        \vskip\baselineskip

        \caption[]
        {\small \label{fig:herschel} \textit{Herschel} images of cepheids and overlaid \textit{Spitzer} FOV. Left and right panels correspond to $70$ and $160\mu m$ respectively. The FOV is 8' large. Short and long green rectangles are SL and LL modules in the \textit{Spitzer} focal plane detector respectively.} 
        \label{fig:models}
    \end{figure}

Observations are presented in Fig.~\ref{fig:herschel} and the derived photometry in Table~\ref{Tab.herschel}. 
From these data we observe important IR excesses at 70$\mu m$ and 160$\mu m$, which are associated to cold environments with temperatures of about 40K and 20K respectively according to Wien's displacement law ($\lambda T = 2900\mu \mathrm{m}.\mathrm{K}$). Cold dusty material around RS Pup  was already observed, while the large cloud observed around V~Cen (see Fig. \ref{fig:v_cen_herschel}) at 160$\mu m$ could be related to a star forming region, since V~Cen is likely a member of the open cluster NGC 5662 \citep{turner1982,claria1991,anderson2013}. \textit{Herschel} is sensitive not only to emission from the star itself and a potential CSE, but also from cold extended emission from the interstellar medium. We thus do not consider the \textit{Herschel} photometry in the following of the paper.

Interestingly, we find a qualitative correlation between silicate absorption features in \textit{Spitzer} data and the presence of extended emission in the \textit{Herschel} images. For instance, both RS~Pup and V~Cen show at the same time strong silicate absorptions and extended emission in the \textit{Herschel} data (see \cref{fig:rs_pup_herschel,fig:v_cen_herschel}), while $\eta$~Aql and $\zeta$~Gem present no obvious silicate absorption and weaker interstellar environment (\cref{fig:eta_aql_herschel,fig:zet_gem_herschel}). 

\begin{table}
\caption{\label{Tab.herschel} \small \textit{Herschel} data set. $F_{70}$ and $F_{160}$ are the flux in Jansky at $70$ and $160\mu m$ respectively integrated within a 1' circle centered on the star. $\Delta \mathrm{mag_{70}}$ and $\Delta \mathrm{mag_{160}}$ are the corresponding IR excess derived following the definition of Eq.~(\ref{eq:mag_spitzer}) but for \textit{Herschel} observations.}
\begin{center}
\begin{tabular}{ccccc}
\hline
\hline
	&	RS Pup	&$\zeta$ Gem &$\eta$ Aql &	V Cen			\\
\hline									
$F_{70}$(Jy)	&11.5$^{+3.4}_{-3.4}$		&	0.7$^{+0.8}_{-0.8}$& 0.7$^{+0.8}_{-0.8}$ & 0.9$^{+0.9}_{-0.9}$				\\[0.15cm]
$F_{160}$(Jy)	&6.7$^{+2.6}_{-2.6}$		&	0.4$^{+0.6}_{-0.6}$	&	0.4$^{+0.6}_{-0.6}$&	5.2$^{+2.3}_{-2.3}$	\\[0.15cm]
\hline
$\Delta \mathrm{mag_{70}}$	&	-6.5$^{+0.3}_{-0.3}$ &-1.9$^{+1.2}_{-1.2}$ 	& -1.7$^{+1.2}_{-1.2}$	&-4.6$^{+1.1}_{-1.1}$	\\[0.15cm]
$\Delta \mathrm{mag_{160}}$	&   -7.7$^{+0.4}_{-0.4}$ &-3.1$^{+1.6}_{-1.6}$  & -2.9$^{+1.6}_{-1.6}$	&-8.3$^{+0.5}_{-0.5}$	\\[0.15cm]
\hline			
\end{tabular}
\normalsize
\end{center}
\end{table}

\section{Correcting the interstellar silicate absorption in \textit{Spitzer} data}\label{deredden}
Absorption of silicates at 9.7$\mu m$ is easily identified in the {\it Spitzer} spectra of RS~Pup, V~Cen and SU~Cyg whereas it is not observed in $\eta$~Aql and $\zeta$~Gem data (see Fig. \ref{fig:data1}). Such absorption has in principle two components, an emission due to a CSE close to the star (if present), which is then absorbed by interstellar environment. Thus, the silicate absorption observed by {\it Spitzer} $A^\mathrm{\textit{Spitzer}}_{9.7}$, if it is corrected from the expected silicate absorption from the ISM ($A^\mathrm{ISM}_{9.7}$), can indicate, in case of residual emission, whether there is a silicate emission from the CSE $E^\mathrm{CSE}_{9.7}$ or not. We thus have the following relation:
\begin{equation}\label{Ecse}
A^\mathrm{\textit{Spitzer}}_{9.7} - A^\mathrm{ISM}_{9.7}  = E^\mathrm{CSE}_{9.7}   
\end{equation}
 All values are magnitudes  at 9.7$\mu m$. We will now determine $A^\mathrm{\textit{Spitzer}}_{9.7}$ (Sect. \ref{sili.obs}) and $A^\mathrm{ISM}_{9.7}$ (Sect. \ref{sili.true}) in order to estimate the CSE emission at 9.7$\mu m$ (Sect. \ref{sect:3.3}) and then, we will extend our correction to the whole wavelength range of  \textit{Spitzer} (Sect. \ref{sect:ext_corr}).

\subsection{Quantifying the observed silicate absorption from \textit{Spitzer}}\label{sili.obs}
\begin{figure}[htp]
    \centering
    \begin{subfigure}[b]{0.36\textwidth}
        \centering
        \includegraphics[width=\textwidth]{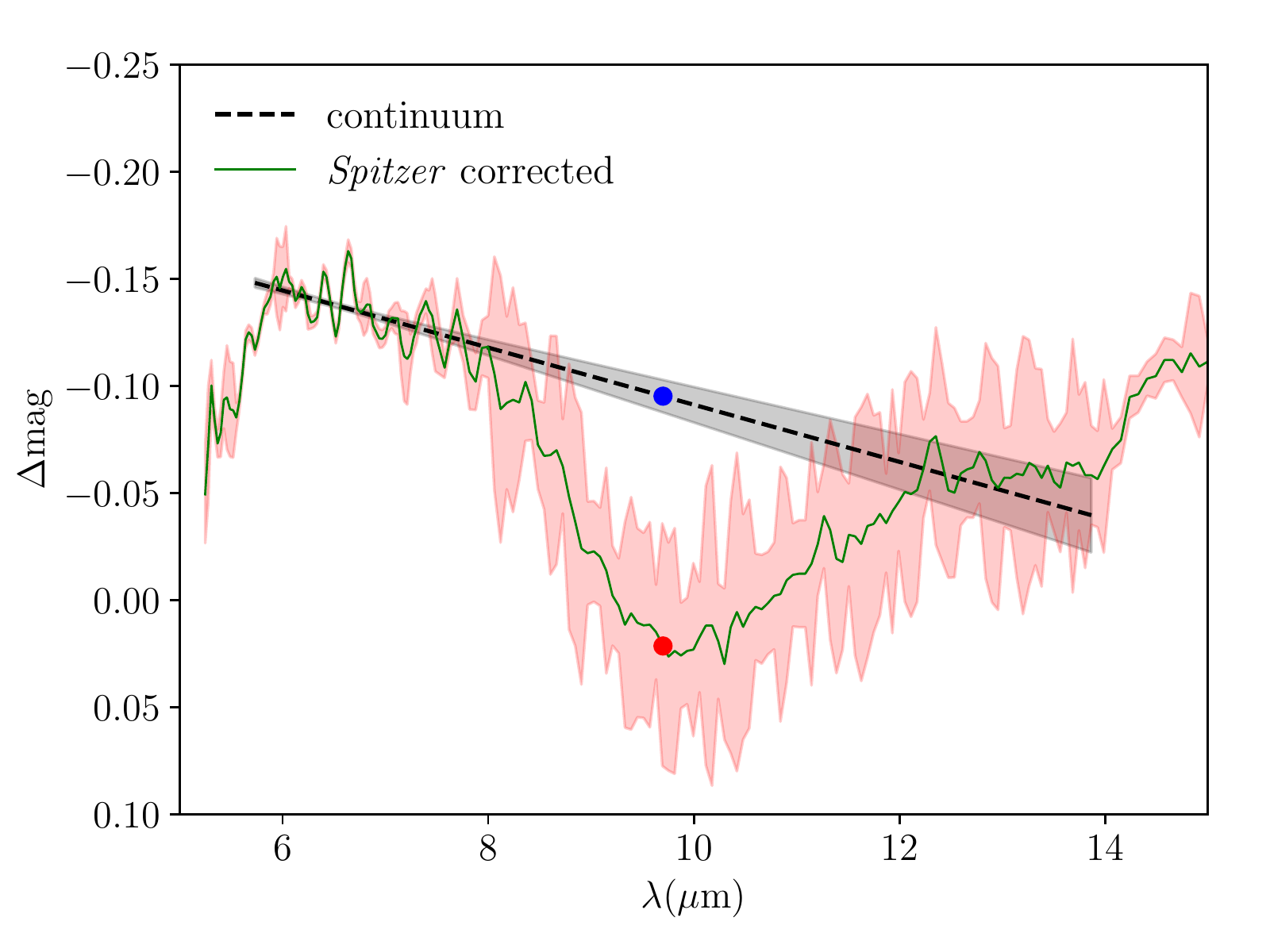}
        \caption[Network2]%
            {{\small RS~Pup}}    
            \label{fig:rs_model}
        \end{subfigure}
        \hfill
                \begin{subfigure}[b]{0.36\textwidth}   
            \centering 
            \includegraphics[width=\textwidth]{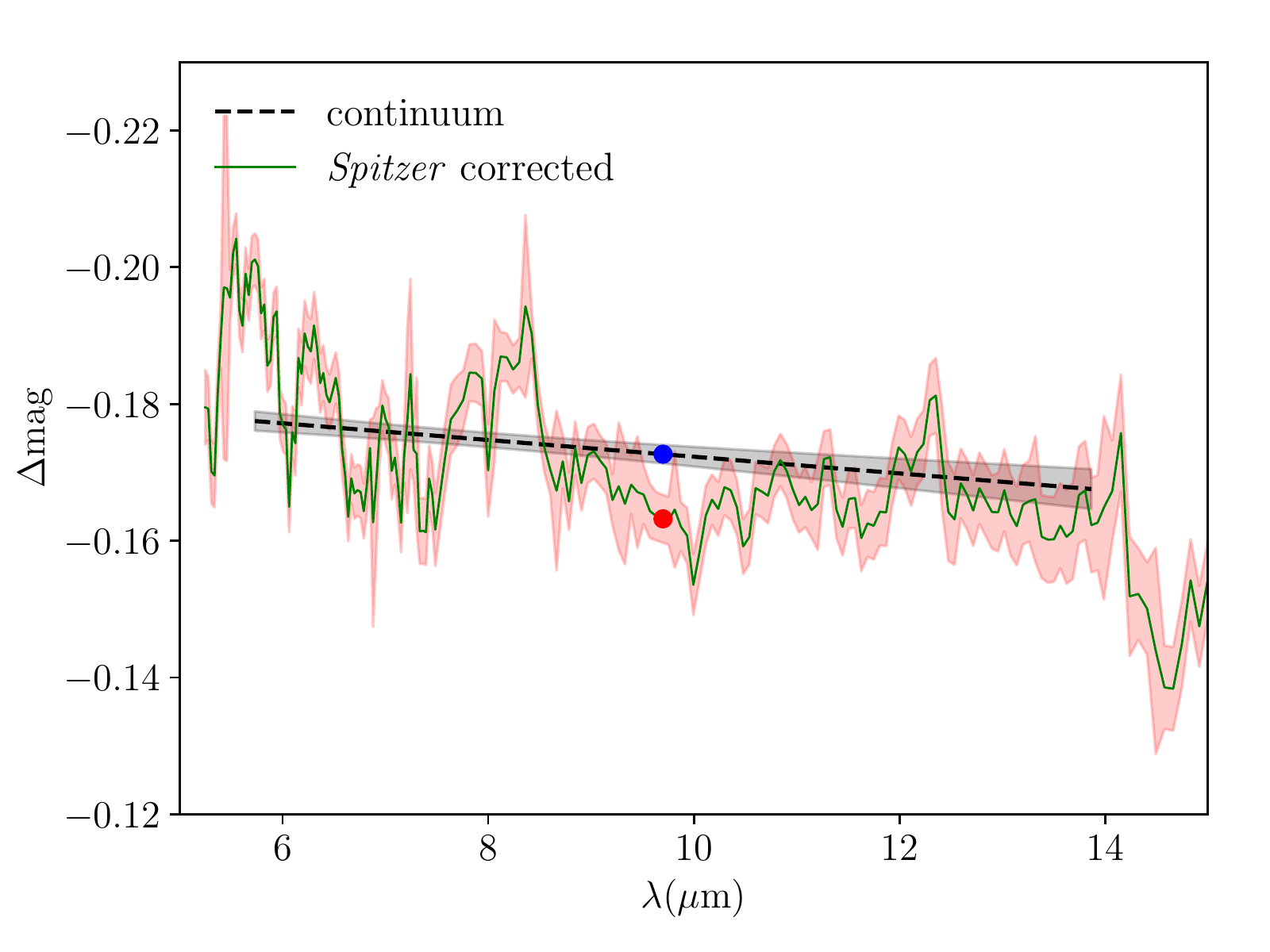}
            \caption[]%
            {{\small $\eta$~Aql}}    
            \label{fig:eta_model}
        \end{subfigure}

        \begin{subfigure}[b]{0.36\textwidth}  
            \centering 
            \includegraphics[width=\textwidth]{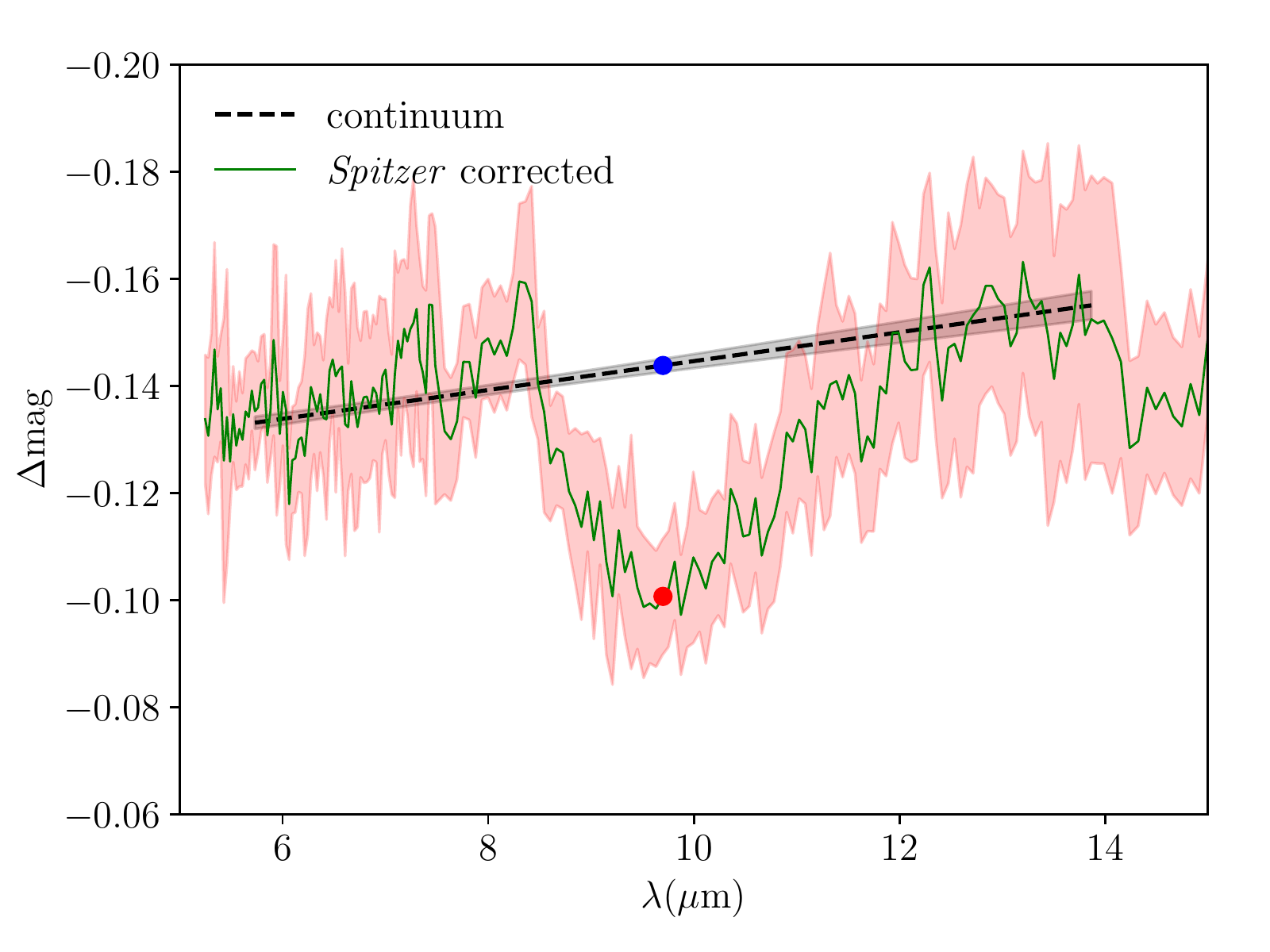}
            \caption[]%
            {{\small V~Cen}}    
            \label{fig:v_cen_model}
        \end{subfigure}

        \begin{subfigure}[b]{0.36\textwidth}   
            \centering 
            \includegraphics[width=\textwidth]{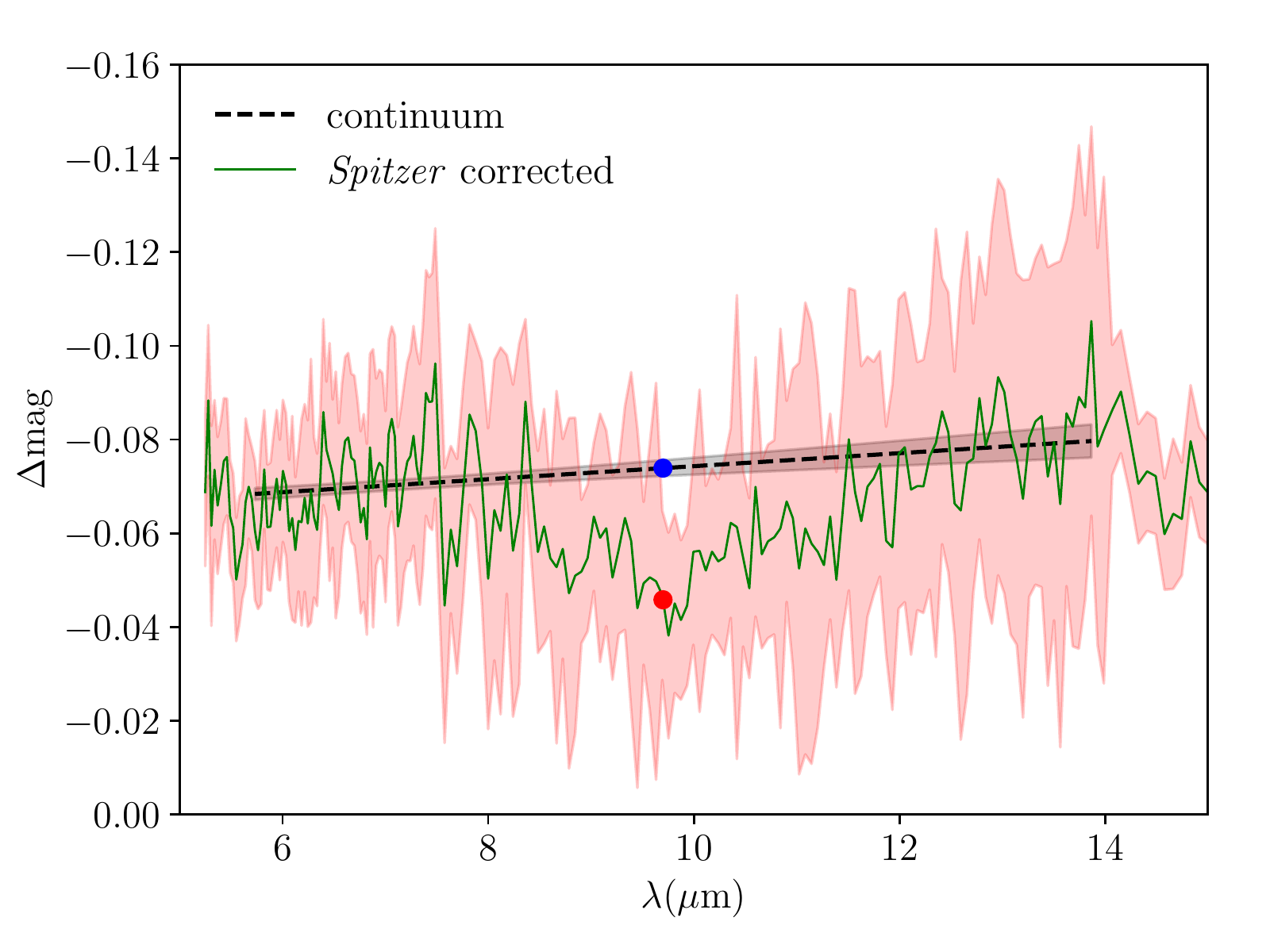}
            \caption[]%
            {{\small SU~Cyg}}    
            \label{fig:su_model}
        \end{subfigure}
        \caption[]
        {\small Silicate absorption features from the \textit{Spitzer} spectra. The dashed black line represents the continuum. Blue and red points are the continuum and \textit{Spitzer} excesses at 9.7$\mu m$ respectively.\label{absorption}. For $\zeta$~Gem no evidence of silicate absorption was found, thus for this star $A^\mathrm{\textit{Spitzer}}_{9.7}$ is set to $0.000\pm0.002$.}
\end{figure}

In order to estimate the apparent silicate absorption at 9.7$\mu m$ ($A^\mathrm{\textit{Spitzer}}_{9.7}$), we have to define the IR excess continuum. For that, we fit the IR excess continuum on each side of the absorption feature over the wavelength range [6,7]$\cup$~[12,13] (dashed line in Fig.~\ref{absorption}), and we interpolate the excess continuum value at 9.7$\mu m$ (blue dot in Fig.~\ref{absorption}). We then subtract to this value the IR excess corresponding to the core of the silicate absorption at 9.7$\mu m$ (red dot). The uncertainty on $A^\mathrm{\textit{Spitzer}}_{9.7}$ is derived by adding in quadrature both the error on the continuum and the error on the \textit{Spitzer} observation at 9.7$\mu m$. For $\zeta$~Gem we do not find any evidence for absorption.

\subsection{Estimation of the interstellar medium (ISM) silicate absorption.}\label{sili.true}
 The ISM extinction theory can be used to estimate the silicate absorption directly from the color excess E(B-V). In order to select the most reliable E(B-V) values we compared for each star (1) the E(B-V) fitted by the SPIPS algorithm, (2) the E(B-V) from the David Dunlap Observatory Database of Galactic Classical Cepheids\footnote{\url{http://www.astro.utoronto.ca/DDO/research/cepheids/}}\citep{DDOP1995} (hereafter DDOD). This database computes the mean extinction and standard error from various values obtained in the literature (3) the E(B-V) obtained with the Stilism 3D map \citep{2017Stilism} using the bayesian inversion method to a wide color excess dataset and parallaxes (4) the 2D map from \citet{1998SFD} (hereafter SFD98) using 100$\mu m$ data from COBE/DIRBE and IRAS. In the latter 2D map we have corrected the extinction by taking into account the star's location using a three-dimensional model of the Milky Way from \citet{drimmel01}. We present these E(B-V) values in Table~\ref{Tab.EBV}. The best agreement is obtained when comparing SPIPS and DDOD values, two totally independent methods, which reinforces the reliability of the SPIPS fitting. 
For the Stilism and SFD98 2D maps, the E(B-V) values are based on models of dust distribution within the Milky Way, and thus cannot consider local over- or under-densities in the vicinity of the Cepheids, which probably explains the large discrepancies found in particular for RS Pup, SU Cyg, and also V Cen. For these stars the uncertainty on E(B-V) is also particularly large in the case of Stilism. The E(B-V) values provided by SPIPS are the most precise by an order of magnitude but uncertainties are underestimated \citep{merand15}. Therefore we decided to adopt an independent and more conservative approach using the DDOD values.
\begin{table*}[]
\caption{\label{Tab.EBV} Comparison of the E(B-V) values of the stars in the sample considering different approaches.} 
\begin{center}
\begin{tabular}{c|c|c|c|c}
\hline
\hline
	&	SPIPS$^a$	&	DDOD$^b\qquad(n)$   		&	Stilism 3D map		&	SFD98 2D map$^c$				\\
\hline
RS Pup	&		$0.550^{+0.005}_{-0.005}$		&	$0.480^{+0.011}_{-0.011}\quad (9)$&	$0.217^{+0.447}_{-0.207}$	&		$0.260^{+0.038}_{-0.038}$\\[0.15cm]

$\zeta$ Gem	&		$0.021^{+0.003}_{-0.003}$&	$0.044^{+0.020}_{-0.020}\quad (6)$	&		$0.013^{+0.024}_{-0.024}$	&		$0.034^{+0.002}_{-0.002}$	\\[0.15cm]

$\eta$ Aql	&		$0.149^{+0.002}_{-0.002}$	&	$0.152^{+0.012}_{-0.012}\quad (12)$	&	$0.155^{+0.050}_{-0.050}$	&	$0.065^{+0.008}_{-0.008}$\\[0.15cm]

V Cen$^d$	&		$0.298^{+0.004}_{-0.004}$&		$0.282^{+0.017}_{-0.017}\quad (8)$	&		$0.283^{+0.480}_{-0.480}$	&		$0.220^{+0.024}_{-0.024}$		\\[0.15cm]

SU Cyg	&		$0.109^{+0.002}_{-0.002}$	&	$0.133^{+0.031}_{-0.031}\quad (6)$	&	$0.257^{+0.113}_{-0.113}$	&	$0.350^{+0.050}_{-0.050}$ \\[0.15cm]
\hline					
\end{tabular}
\normalsize
\end{center}
\begin{tablenotes}
\small
\item $^a$ E(B-V) values are calculated from the SPIPS algorithm.
\item $^b$ David Dunlap Observatory Database. Standard error is the deviation of the number ($n$) of measurements present in the literature.
\item $^c$ Values were corrected by taking into account the star location using three-dimensional model for the Milky Way.
\item $^d$ Extinction values are in agreement with extinction of others members of the open cluster NGC 5662 giving in average $E(B-V)=0.31\pm0.04$ \citep{claria1991}.
\end{tablenotes}
\end{table*}

 As a second step, we derive the visible absorption $A_\mathrm{v}$ assuming an extinction law $A_\mathrm{v}=R_\mathrm{v}E(B-V)$ with a ratio of total-to-selective extinction of $R_\mathrm{v}=3.1$ which corresponds to a diffuse ISM along the line of sight \citep{mathis1979}. Then we used the relation $A_\mathrm{v}/\tau_{9.7}=18.5$ \citep{Roche1984} which is suited to diffuse ISM in the solar vicinity, in order to derive $\tau_{9.7}$. This was done for each star except RS Pup. Indeed, RS~Pup is known for being embedded in a large environment which is most probably the remnant of the molecular clouds in which the star formed \citep{Kervella2009,kervella12}. Hence we treated RS~Pup as a special case since the extinction towards dense clouds is different from diffuse ISM \citep{whittet1988,indebetouw2005,flaherty2007,chiar2007,vanbreemen2011}. We assumed most of the extinction is due to a dense cloud and we used $R_\mathrm{v}=5$ and $A_\mathrm{v}/\tau_{9.7}=11.46$ following the empirical work in star forming regions  by \citet{clure2009}. 
Finally, the extinction $A_\lambda$ is given by the intensity absorption along the line of sight considering an optical depth $\tau_\lambda$. 
Combining both the Beer-Lambert law and absorption definitions for any wavelength $\lambda$ we have $A_\lambda=1.086\tau_\lambda$, thus $A_{9.7}=1.086\tau_{9.7}$ at the specific wavelength 9.7$\mu$m.
This approach can be finally summarize using using two equations :
\begin{equ}[ht!]
\begin{equation}\label{eq.ism}
A^\mathrm{ISM}_{9.7} = 1.086\frac{3.1}{18.5}=0.182\,E(B-V),
\end{equation}
\caption{Assuming silicate reddening law for diffuse ISM.}
\end{equ}

\begin{equ}[ht!]
\begin{equation}\label{eq.dense}
A^\mathrm{ISM}_{9.7} =1.086\frac{5}{11.46}= 0.474\,E(B-V),
\end{equation}
\caption{Assuming silicate reddening law for dense clouds.}
\end{equ}
for stars in diffuse ISM (4)  and for RS Pup (5).

The final values of $A^\mathrm{ISM}_{9.7}$ we consider are listed in Table~\ref{Tab.result}, with their corresponding uncertainties.


\subsection{Residual silicate CSE emission at 9.7$\mu m $ \label{sect:3.3}}
\begin{figure}[]
\begin{center}
\resizebox{1.0\hsize}{!}{\includegraphics[clip=true]{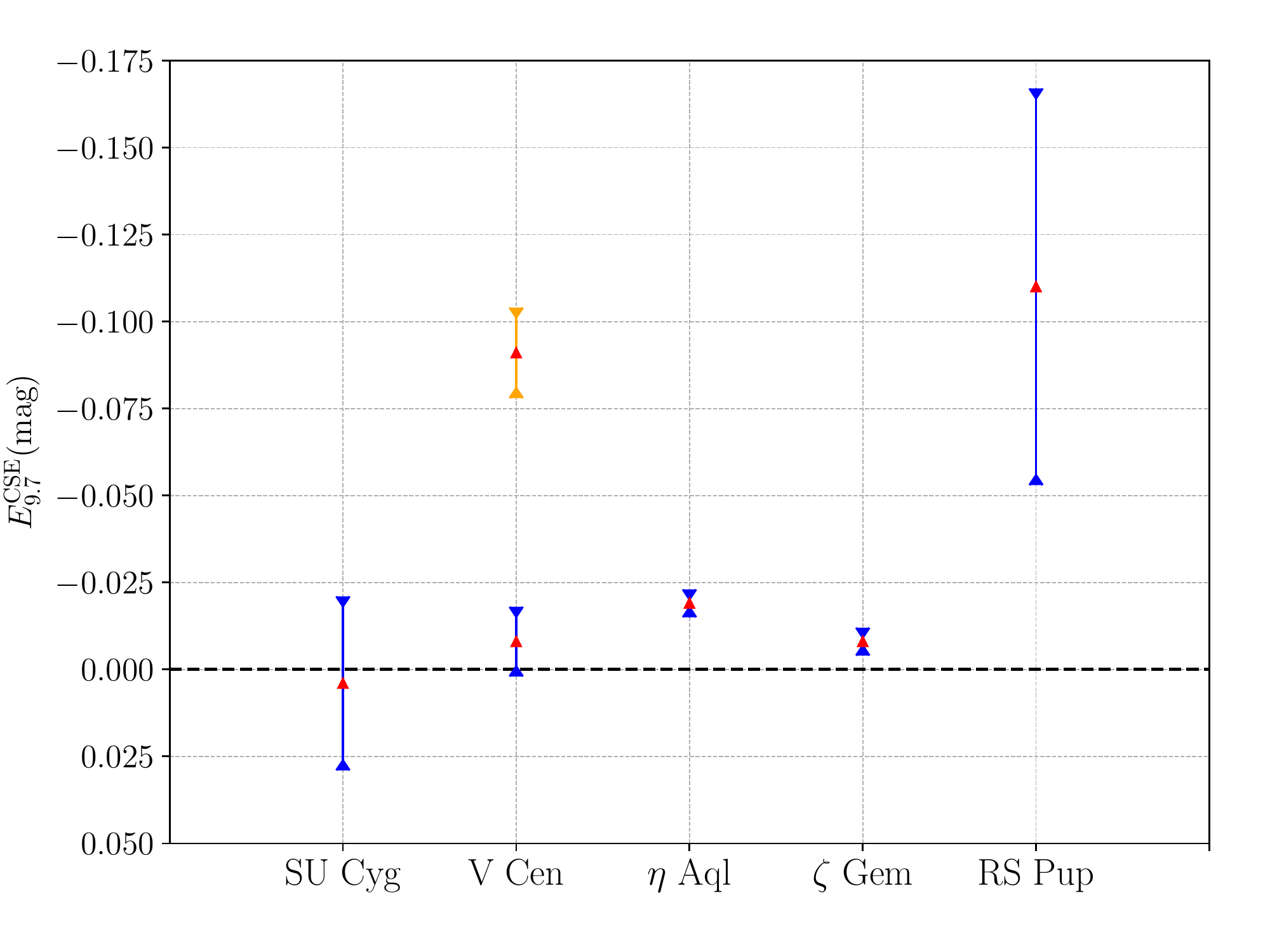}}
\end{center}
\caption{\small The CSE emission at 9.7$\mu m$ is calculated from Eq.~(\ref{Ecse}) and indicated for each star in the sample (in magnitude). Cepheids are ordered by increasing pulsation period. If one uses Eq.~(\ref{eq.dense}) for V~Cen, instead of Eq.~(\ref{eq.ism}), than the $E^\mathrm{CSE}_{9.7}$ value is larger (orange bar in the figure). CSE emission appears when $E^\mathrm{CSE}_{9.7} < 0$.}\label{plot.Ecse}
\end{figure}

Using values of $A^\mathrm{\textit{Spitzer}}_{9.7}$ (Sect. 3.1) and $A^\mathrm{ISM}_{9.7}$ (Sect. 3.2), we now calculate $E^\mathrm{CSE}_{9.7}$ using Eq.~(\ref{Ecse}). 
Results are illustrated in Fig.~\ref{plot.Ecse}.
Using this method, silicate emission from CSE exists only when $E^\mathrm{CSE}_{9.7} < 0$. We find no residual CSE emission for SU~Cyg and V~Cen, and significant but weak emissions for $\zeta$~Gem ($-0.008\mathrm{~mag}\pm 0.004$), $\eta$~Aql ($-0.019\pm 0.004\mathrm{~mag}$) and RS Pup ($-0.110 \pm 0.057\mathrm{~mag}$). In the case of V~Cen, the
emission is likely underestimated when using Eq.~(\ref{eq.ism}). If we assume instead the presence of dense cloud on the line of sight using Eq.~(\ref{eq.dense}), we obtain a silicate emission of $-0.091\pm0.013\mathrm{~mag}$ (see Fig. \ref{plot.Ecse}). 

From this figure (and Table~\ref{Tab.EBV}), it is difficult to conclude whether there is a dusty CSE of silicate around the Cepheids in the sample or not, in particular because our method is sensitive to the Equation used (\ref{eq.ism} or \ref{eq.dense}) when correcting the silicate ISM absorption (case of V Cen for instance), and to various sources of uncertainties, in particular in the E(B-V) estimate. We also assume that the IR excess is not varying with time, as suggested by the SPIPS analysis of Sect. \ref{spips.photometry}. But still, with this method, we find residual dust emission at 9.7$\mu$m for the long-period Cepheids in the sample ($\eta$~Aql, $\zeta$~Gem and RS Pup).


\begin{table*}[]
\caption{\label{Tab.result} \small Result of the residual silicate CSE emission at 9.7$\mu m$. The silicate absorption due to the interstellar extinction $A^\mathrm{ISM}_{9.7}$ is calculated (using Eq.~(\ref{eq.ism})) based on reddenings from DDOD. The observed absorption of silicate from \textit{Spitzer} $A^\mathrm{\textit{Spitzer}}_{9.7}$ is used together with Eq.~(\ref{Ecse}) in order to derive $E^\mathrm{CSE}_{9.7}$, the silicate emission due to a CSE. } 
\begin{center}
\begin{tabular}{c|c|c|c|c}
\hline
\hline
	&		DDOD$^a$ & Sect. 3.2 & Observation$^b$	&	Result$^c$	\\
\hline																			
	&		$E(B-V)$	&	$A^\mathrm{ISM}_{9.7}$	&$A^\mathrm{\textit{Spitzer}}_{9.7}$	&	$E^\mathrm{CSE}_{9.7}$	\\[0.15cm]
\hline
RS Pup		&	$0.480^{+0.011}_{-0.011}$	&	$0.227^{+0.005}_{-0.005}$&	$0.117^{+0.057}_{-0.057}$	&$-0.110^{+0.057}_{-0.057}$	\\[0.15cm]

$\zeta$ Gem	&			$0.044^{+0.020}_{-0.020}$	&	$0.008^{+0.004}_{-0.004}$	&	$0.000^{+0.002}_{-0.002}$	&$-0.008^{+0.004}_{-0.004}$	\\[0.15cm]

$\eta$ Aql	&	$0.152^{+0.012}_{-0.012}$	&$0.028^{+0.002}_{-0.002}$		 & $0.009^{+0.003}_{-0.003}$	&	$-0.019^{+0.004}_{-0.004}$	\\[0.15cm]

V Cen$^d$	&			$0.282^{+0.017}_{-0.017}$	&	$0.051^{+0.003}_{-0.003}$&	$0.043^{+0.010}_{-0.010}$	&$-0.008^{+0.010}_{-0.010}$	\\[0.15cm]

SU Cyg	&		$0.133^{+0.031}_{-0.031}$	&	$0.024^{+0.005}_{-0.005}$	 & $0.028^{+0.025}_{-0.025}$	&	$0.004^{+0.025}_{-0.025}$	\\[0.15cm]
\hline					
\end{tabular}
\normalsize
\end{center}
\begin{tablenotes}
\small
\item $^a$ David Dunlap Observatory Database. Standard error is the deviation of the measurement.
\item $^b$ Errors correspond to \textit{Spitzer} uncertainties (see error bars in Fig. \ref{absorption})
\item $^c$ Errors are given by summing quadratically the precedent absorption uncertainties.
\item $^d$ If Eq.~(\ref{eq.dense}) for V~Cen is considered we obtain $A^\mathrm{ISM}_{9.7}=0.134^{+0.008}_{-0.008}$ and $E^\mathrm{CSE}_{9.7}= - 0.091^{+0.013}_{-0.013}$ (see Fig. \ref{plot.Ecse}).
\end{tablenotes}
\end{table*}

\subsection{Extending the correction of the silicate absorption from 9.7$\mu$m to the whole wavelength range of \textit{Spitzer}.}\label{sect:ext_corr}
We correct the entire \textit{Spitzer} observations by subtracting a synthetic interstellar medium composed of silicates. Since we assumed an averaged ISM temperature of 20K, the dust emission is negligible in the \textit{Spitzer} wavelength range according to Wien's law. Thus, we simply derive the absorption $A^{\mathrm{ISM}}_{\lambda}$ analytically using Mie theory. In the calculation we take into account only the effective absorption cross-section $C^{\mathrm{abs}}_{\lambda}$ and we neglect the scattering effects since the radius of grain $a$ is small compared to the mid-IR wavelength. Hence we adopted the following expression for $\lambda$ between 5 and 30 $\mu m$
\begin{equation} \label{eq:abs}
	A^{\mathrm{ISM}}_{\lambda} \propto \kappa_\lambda=\int C^{\mathrm{abs}}_{\lambda}(a)\pi a^2 n(a)\mathrm{d}a
\end{equation}
We first derived $C^{\mathrm{abs}}_{\lambda}$ using complex refractive index for silicates from \citet{DL} (hereafter DL84) assuming an uniform distribution of ellipsoidal shapes given by \citet{Bohren1983}. Then we derived the absorption coefficient $\kappa_\lambda$ by taking into account a standard grain size distribution $n(a)\propto a^{-3.5}$ \citep{MRN}. Finally we normalize $A^{\mathrm{ISM}}_{\lambda}$ using its specific value $A^{\mathrm{ISM}}_{9.7}$ at 9.7$\mu m$ we already derived in Sect. \ref{sili.true}.






\begin{figure*}[htbp]
\centering
\begin{tabular}{cc}
\includegraphics[width=0.45\textwidth]{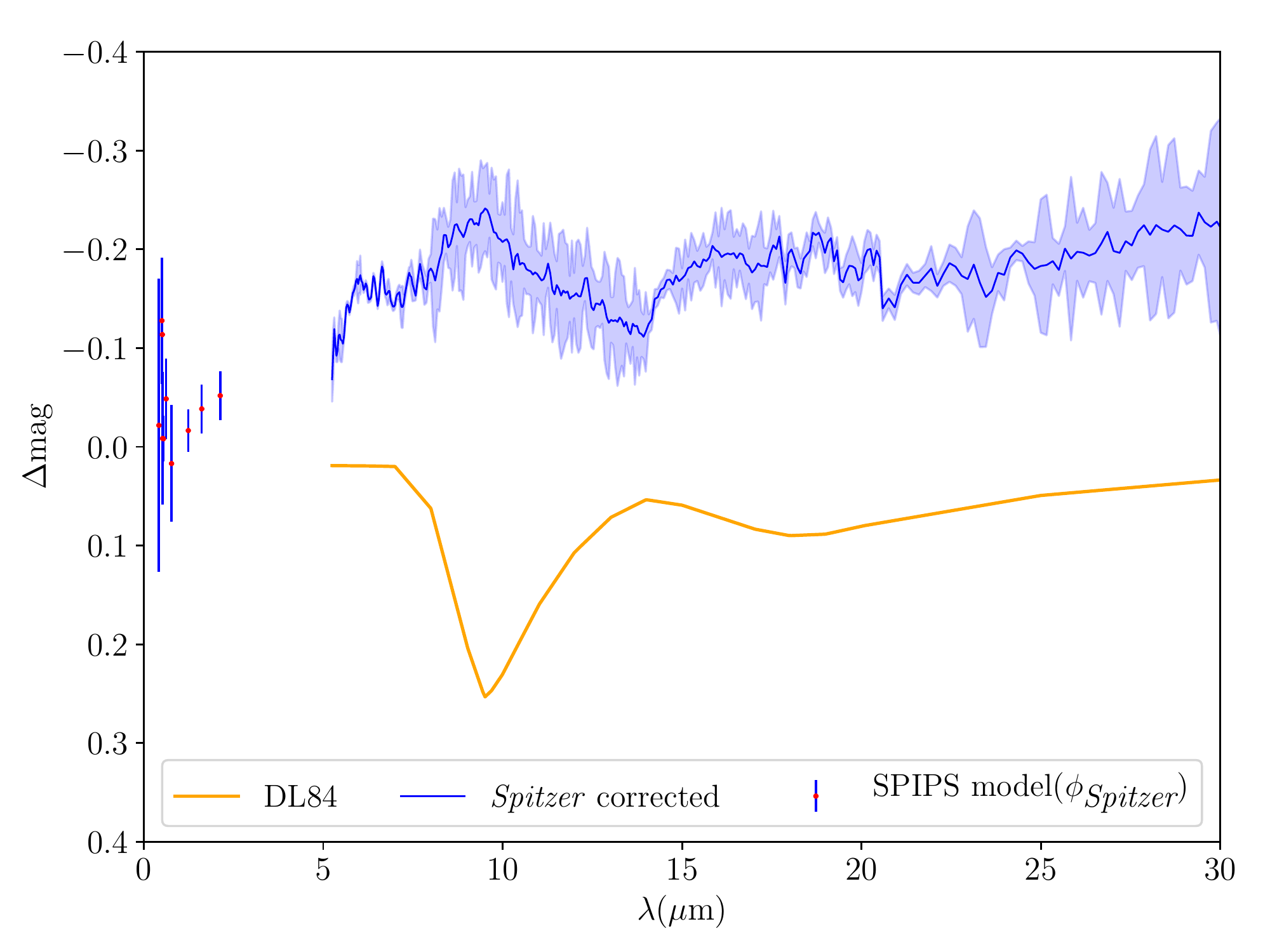}\\
(a) RS Pup, $\phi=0.938$  
\end{tabular}
\begin{tabular}{cc}
\includegraphics[width=0.45\textwidth]{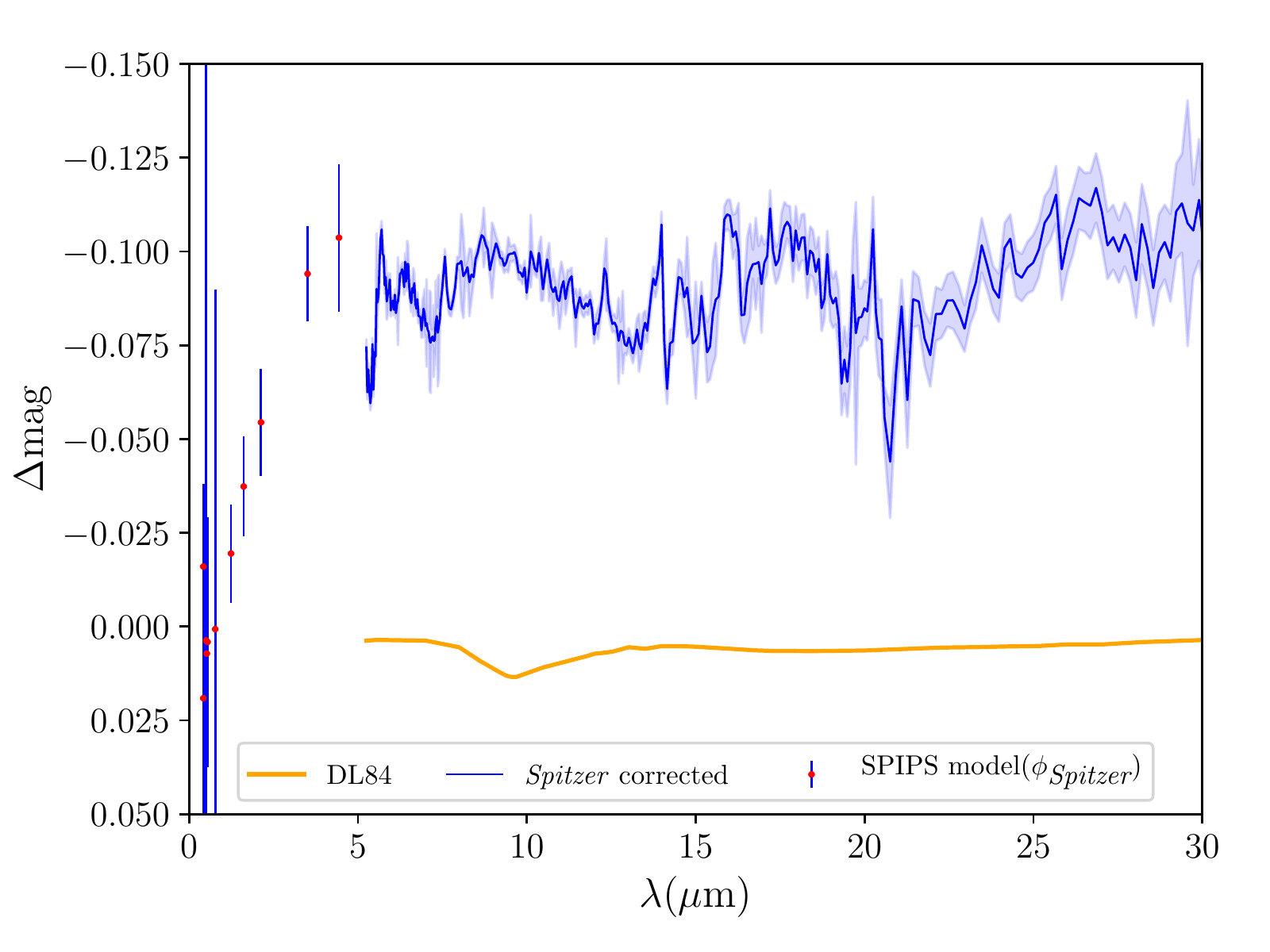}&
\includegraphics[width=0.45\textwidth]{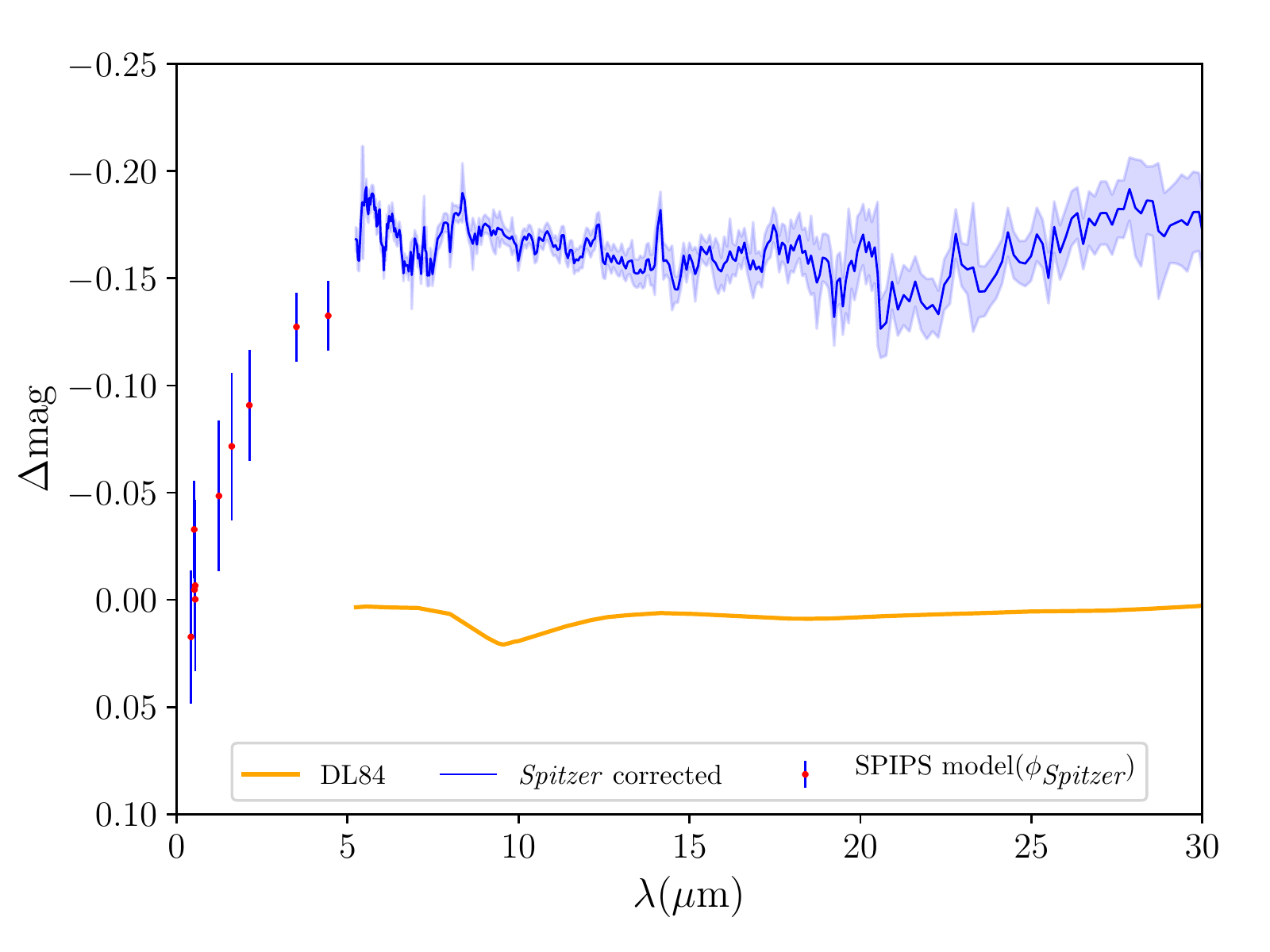}
\\
(b) $\zeta$ Gem, $\phi=0.410$ & (c) $\eta$ Aql, $\phi=0.408$ 
\end{tabular}
\begin{tabular}{cc}
\includegraphics[width=0.45\textwidth]{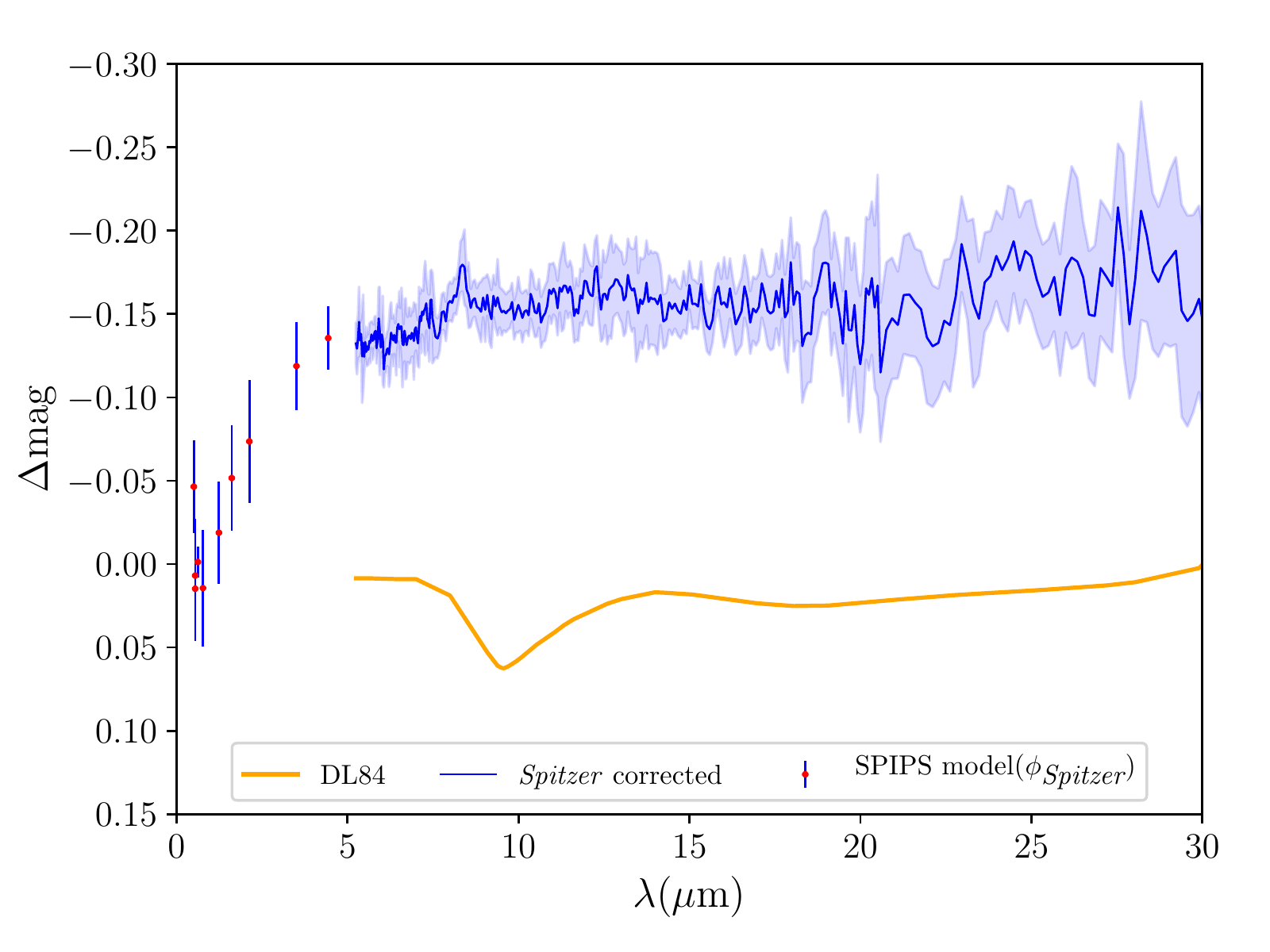}&
\includegraphics[width=0.45\textwidth]{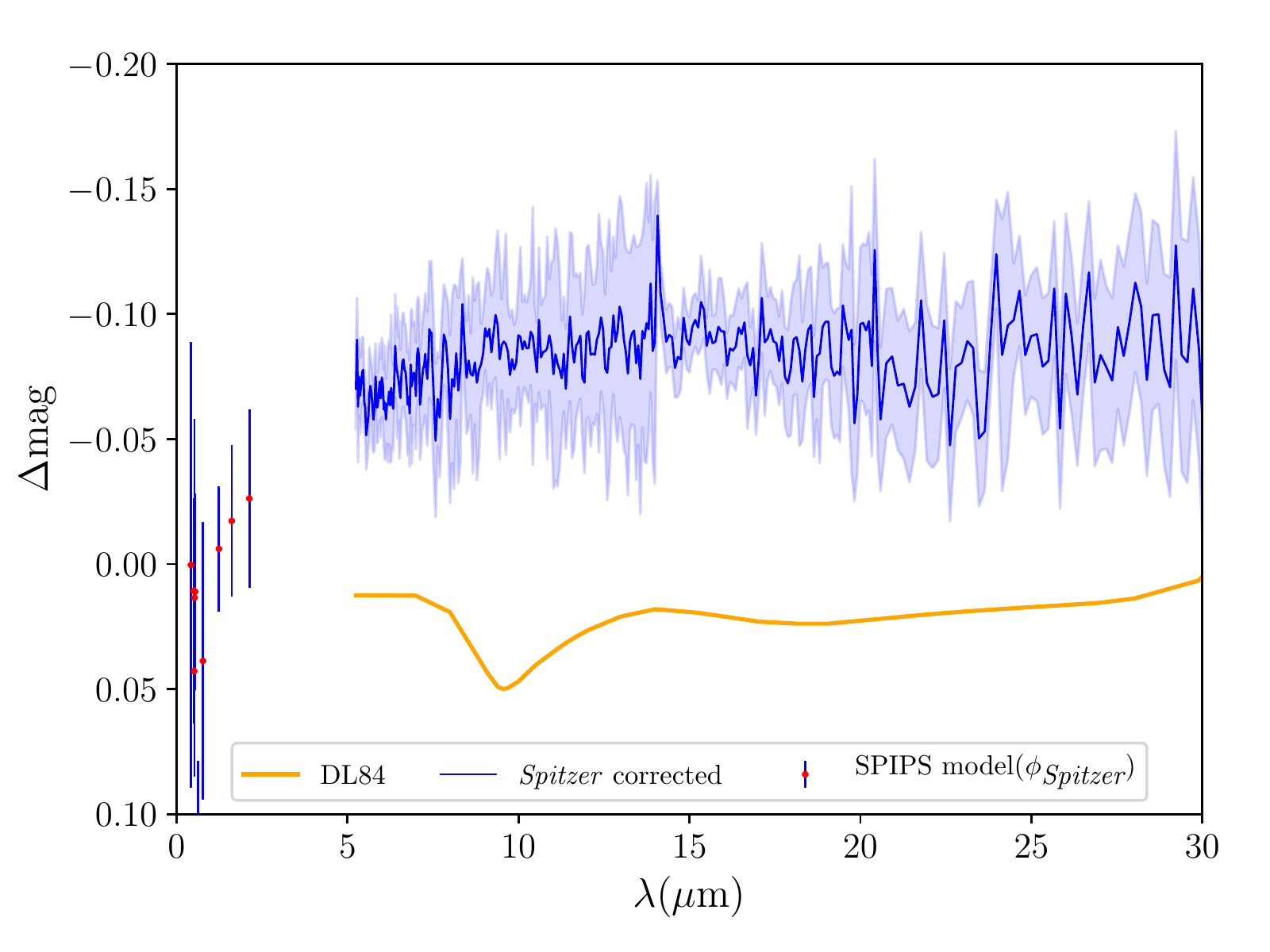} \\
(d) V Cen, $\phi=0.960$ & (e) SU Cyg, $\phi=0.013$ 
\end{tabular}

\caption{\small \label{fig:dereddened} The IR excess for all the stars in the sample at a specific phase is presented, including (1) the interpolated IR excess model from SPIPS, (see Sect.~\ref{spips.photometry}), (2) the \textit{Spitzer} observations cleaned from different camera effects (see Sect. \ref{spitzer.data}) and also corrected from the silicate absorption due to the ISM (orange curve, see Sect. \ref{sect:ext_corr}).}
\end{figure*}
For all the stars in the sample, we plot the IR excess (found to be constant with the pulsation phase, see Sect. \ref{spips.photometry}) corrected from the ISM silicate absorption in Fig. \ref{fig:dereddened}. In the following sections we first show that a dust envelope cannot reproduce the IR continuum excess, then we explore the possibility of a thin envelope of partially ionized gas in order to model the IR excess continuum.

\section{Incompatibility of dust CSE model to explain IR excess continuum.}\label{dust}
\begin{figure}[]
\begin{center}
\resizebox{1.0\hsize}{!}{\includegraphics[clip=true]{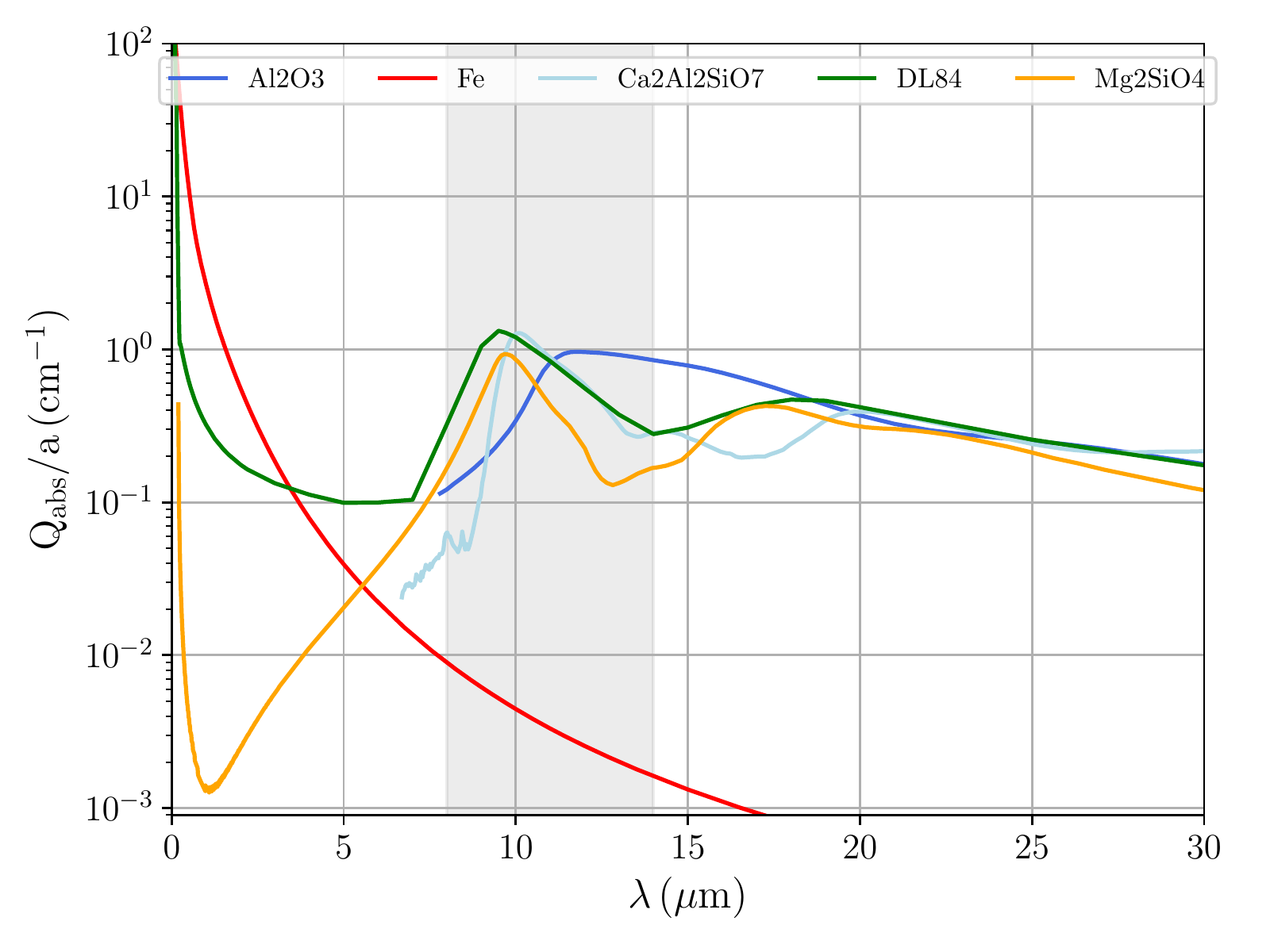}}
\end{center}
\caption{\small Opacity efficiencies of typical dust encountered in circumstellar envelopes of oxygen-rich stars. These opacities are from aluminium oxide $\mathrm{Al_2O_3}$ \citep{begemann1997}, iron Fe \citep{henning1996}, gelhenite $\mathrm{Ca_2Al_2SiO_7}$ \citep{mutschke1998}, astronomical silicate from \citep{DL}(DL84), and forsterite $\mathrm{Mg_2SiO_4}$ \citep{jager2003}. The N-band between 8 and 14$\mu m$ is represented with a grey strip and highlights the silicates vibrational modes.\label{Fig.opacity}}
\end{figure}

\begin{figure}[]
\begin{center}
\resizebox{1.1\hsize}{!}{\includegraphics[clip=true]{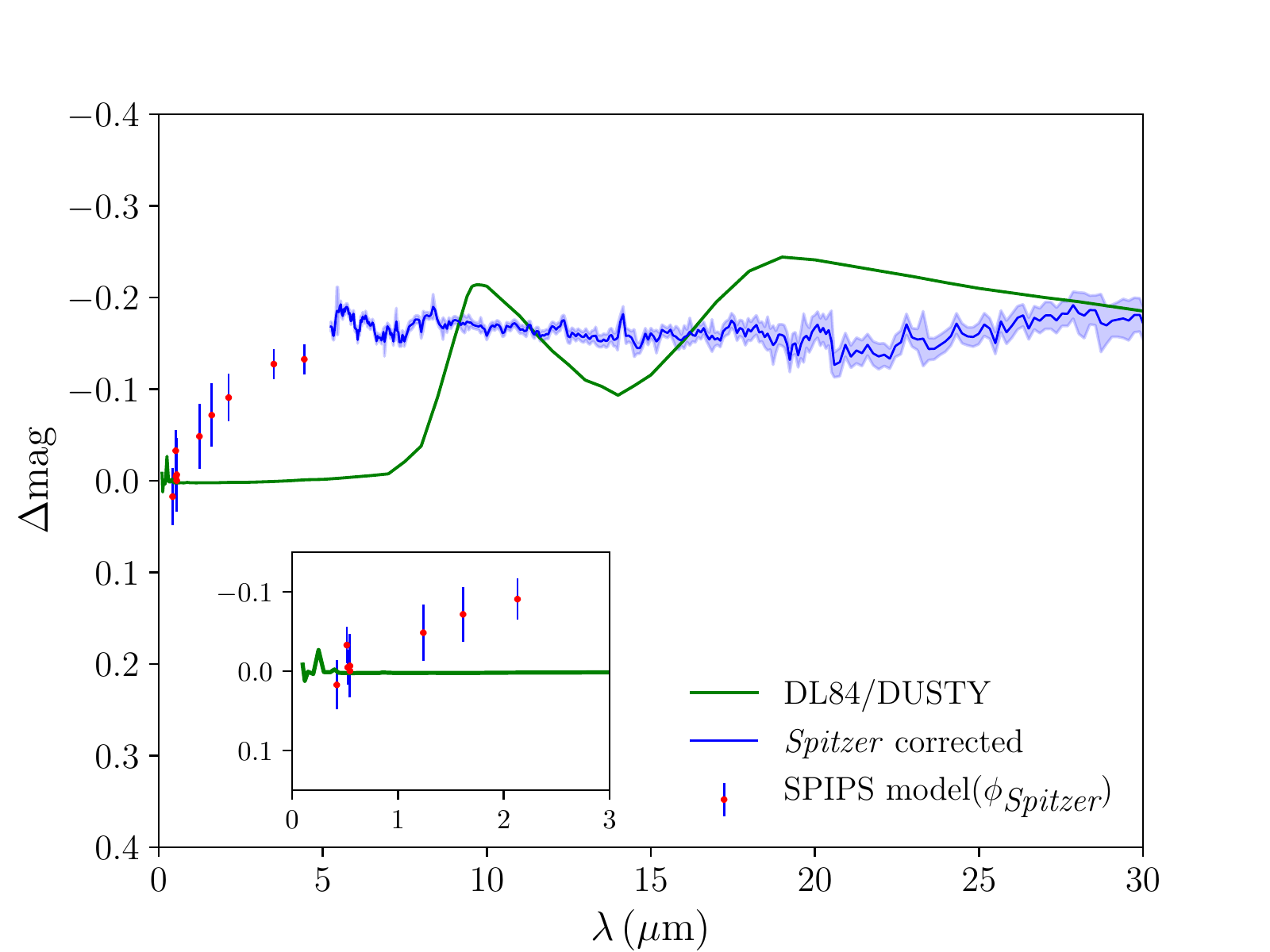}}
\end{center}
\caption{\small \label{fig:eta_silicate} IR excess of a silicate circumstellar envelope modeled with DUSTY around $\eta$~Aql. Green curve is the silicate model with silicate grains from DL84.}
\end{figure}

Cepheids are oxygen-rich stars, and most of the carbon in their envelopes is thus locked in CO molecules. In the condensation sequence described by \citet{Gail1999} corundum ($\mathrm{Al_2O_3}$) is expected to form first because of its high condensation temperature of about $1400$K in typical pressures encountered in circumstellar shells. For lower temperatures, $\mathrm{Al_2O_3}$ is depleted and silicates such as gehlenite ($\mathrm{Ca_2Al_2SiO_7}$) and forsterite ($\mathrm{Mg_2SiO_4}$) can form. All these components present emission components mostly observed in the N band (8-13 $\mu$m, see Fig. \ref{Fig.opacity}), with a width of at most a few microns. The IR excess we observe in \textit{Spitzer} and SPIPS data does not present any clear spectral feature and is broader than 20 microns.  Fig. \ref{fig:eta_silicate} shows the best fit we obtain with silicate dust. Silicate dust features are clearly unlikely to explain the observed IR excess continuum from near- to mid-IR. This conclusion is in agreement with the work made by \cite{Schmidt2015} who has found that CSE made of silicate dust cannot explains IR excess of 132 classical and type-II Cepheids. However, since the opacity of iron exhibit no particular feature but rather a continuum (see Fig. \ref{Fig.opacity}) we investigate whether a warm dust envelope of iron could explain the IR excess continuum observed with SPIPS and \textit{Spitzer} dataset.

\begin{figure}[]
\begin{center}
\resizebox{1\hsize}{!}{\includegraphics[clip=true]{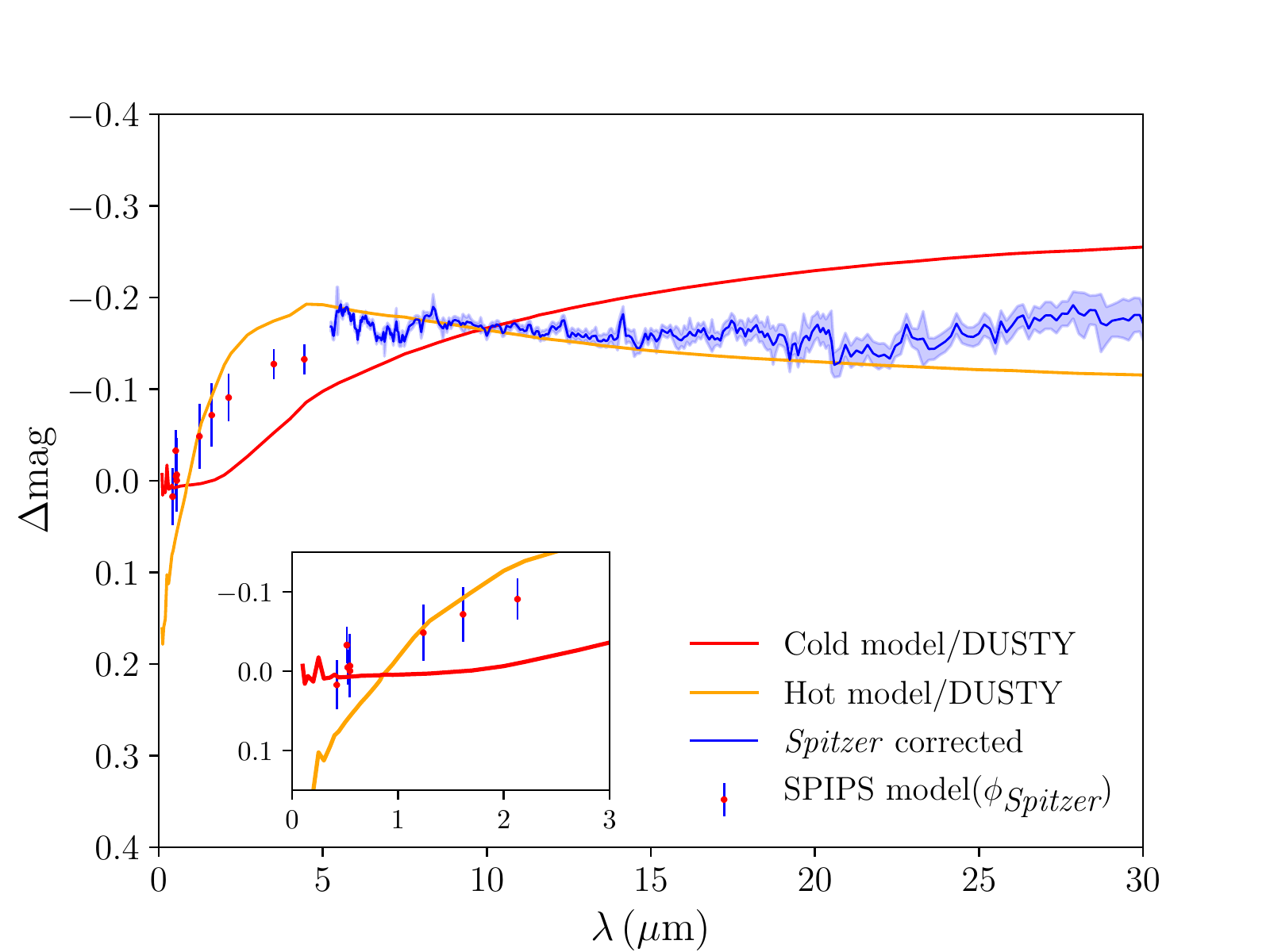}}
\end{center}
\caption{\small IR excess of an iron circumstellar shell modeled with \texttt{DUSTY} around $\eta$~Aql.}\label{Fig.warm}
\end{figure}

The CSEs were modeled using \texttt{DUSTY} \citep{DUSTY}, which solves the radiative transfer equations in a dusty environment. The method is based on a self-consistent equation for the spectral energy density, including dust scattering, absorption, and emission. We present 2 CSEs models with different inner shell temperatures. Iron is a component in oxygen-rich star mineralogy with a condensation temperature of $\approx 1000$K for circumstellar pressures. Hence, the cold model takes into account the typical condensation temperature of iron ($\approx$1000K) as the inner shell temperature which is thought to be realistic. A reduced $\chi^2$ fitting is applied to adjust the optical opacity $\tau_{0.55}$. 
 The hot model lets both the inner shell temperature and the optical opacity as free parameters during $\chi^2$ fitting. In both hot and cold model we used a standard MRN size distribution \citep{MRN}, and we computed the density distribution in the case of an envelope expansion which is driven by radiation pressure on the dust, the wind structure is derived taking into account the dust drift and the star's gravitational attraction. The outer shell radius is set to 500 times the inner shell radius. We test the consistency of these models on $\eta$~Aql using interpolated ATLAS9 atmospheric models (see Table~\ref{Tab.param} in Sect. \ref{spitzer.data}) as a central source's radiation in the \texttt{DUSTY} computation.
The results of the computation are summarized in Table~\ref{dusty} and the computed IR excess is presented Fig. \ref{Fig.warm}. 
Both models fails to reproduce the IR excess continuum modeled with SPIPS and observed with \textit{Spitzer}. The cold model is well below the observed IR excess and the temperature required for iron in the hot model is much higher than iron condensation temperature for circumstellar shell pressures. In that case  solid iron would not form or would be sublimated. Last, we can also argue that a CSE made of iron exclusively for each Cepheid is very unlikely. In conclusion we do not expect a warm dust envelope to cause the IR excess continuum of Cepheids  in the 1 - 30 $\mu m$ range. 

\begin{table}
\begin{center}
\begin{threeparttable}
\caption{\small Warm dust envelope parameters. $T_\mathrm{in}$ and $\theta_\mathrm{in}$ are the inner temperature and diameter in milliarcseconds respectively. $\tau_{0.55}$ is the opacity at $0.55\mu m$. Mass loss $\dot{\mathrm{M}}$ derived by \texttt{DUSTY} is also indicated.}
\label{dusty}
\begin{tabular}{cccc}
\hline
\hline
	&	Hot model	&	Cold model	\\
\hline					
Grain size	&	MRN	&	MRN	\\
$T_{in}$(K)$^a$	&	2238$^a$		&	1000	\\
$\tau_{0.55}$	& 0.038$^a$			&0.006$^a$			\\
$\theta_{in}$(mas)	& 10 & 109			\\
$\dot{\mathrm{M}}(\Mo/yr)$	& $2.04.10^{-7}$ & $9.72.10^{-8}$			\\
\hline
$\chi^2$	&	7.3	&	50.3	\\
\hline														
\end{tabular}
\normalsize
\begin{tablenotes}
\small
\item $^a$ Fitted parameters.
\end{tablenotes}
\end{threeparttable}
\end{center}
\end{table}

\section{The IR excess from a thin shell of ionized gas}\label{ionized}
We investigate hereafter if an ionized, spherical gas shell can explain at the same time the near- ($\sim1 \mu$m < $\lambda$ < 5 $\mu$m) and mid (5 $\mu$m < $\lambda$ < 30 $\mu$m) IR excess of Cepheids. As a first step, this study is made only at the specific pulsation phase of {\it Spitzer} data. We discuss the possible time-variability of such shell in Sect. \ref{perspectives}

\begin{figure}[]
\begin{center}
\resizebox{1.0\hsize}{!}{\includegraphics[clip=true]{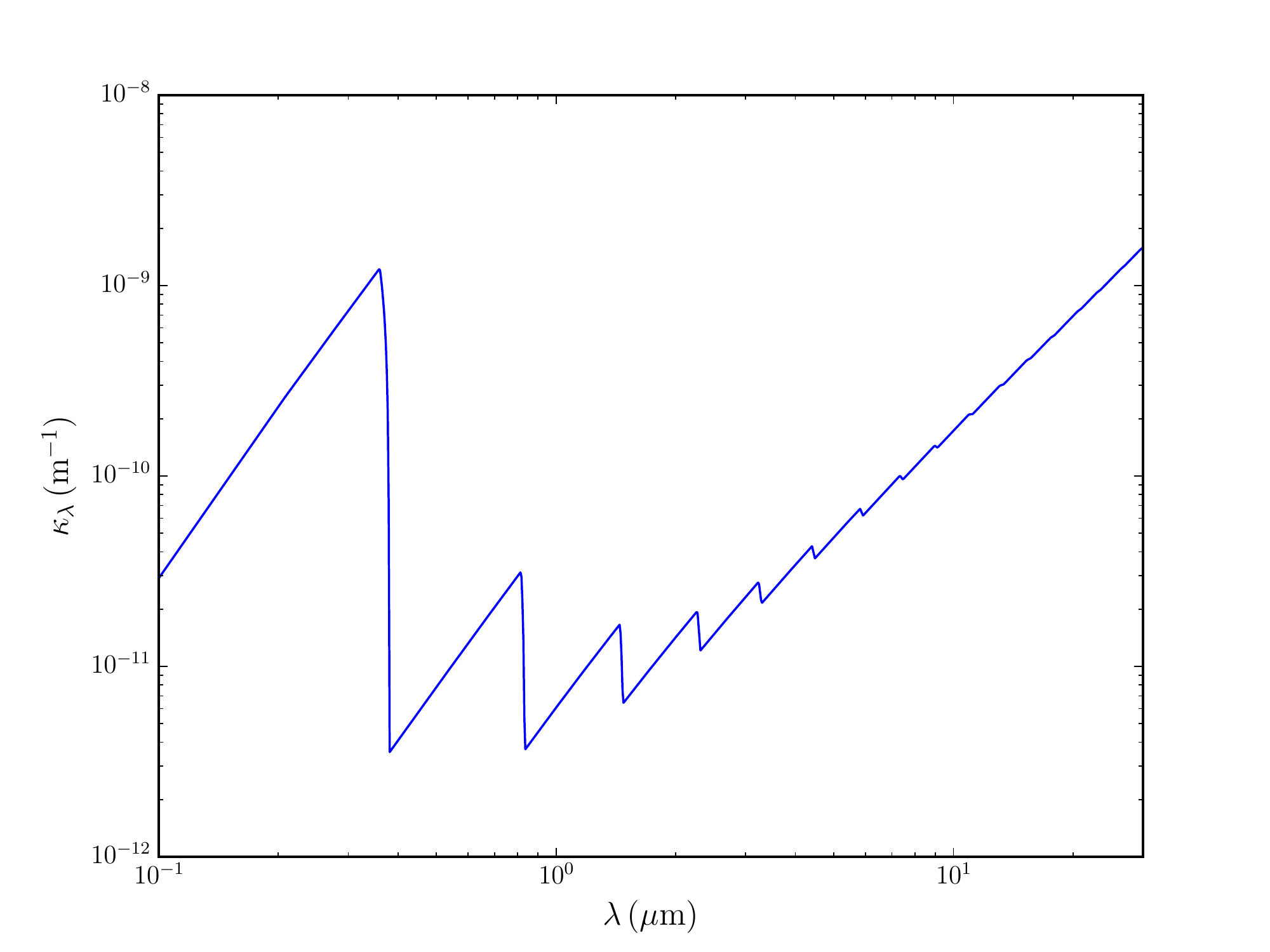}}
\end{center}
\caption{\small \label{fig:kappa} Bound-free and free-free absorption coefficient $\kappa_\lambda$ computed for V~Cen from 0.1$\mu$m to 30$\mu$m. Bound-free absorption is characterized by saw teeth shapes which are sharper for shorter wavelength. Free-free absorption dominates wavelength above 1$\mu$m.}
\end{figure}

We consider the emission of a thin gas shell around the star with constant density and temperature for the sake of simplicity. The shape of the mid-IR excess, saturating to a constant flux ratio at large wavelengths (see Fig.~\ref{fig:dereddened}), suggests an opacity source increasing with wavelength. We used the free-free and bound-free opacities for a pure H shell presenting such behaviour. The combined absorption coefficient (in m$^{-1}$; SI(MKS) unit system) for these two opacities sources is given by \citep[e.g.][]{rybicki2008}



\begin{equation} \label{eq:1}
\begin{split}
	\kappa_\lambda=3.692\times10^{-2}\left[1-\e^{-\frac{hc}{\lambda k \Ts}}\right] \Ts^{-1/2}\times(\lambda/c)^3\\
	(\gamma \rho/\mH)^2[\gff(\lambda,\Ts)+\gbf(\lambda,\Ts)],
\end{split}
\end{equation}

where h, c, and k have they usual meanings, $\gamma$ is the degree of ionization (between 0 and 1), $\Ts$ the temperature of the shell, $\mH$ the hydrogen mass and $\gff$ and $\gbf$ are the free-free and bound-free Gaunt factors respectively. These factors were computed mainly from approximation formulas given by \citet{Brussaard1962_v34p507-520}, \citet{Hummer1988_v327p477-484}, and references therein. As an example we present a typical shape of the absorption coefficient $\kappa_\lambda$ computed for V~Cen in Fig. \ref{fig:kappa} using the shell parameters from Table~\ref{result.shell}.

\begin{figure*}[]
\centering
\begin{tabular}{cc}
\includegraphics[width=0.4\textwidth]{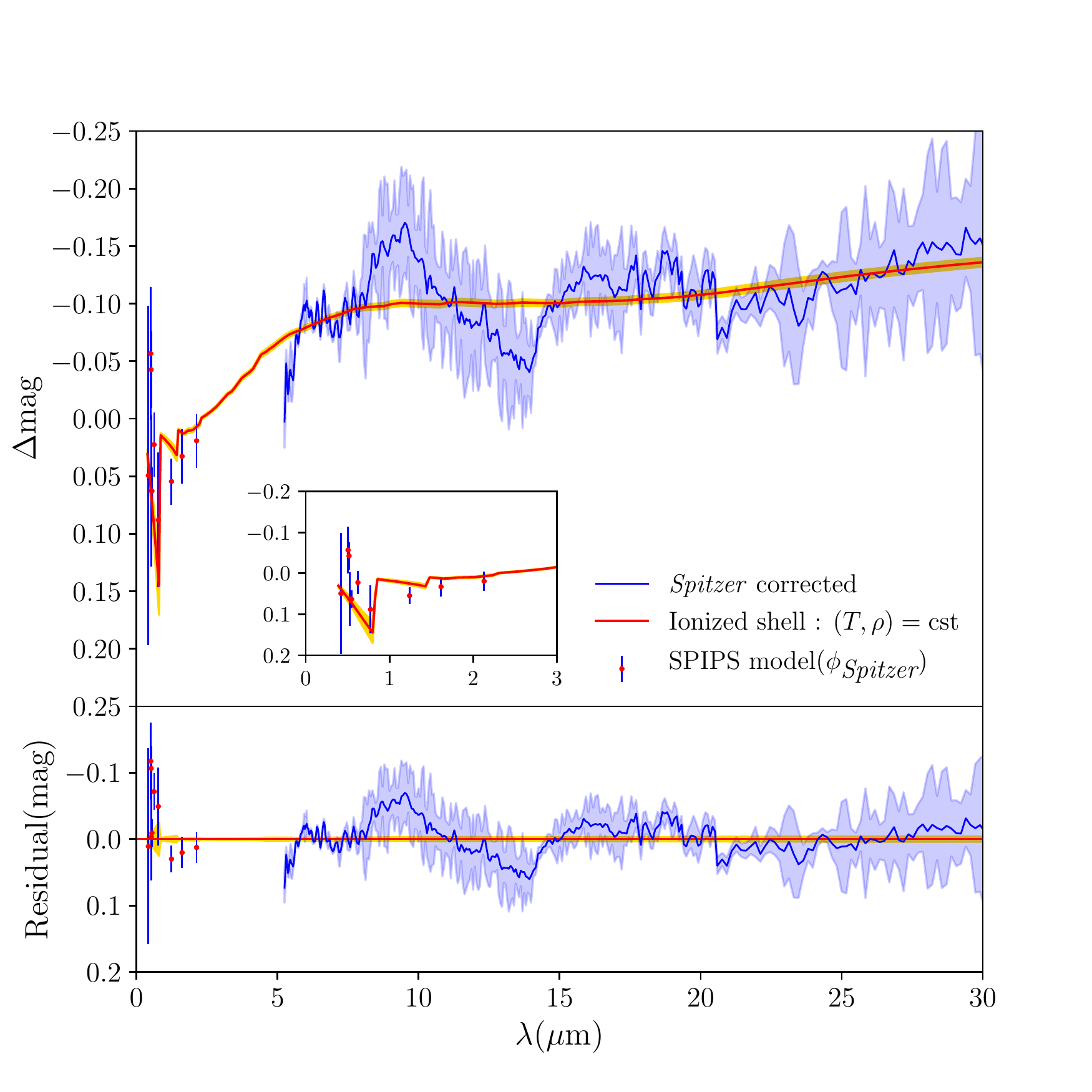}\\
(a) RS Pup, $\phi=0.938$  
\end{tabular}
\begin{tabular}{cc}
\includegraphics[width=0.4\textwidth]{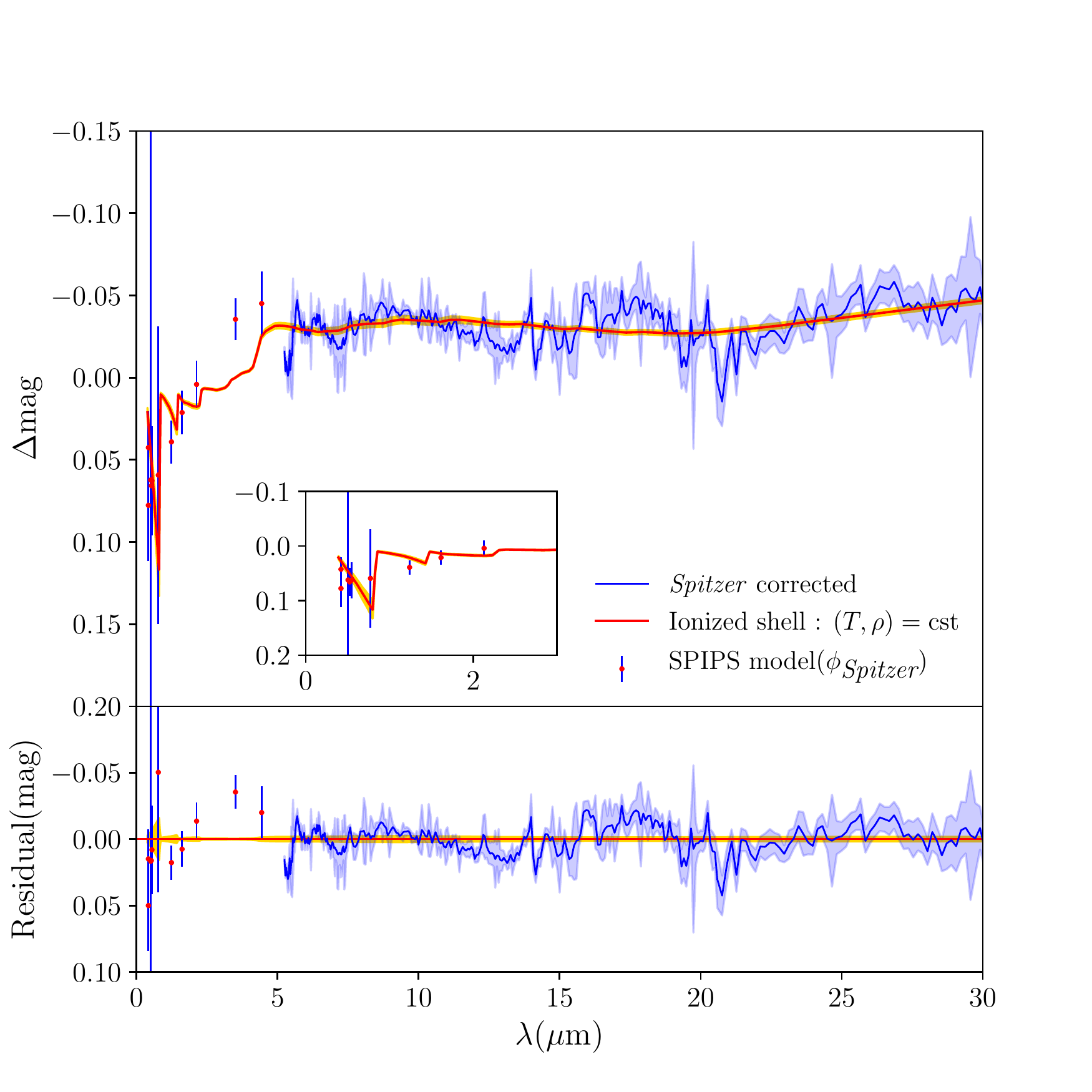}&
\includegraphics[width=0.4\textwidth]{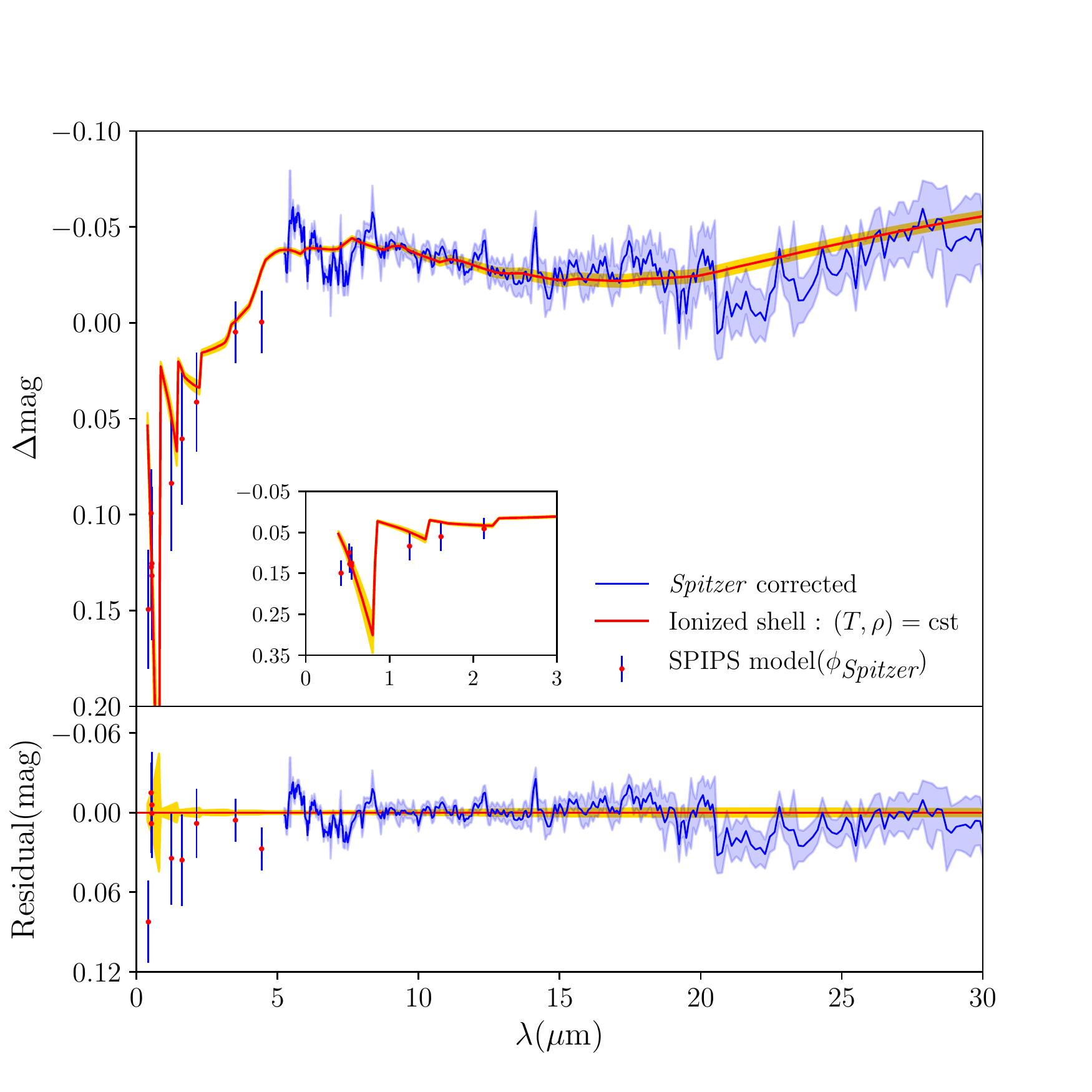}
\\
(b) $\zeta$ Gem, $\phi=0.410$ & (c) $\eta$ Aql, $\phi=0.408$ 
\end{tabular}
\begin{tabular}{cc}
\includegraphics[width=0.4\textwidth]{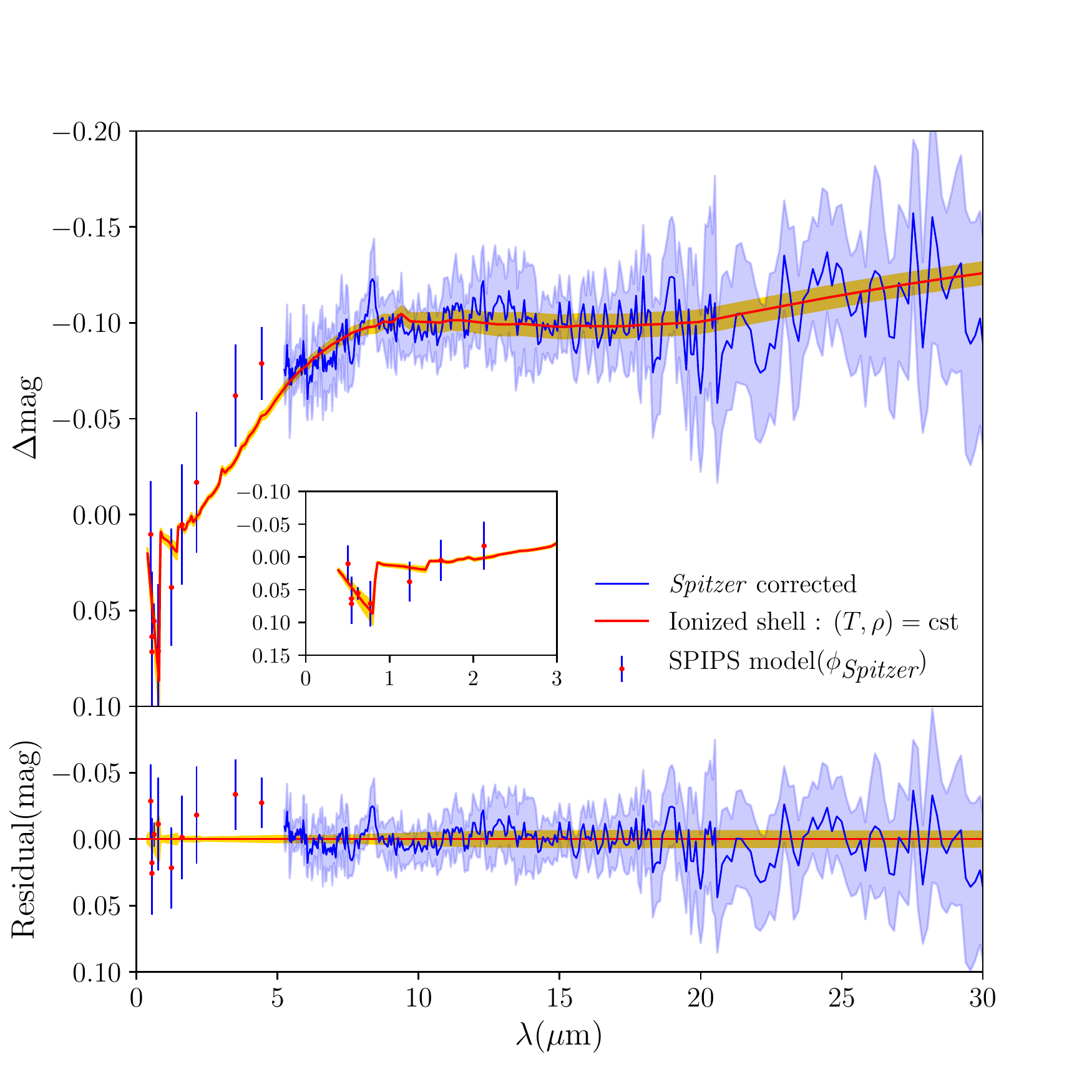}&
\includegraphics[width=0.4\textwidth]{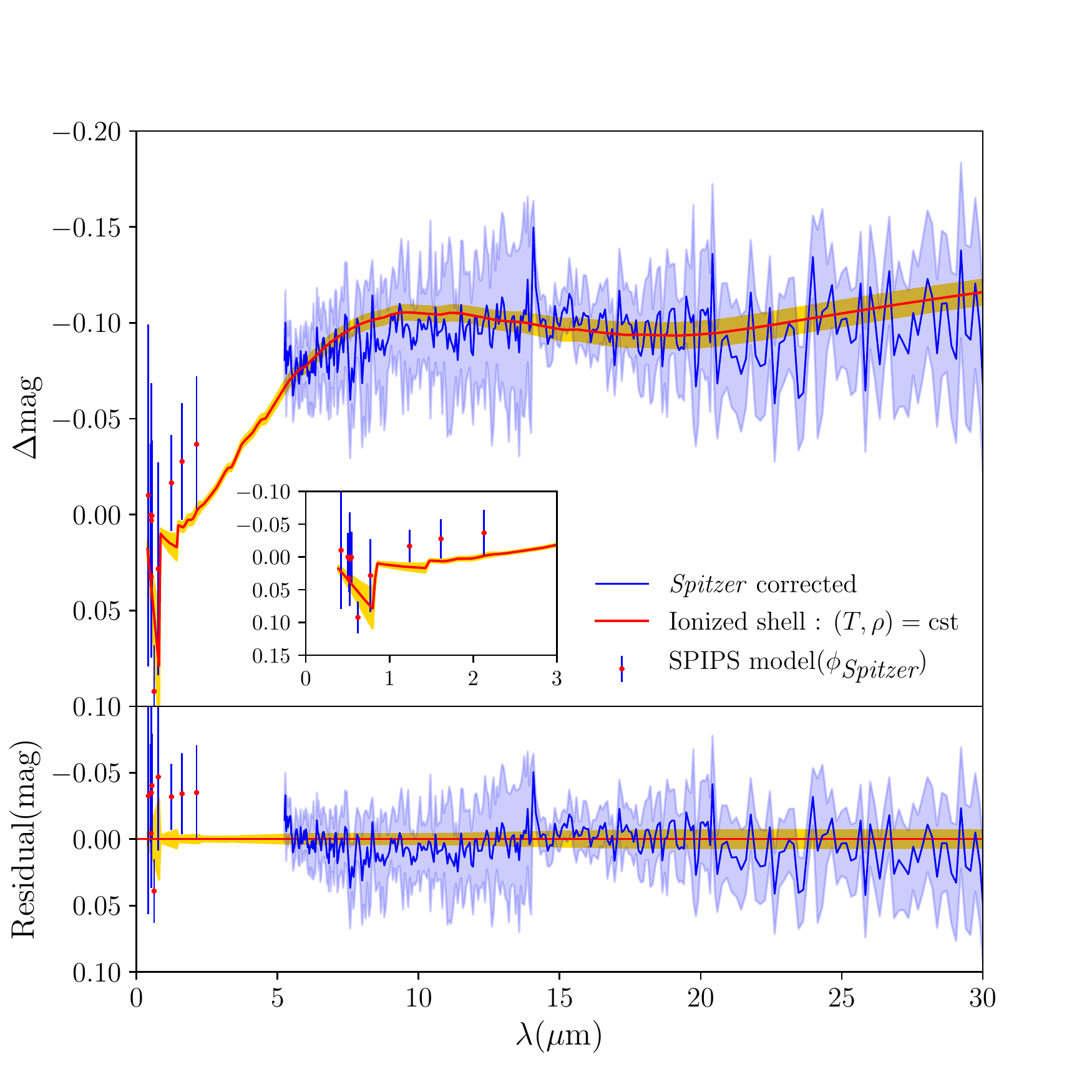}
 \\
(d) V Cen, $\phi=0.960$  & (e) SU Cyg, $\phi=0.013$
\end{tabular}

\caption{\small \label{kromo.result} IR excess fitting results of a shell of ionized gas (red curve, see Sect. \ref{ionized} and Appendix \ref{app:thin_gas_shell}) presented with residuals. Yellow region is the error on the magnitude obtained using the covariance matrix of the fitting result. Pulsation phase $\phi$ of \textit{Spitzer} observations and SPIPS interpolation is indicated.}
\end{figure*}

We computed the SED of the star plus the gas shell as described in Appendix~\ref{app:thin_gas_shell}, taking into account the latter absorption coefficient $\kappa_\lambda$. In order to match the SPIPS photometries plus corrected \textit{Spitzer} spectra presented in Sect~\ref{deredden} we performed a $\chi^2$ fitting using the Levenberg-Marquardt method. We used filters bandpass to convert the flux of the physical model into the corresponding SPIPS photometries for the fitting procedure. In addition, since the SPIPS fitting assumes that there is no excess in the visible domain i.e $\Delta \mathrm{m}=0$ for $\lambda$ < 1.2 $\mu m$ (see Sect. \ref{spips.photometry}) it is necessary to relax this assumption to allow the data to present deficit or excess in the visible depending on the physical behaviour of the ionized shell. Soften this assumption is equivalent to suggest an improvement in the SPIPS physical treatment of the circumstellar environment. As a first step, we consider a simple thin, spherical, and partially ionized gas shell in order to reproduce the IR excess of the Cepheids in the sample. Hence, we fitted 4 parameters: three parameters from the gas shell i.e the ionized shell mass $\gamma\Ms$, its temperature $\Ts$ and radius $\Rs$ plus one parameter corresponding to the IR excess offset corresponding to $\Delta \mathrm{m}\neq0$ for $\lambda < 1.2 \mu m$.
Since \textit{Spitzer} data have a higher statistical weight than SPIPS data we first find the different parameters by hand to use it as first guesses. The results are presented in Table~\ref{result.shell} and in Fig.~\ref{kromo.result}.

In Fig. \ref{kromo.result} we observe discontinuities at short wavelengths, from visible to near-IR, which are due to bound-free opacities (see Fig. \ref{fig:kappa}). Bound-free opacity decreases for longer wavelength whereas free-free opacity increases. Beyond 5 $\mu$m the bound-free contribution is negligible compared to the free-free contribution which explains the observed smooth continuum. For the Cepheids in the sample, we obtained shell temperatures between 3500 and 4500K and an envelope thickness of $\simeq$ 15\% of the radius of the star, while $\gamma \Ms $ is ranging from $10^{-9}-10^{-7}$\Msolar. The IR excess correction offsets found have positive values which means $\Delta \mathrm{m}\geqslant0$ for $\lambda < 1.2 \mu m$ for all stars except for SU~Cyg which presents a slight negative value. Indeed the several computed models present absorption in the visible domain (see Fig. \ref{kromo.result}), thus correction offsets have positive values to allow a deficit in the visible due to ionized shell absorption.

\begin{table*}[htbp]
\caption{\label{result.shell} \small Fitted parameters of the thin circumstellar shell of partially ionized gas, with constant temperature and density.}
\begin{center}
\begin{tabular}{c|ccccc}
\hline
\hline
	&	$T_\mathrm{shell}$ (K)	&	$R_\mathrm{shell}$ ($R_\mathrm{star}$)	&	$\gamma M_\mathrm{shell}$ (\Msolar)	&	Offset (mag) &	$\chi^2$ \\
\hline											
RS Pup	&	$4011\pm88$	&	$1.170\pm0.014$	&	$2.13.10^{-7}\pm1.1.10^{-8}$	&	$0.071\pm0.002$	&	3.46\\

$\zeta$ Gem	&	$3780 \pm 85$	&	$1.087 \pm 0.013$	&	$9.50.10^{-9}\pm9.0.10^{-10}$	&	$0.059 \pm 0.001$&	3.23\\

$\eta$ Aql	&	$3569 \pm61	$&	$1.164 \pm 0.010$	&	$1.17.10^{-8}\pm4.0.10^{-10}$	&$0.132 \pm 0.002$&	2.73\\

V Cen	&	$4353 \pm 106$	&	$1.156 \pm0.012$	&	$3.61.10^{-9}\pm2.0.10^{-10}$	&	$0.057 \pm 0.004$&	0.28	\\

SU Cyg	&	$4402 \pm 204$	&	$1.176 \pm 0.027$	&	$7.89.10^{-9}\pm8.6.10^{-10}$	&	$-0.010 \pm 0.004$&	0.36\\
\hline
\end{tabular}
\normalsize
\end{center}
\end{table*}
In the prospect of JWST observations, we provide the synthetic IR excess anticipated from the shell of ionized gas in several bands (Table \ref{tab.jwst}). We used filter transmissions of wide filters: F070W and F200W from Near-Infrared Camera (NIRCam)\footnote{\url{https://jwst-docs.stsci.edu/display/JTI/NIRCam+Filters}} and F560W and F1000W from Mid-Infrared Instrument (MIRI)\footnote{\url{https://jwst-docs.stsci.edu/display/JTI/MIRI+Filters+and+Dispersers}}. These filters, centered on 0.70$\mu m$, 2$\mu m$, 5.6$\mu m$ and 10$\mu m$ respectively, are suitable to measure the IR excess continuum from the ionized shell model.
\begin{table*}[htbp]
\caption{\label{tab.jwst} \small IR excess from the ionized shell models in various bands of the JWST. Wide filters F070W, F200W, F560W and F1000W centered respectively on 0.70$\mu m$, 2$\mu m$, 5.6$\mu m$ and 10$\mu m$ are presented. Values are given in magnitudes.}
\begin{center}
\begin{tabular}{c|cccc}
\hline
\hline
	&	$\Delta$ F070W	&	$\Delta$ F200W & $\Delta$ F560W	&	$\Delta$ F1000W	 	\\
\hline											
RS Pup	&	0.110	&	0.009	&	-0.074	&	-0.100\\

$\zeta$ Gem	&	0.090	&	0.016	&	-0.029	&	-0.034	\\

$\eta$ Aql	&	0.220 &	0.030	&	-0.038	& -0.037	\\

V Cen	&	0.069	&	0.004	&	-0.071	& -0.101	\\

SU Cyg	&	0.063	&	0.001	&	-0.071	&	-0.105	\\
\hline
\end{tabular}
\normalsize
\end{center}
\end{table*}
\section{Discussion}\label{Discussion}
\subsection{The IR excess and dust environment \label{sect:6.1}}
We estimated the infrared excess of Cepheids using SPIPS algorithm ($\lambda <5\mu$m) and {\it Spitzer} data ($ 5 < \lambda < 30\mu$m). The observed IR excess presented in Fig~\ref{fig:dereddened} is thus calculated at the specific phases of {\it Spitzer} and makes the assumption that there is no excess and/or deficit in the visible domain due to the CSE ($\Delta \mathrm{m}=0$ for $\lambda < 1.2 \mu m$). This leads to the following conclusions. 

First, these IR excess emissions show a continuum which is consistent across the wavelength range for all stars. It ranges in the data from about 2$\mu$m to 30$\mu$m and corresponds to differences of magnitudes of up to -0.1 and even -0.2 magnitudes for RS Pup and $\eta$ Aql, between visible and far-infrared. Importantly, this IR excess rises in the near-infrared (around 2$\mu$m) from about 0 magnitude of differences (assuming $\Delta \mathrm{m}=0$ for $\lambda < 1.2 \mu m$) to -0.1 magnitude around 5$\mu$m, for each star, which brings strong constrains on the models and in particular invalidate a pure CSE of dust to explain the IR excess continuum. 

Second, in order to unveil CSE emission in the N-band we performed an independent correction of the ISM silicate absorption (Sect. \ref{deredden}) which seems to fill almost perfectly the silicate absorption seen in the {\it Spitzer} data. We have determined the excess at 9.7$\mu m$ and we have found that there is no emission within the uncertainty for SU~Cyg and V~Cen, while weak emissions are found for $\zeta$~Gem, $\eta$~Aql and RS Pup. 
The slight residuals that we found from our spectroscopic analysis at 9.7$\mu$m could be attributed to faint dusty CSE around Cepheids with a flux contribution (compared to the stellar flux) of few percents, which is in average ten times less than what was predicted by other studies based on interferometry \citep{Gallenne2012,gallenne13b}, and consistent with \cite{Schmidt2015} who found no silicate emission for a large sample of Cepheids.

Third, using photometric bands and basically the same approach but without considering the pulsation of the Cepheids, \cite{Schmidt2015} found that 21 over 132 classical and type 2 Cepheids in his sample has a clear or weak IR infrared excess. Moreover, these 21 Cepheids have periods larger than 11 days. Conversely, we have 4 short periods Cepheids in the sample (except RS Pup) and all of them show a clear IR excess. However, there is no star in common between the two studies to go deeper in the analysis. 

Four, we find cold, large and inhomogeneous circumstellar environment around the Cepheids RS Pup, V Cen, which is seen in the \textit{Herschel} images. As Cepheids are relatively young stars, they are still likely close to the cloud where they formed and such environment can contribute to the overall IR excess, either in absorption or emission. This could also affect the mid-infrared photometry of Cepheids, and thus potentially estimations of the PL relation when using instruments such as the JWST and forthcoming mid-infrared instruments on the future 40 meter-class telescopes.

\subsection{The IR excess explained by a thin shell of ionized gas}
We show that a thin shell of ionized gas with a temperature ranging from 3500 to 4500K depending on the Cepheid considered, and with a width of typically $\simeq$ 15\% of the radius of the star can reproduce the IR excess. Up to now, the only attempt to detect ionized material were carried out by the Very Large Array (VLA) at 5GHz on $\eta$~Aql and 4 others classical Cepheids \citep{welch1988}. Since no $3\sigma$ detection has been reported, only upper limits on flux density were derived. From the ionized shell model presented in this paper, we derived a flux density between 0.01 and 0.1 $\mu$Jy at 5GHz ($\approx$20\% above the star continuum) which is below the upper limit on flux density of $\approx$100$\mu$Jy estimated by \cite{welch1988}.

On the other hand interferometric observations have resolved CSEs around Cepheids. The first detection was reported around l~Car \citep{kervella06a} followed by $\delta$~Cep and Polaris \citep{merand07}. These CSEs were modeled with a ring at a distance of 2 to 3$R_\star$, i.e. close to the star, in a region where the temperature is high enough ($>2000K$) to prevent dust condensations \citep{Gail1999}. Thus, these observations are more likely explained by a shell of partially ionized gas. This occurence in detection by interferometry plays in favor of a widespread phenomenon amongs classical Cepheids.

Also, extensive studies of H$\alpha$ profiles in the atmosphere of short-, mid- and long-period Cepheids have shown that strong increases of turbulence occurs when the atmosphere is compressed during its infalling movement, or because of shock waves dynamics \citep{breitfellner93a,breitfellner93b,breitfellner93c,fokin96}. In the case of long period Cepheids, several shock waves can be observed and P Cygni profiles show that there is an expanding shell of H$\alpha$ emission which is detached from the photosphere \citep{gillet14}. Besides, analytical work of \cite{Neilson&Lester2008} have shown that mass loss is enhanced by pulsations and shocks in the atmosphere. The present paper suggests this mass loss could be in the form of partially ionized gas.


The model of the shell of ionized gas could also be linked to the chromospheric activity of Cepheids. \cite{sasselov94a} report HeI$\lambda$10830 observation on seven Cepheids providing the evidence of a high temperature plasma and steady material outflow in the highest part of the atmosphere. In addition, high resolution profiles of Mg II h and k lines (2900A) of five Cepheids with International Ultraviolet Explorer instrument \citep{schmidt1984II} revealed extended dynamical chromospheres up to tenth of stellar radii composed of rising and falling material. Such extensions could be compatible with the shell thickness presented in this paper. Outflowing materials, with velocity of 50 to 100 km/s, can in turn eject material to several stellar radii if the velocity lasts through the pulsation cycle. These strong mass transfers could explain the enigmatic X-rays detections with Chandra and XMM Newton \citep{engle2017}.


In parallel, it is interesting to compare Cepheids to very long period Mira stars for which a radiosphere near $2R_\star$ due to free-free emission has been reported \citep{reid1997}.
 
\subsection{Limit of the model and perspectives \label{perspectives}}
We have found relatively low temperatures for the hydrogen ionization (between 3500 and 4500K see Table~\ref{result.shell}) and only a small fraction of the gas should be ionized according to Saha equation. In particular under 4000K the number of free electrons should be mostly provided by metals such as iron or aluminum which have low ionization potential. This effect could in turn produce a fainter ionized shell for low-metallicity Cepheids in the Magellanic clouds. Thus, it would be interesting to quantify theoretically the impact of metallicity on the shells of ionized gas, and the PL relation. Moreover, our model does not take into account temperature nor density gradients in the star's atmosphere, and in particular compression and/or shock waves which could also heat up the shell and ionize the gas. 

There are indications from SPIPS analysis that the IR excess of Cepheids is not or only slightly time-dependent. Evidences of slight cycled variations exist due to the opacity change at 4.5$\mu m$ caused by periodic formation and destruction of CO molecules in the atmosphere \citep{scowcroft2016}. As we have {\it Spitzer} data only at a specific phase of pulsation for each Cepheid, we cannot firmly conclude for the time-dependency of the mid-IR excess. Nevertheless, if we assume that the IR excess is constant, then, as our thin shell of ionized gas is close to the star ($\simeq$ 15\% of the radius of the star), it is supposed to be dynamic and its parameters should vary with time. Doing the test for $\eta$ Aql we find a rather stable relative size of the shell ($\simeq$ 15 $\pm$2 \%) and a temperature variation from about 3500 to 4500K, which is similar to the values we obtained from one star to the other in the sample.

Last, our simple model suggests that the thin gas shell absorbs the light coming from the star in the visible domain (from 0.01 up to 0.13 mag), which invalidates the initial assumptions of $\Delta \mathrm{m}=0$ for $\lambda < 1.2 \mu m$ in the SPIPS algorithm. We obtain satisfactory results only if we apply a correction offset $\Delta \mathrm{m}>0$ for all wavelengths (except for SU Cyg with a slight negative offset). To explain this shift, we suggest that the distance found by the SPIPS algorithm might be too large by few percents (factor $10^{\Delta \mathrm{m}/2.5}$), all other parameters being unchanged. In other words, if obscured by a shell of ionized gas, Cepheids could be slightly closer than expected by SPIPS.
On the interferometric side, the angular diameter of the star would be also lower by few percents, but this can be compensated by the effective size of shell, and would have little impact in the SPIPS fitting. Thus, a spatial and chromatic analysis of the shell including interferometric constrains in all available bands with in particular VEGA/CHARA (visible), PIONIER/VLTI (infrared) and MATISSE/VLTI (L, M, N bands) is still necessary to better understand the environment of Cepheids, and eventually, check the impact on the PL relation.

\section{Conclusion}\label{Conclusion}
\begin{enumerate}
\item For the five Cepheids, we report a continuum IR excess increasing up to $\approx$-0.1/-0.2 magnitudes at 30$\mu$m, which cannot be explained by a hot or cold dust model of CSE.
\item Within the limits of our assumptions, we do not firmly conclude about the presence of CSE emission in the N-band but it is likely weak (> -0.1mag) according to our results. 
\item We show for the first time that IR excess of Cepheids can be explained by a free-free emission of a thin shell of ionized gas with a thickness of $\simeq$8-17\% star radius, an ionized mass of $10^{-9}-10^{-7}$\Msolar and a temperature of 3500-4500K. In this simple model, density and temperature have a constant radial distribution.
 \item The presence of a thin shell of partially ionized gas around Cepheids has to be tested with interferometers operating in visible, in the mid-IR or in the radio domain. The impact of such CSEs of ionized gas on the PL relation of Cepheids needs also more investigations.
\end{enumerate}

\begin{acknowledgements}
The research leading to these results has received funding
from the European Research Council (ERC) under the European Union’s
Horizon 2020 research and innovation programme under grant agreement No
695099 (project CepBin). The authors acknowledge the support of the French Agence Nationale de la Recherche (ANR), under grant ANR-15-CE31-0012- 01 (project UnlockCepheids). W.G. and G.P. gratefully acknowledge financial support for this work from the BASAL Centro de Astrofisica y Tecnologias Afines (CATA) AFB-170002.We acknowledge financial support from ``Programme National de Physique Stellaire'' (PNPS) of CNRS/INSU, France. This project was partially supported by the Polish Ministry of Science grant Ideas Plus. This research made use of the SIMBAD and VIZIER\footnote{Available at \url{ http://cdsweb.u- strasbg.fr/}} databases at CDS, Strasbourg (France) and the electronic bibliography maintained by the NASA/ADS system. This work is based in part on observations made with the Spitzer Space Telescope, which is operated by the Jet Propulsion Laboratory, California Institute of Technology under a contract with NASA. PACS  has  been  developed  by  a  consortium  of  institutes  led  by  MPE  (Ger-many) and including UVIE (Austria); KU Leuven, CSL, IMEC (Belgium); CEA,LAM (France); MPIA (Germany); INAF-IFSI/OAA/OAP/OAT, LENS, SISSA(Italy); IAC (Spain). This development has been supported by the funding agen-cies BMVIT (Austria), ESA-PRODEX (Belgium), CEA/CNES (France), DLR(Germany),  ASI/INAF  (Italy),  and  CICYT/MCYT  (Spain). 
This research also made use of SAOImage DS9,  an  astronomical  imaging  and  data  visualization  application \citep{ds92000} and  Astropy, a  community-developed corePython package for Astronomy \citep{astropy2018}.
\end{acknowledgements}

\bibliographystyle{aa}  
\bibliography{draft_vh_paperI} 

\newpage
\begin{appendix} 
\section{SPIPS data set and fitted pulsational model of the star sample.\label{app:spips}}
The plots are organized as follows: the pulsational velocity,  the effective temperature and the angular diameter curves according to the pulsation phase are shown on the left panels while  the right panels display photometric data in various bands. Above each figure is indicated the projection factor set to $p=1.270$, the fitted distance $d$ using parallax-of-pulsation method, the fitted color excess E(B-V), and the \textit{ad-hoc} IR excess law. In the photometric panels, the gray dashed line corresponds to the magnitude of the SPIPS model without CSE. It actually corresponds to the magnitude of a Kurucz atmosphere model, $m_\mathrm{kurucz}$, obtained with the ATLAS9 simulation code from \cite{castelli2003} with solar metallicity and a standard turbulent velocity of 2km/s. The gray line corresponds to the best SPIPS model, which is composed of the latter model without CSE plus an IR excess model. Note that for WISE, MSX and IRAS filters observations above 5 microns, only one data point is obtained without information on the phase. Hence it is represented by a horizontal gray strip for which the vertical width is the uncertainty of the measurement. In the angular diameter panels the gray curve corresponds to limb-darkened (LD) angular diameters.
For metallic stars, when effective temperature is low enough, CO molecules can form in the photosphere and absorb light in the CO band-head at 4.6 $\mu m$ \citep{scowcroft2016}. This effect is observed in the \textit{Spitzer} I2 IRAC dataset of $\zeta$~Gem and V Cen (see Fig. \ref{spips_fit_zet_gem} and \ref{spips_fit_v_cen}). In this case, these data were ignored during the fitting of SPIPS. When no effective temperatures and no angular diameters are included in the SPIPS model, there is a degeneracy between the mean temperature and E(B-V). We estimate that SPIPS can make an error of +/- 200K on the effective temperature and +/-~0.05 on E(B-V). Only V~Cen has no data for both effective temperature and angular diameter, nevertheless the fitting of SPIPS is thought to be reliable (for example see E(B-V) value compared with literature in Table~\ref{Tab.EBV}).

\begin{figure*}[htb]
\begin{center}
\includegraphics[width=\hsize]{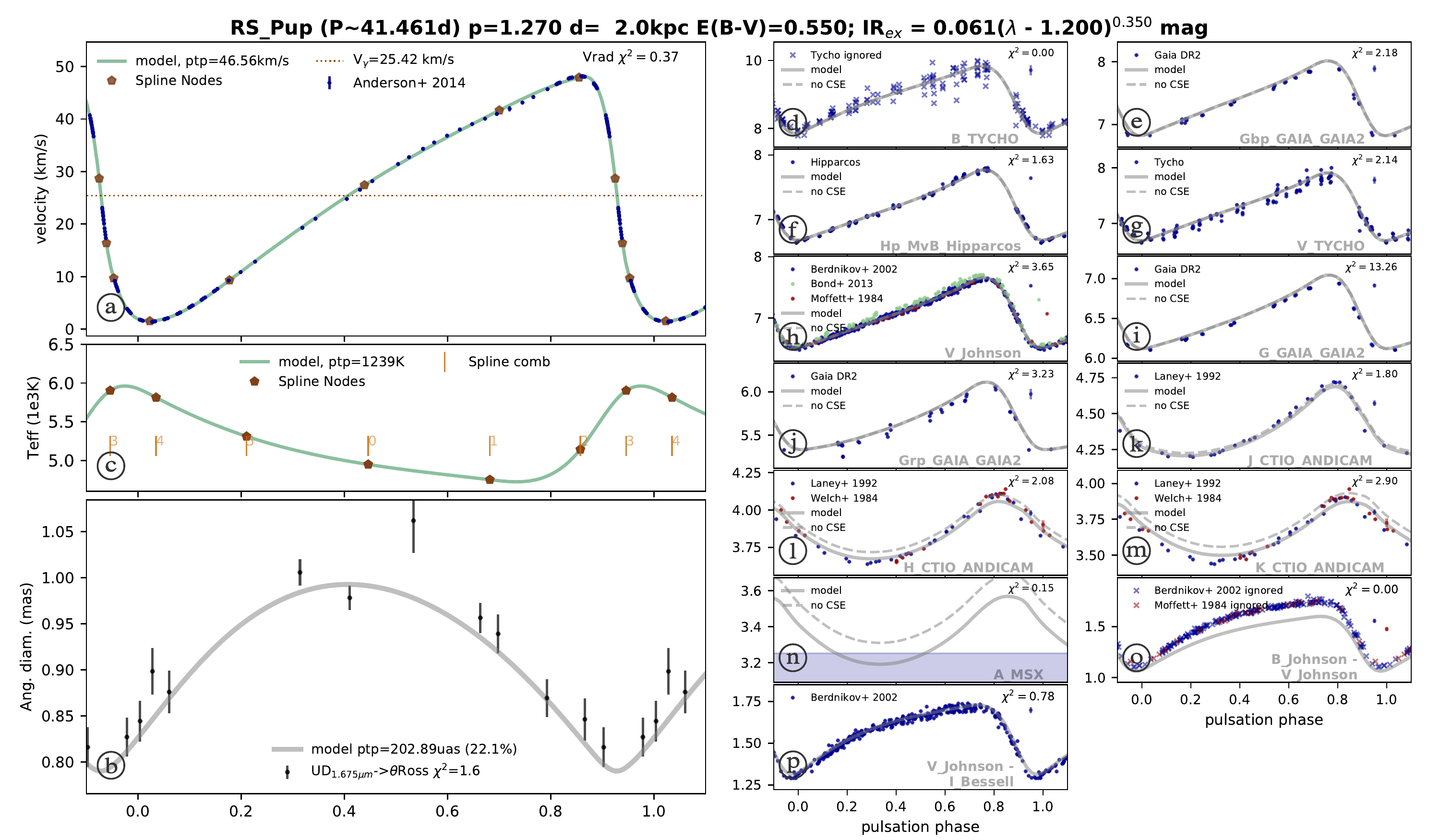}
\caption{\label{rs_pup_spips}\small The SPIPS results of RS Pup. \textbf{Velocity:} \cite{anderson14}. \textbf{Effective temperature:} No data. \textbf{Points are nodes for a spline interpolation.} \textbf{Angular diameter:} \cite{kervella17}. \textbf{Photometry:} \cite{moffett84}, \cite{Welch1984}, \cite{laney1992}, \cite{hipparcos1997}, \cite{Price2001}, \cite{berdnikov2002}, \cite{GaiaDR2}.}
\end{center}
\end{figure*}

\begin{figure*}[htb]
\begin{center}
\includegraphics[width=\hsize]{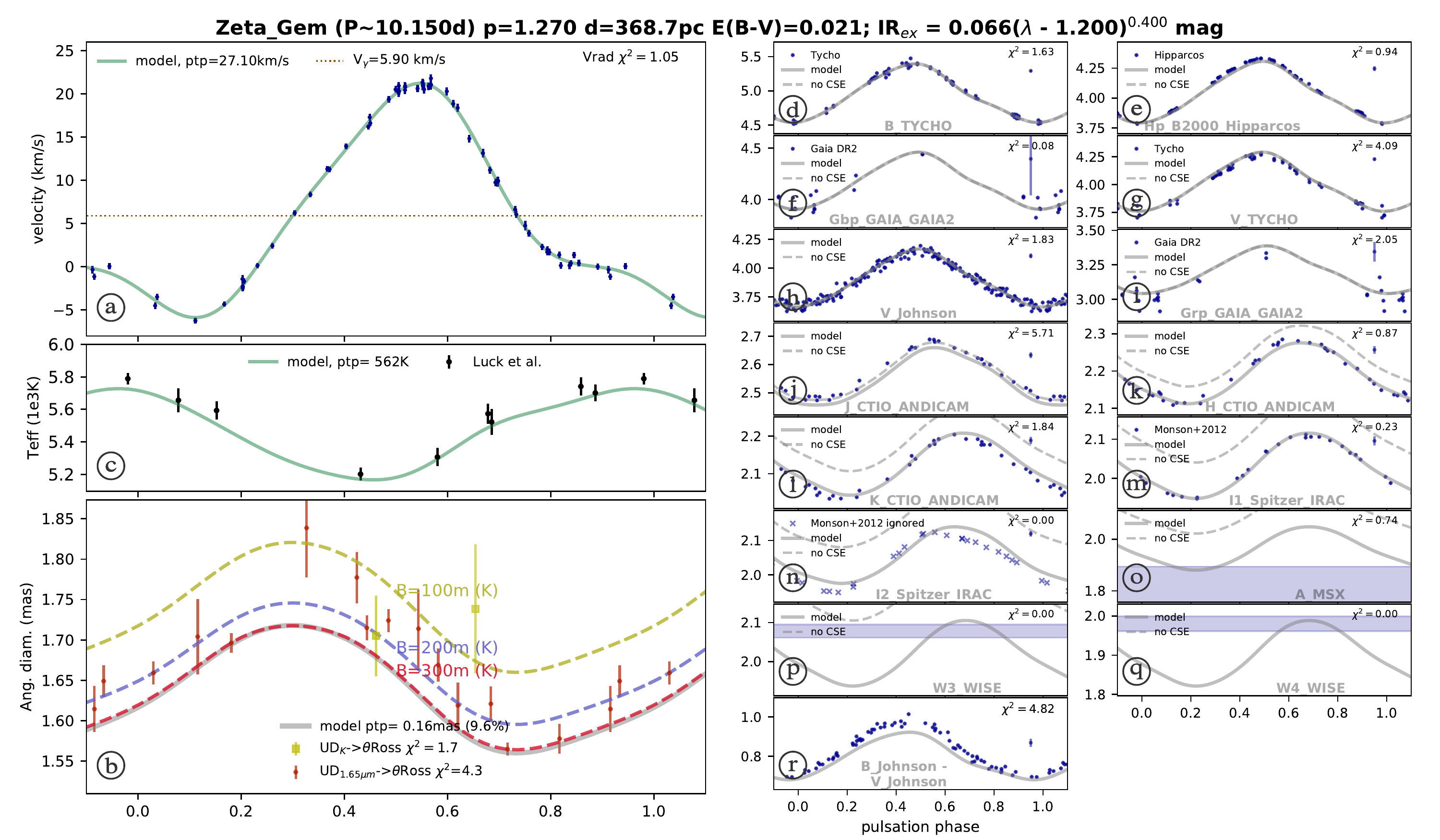}
\caption{\small \label{spips_fit_zet_gem}The SPIPS results of $\zeta$ Gem. \textbf{Velocity:} \cite{bersier94b}. \textbf{Effective temperature:} \cite{Luck2008}. \textbf{Angular diameter:} \cite{lane2002}, \cite{kervella04b}. \textbf{Photometry:} \cite{hipparcos1997}, \cite{vanLeeuwen1997}, \cite{Price2001}, \cite{WISE2010}, \cite{berdnikov2002}, \cite{Feast2008}, \cite{Monson2012}, \cite{GaiaDR2}.}
\end{center}
\end{figure*}

\begin{figure*}[htb]
\begin{center}
\includegraphics[width=\hsize]{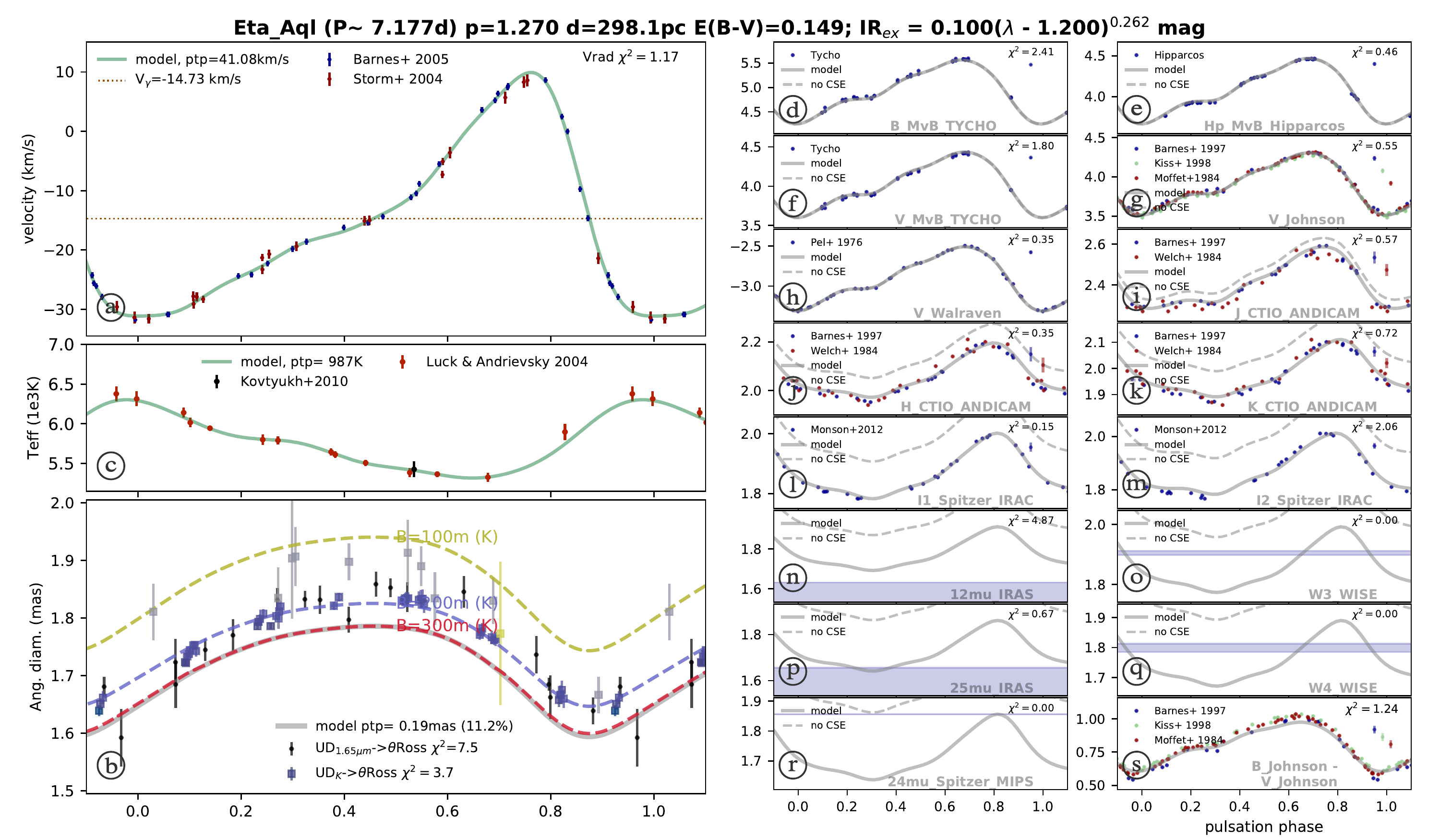}
\caption{\small The SPIPS results of $\eta$ Aql. \textbf{Velocity:} \cite{Storm2004}, \cite{Barnes2005}. \textbf{Effective temperature:} \cite{Luck&Andrievsky2004}, \cite{kovtyukh2010}. \textbf{Angular diameter:} \cite{lane2002}, \cite{kervella04b}. \textbf{Photometry:} \cite{Pel1976}, \cite{Welch1984}, \cite{IRAS1984}, \cite{Barnes1997}, \cite{hipparcos1997}, \cite{vanLeeuwen1997}, \cite{Kiss1998}, \cite{MIPS2004}, \cite{WISE2010}, \cite{Monson2012}.}
\end{center}
\end{figure*}

\begin{figure*}[htb]
\begin{center}
\includegraphics[width=\hsize]{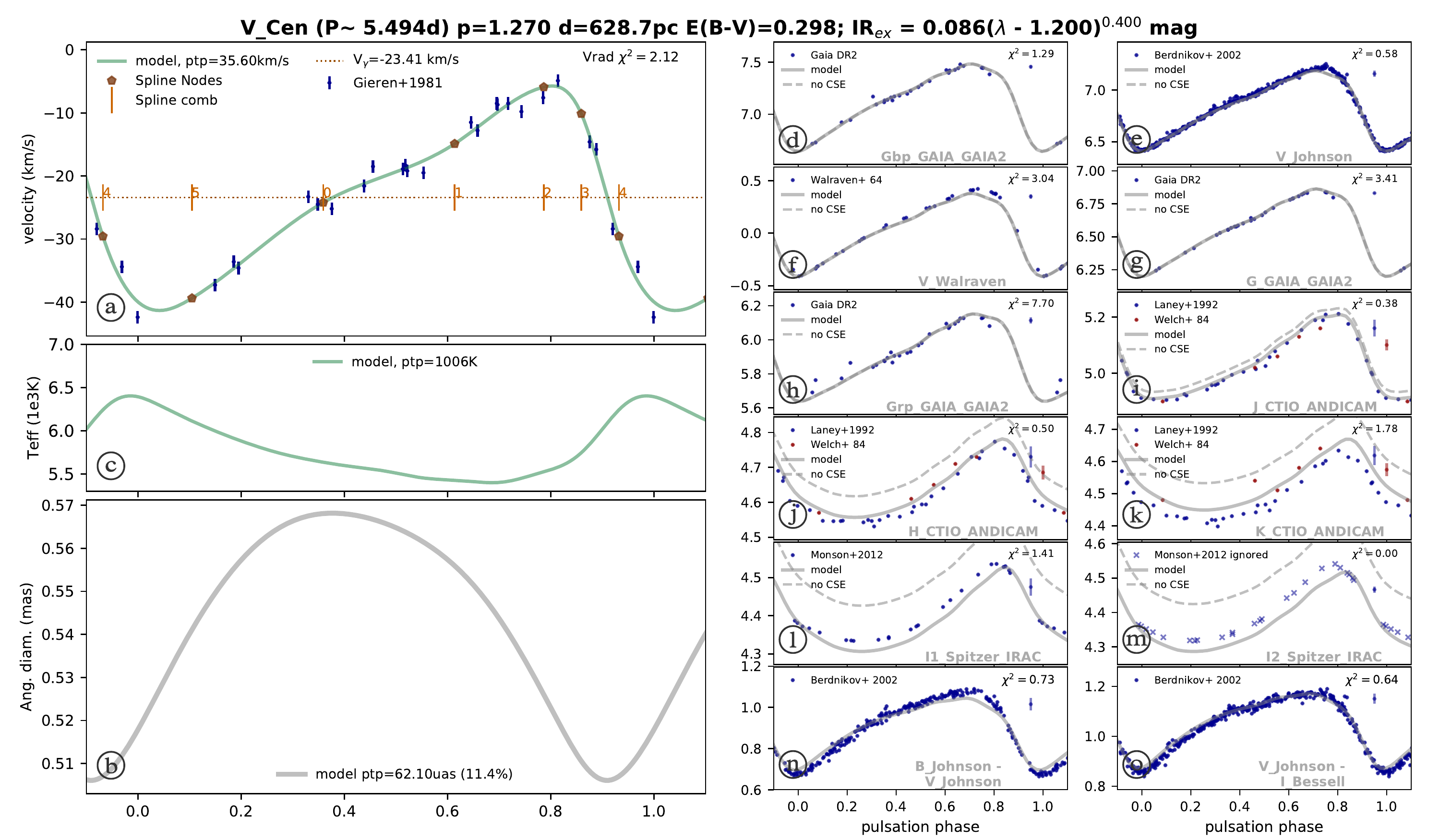}
\caption{\label{spips_fit_v_cen} \small The SPIPS results of V Cen. \textbf{Velocity:} \cite{Gieren1981v}. \textbf{Effective temperature:} No data. \textbf{Angular diameter:} No data. \textbf{Photometry:} \cite{walraven64}, \cite{Welch1984}, \cite{laney1992}, \cite{berdnikov02}, \cite{Monson2012}, \cite{GaiaDR2}.}
\end{center}
\end{figure*}

\begin{figure*}[htb]
\begin{center}
\includegraphics[width=\hsize]{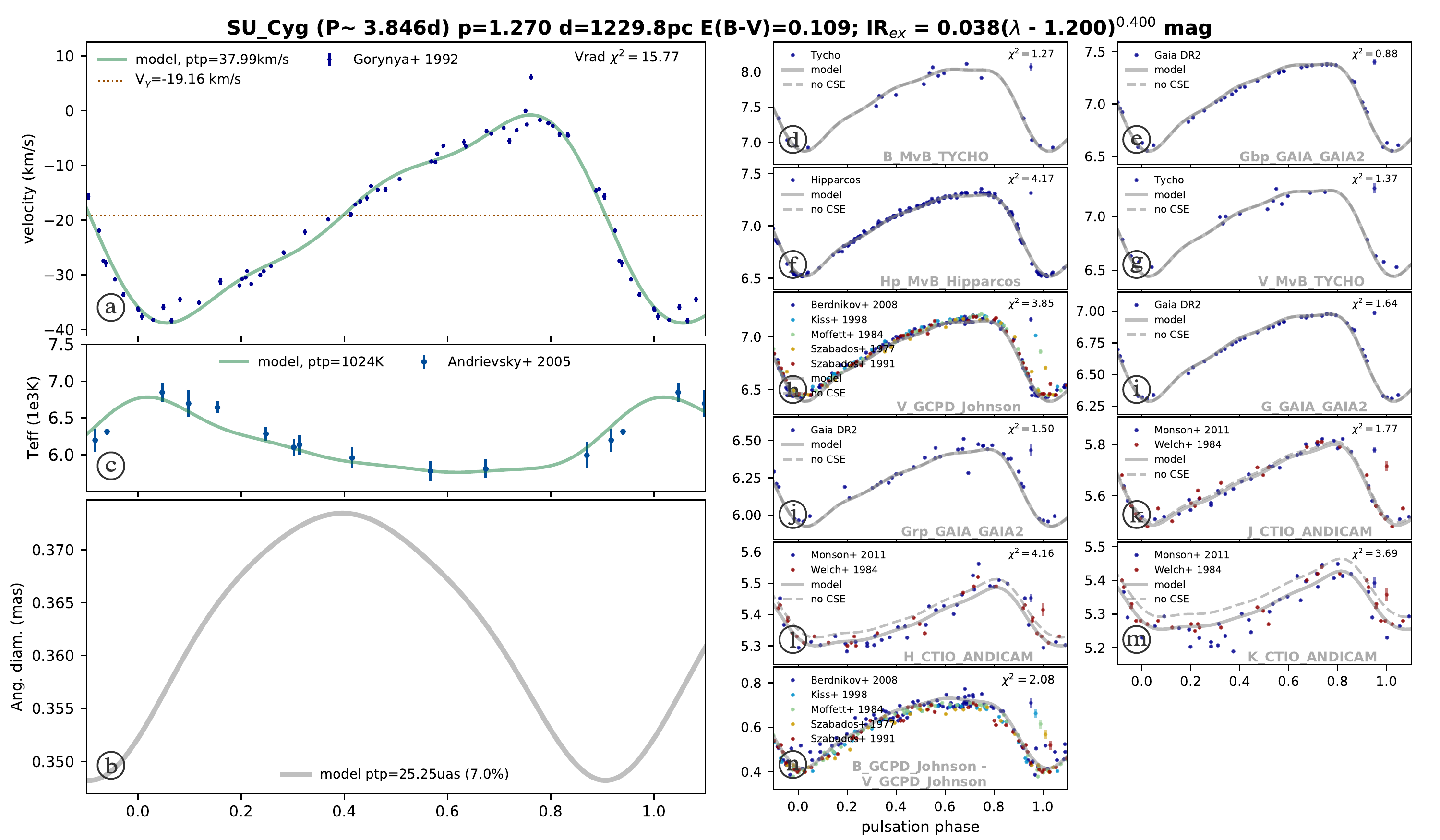}
\caption{\small The SPIPS results of SU Cyg. \textbf{Velocity:} \cite{Gorynya1992} , \textbf{Effective temperature:} \cite{Andrievsky2005}. \textbf{Angular diameter:} No data. \textbf{Photometry:}  \cite{Szabados1977}, \cite{Welch1984}, \cite{moffett84}, \cite{Szabados1991}, \cite{hipparcos1997}, \cite{vanLeeuwen1997}, \cite{Kiss1998}, \cite{Berdnikov2008}, \cite{Monson2011}, \cite{GaiaDR2}.}
\end{center}
\end{figure*}



\section{The IR excess of a thin gas shell at constant temperature and density \label{app:thin_gas_shell}}

\begin{figure}[H]
\begin{center}
\resizebox{1\hsize}{!}{\includegraphics[clip=true]{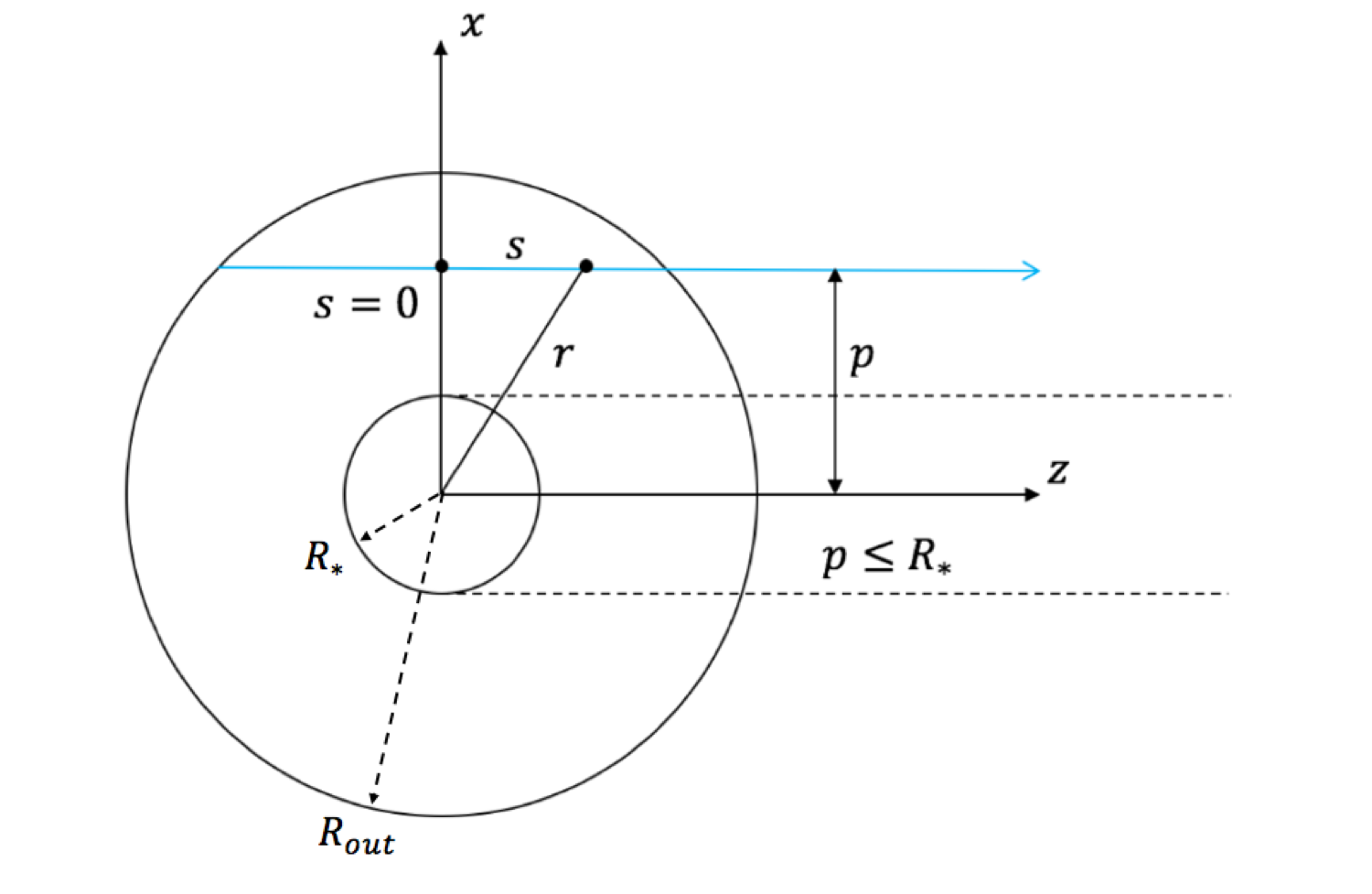}}
\end{center}
\caption{\small Circumstellar shell model. The blue line represents a ray along which the radiative transfer equation is integrated. $p$ is the corresponding impact parameter. $s$ is the distance along the ray $\Rstar$ is the stellar radius. $R_\mathrm{out}$ is the external radius of the shell.\label{frame}}
\end{figure}

The shell emission is obtained integrating the radiative transfer equation along rays defined by their impact parameter $p$ (see Fig.~\ref{frame}) following the method described in \citet{panagia1975}.

We assume a constant shell temperature $\Ts$ and density $\rhos$. The gas opacities (see Eq.~(\ref{eq:1})) can be written under the form $\kappa(\lambda,\Ts)=\rhos^2\,\chi_\lambda(\Ts)$.

According to Fig.~\ref{frame} we have to take into account two cases corresponding to the impact parameter $p$ respectively larger and smaller than the stellar radius $\Rstar$.

For $p\geqslant \Rstar$, taking into account the symmetry of the problem the optical depth along the ray is given by
\begin{equation}
\tau_{\lambda}(p)=2\int_0^{\sqrt{R^2_\mathrm{out}-p^2}}\kappa(\lambda,\Ts)ds \ ,
\end{equation}
which, for a constant $\Ts$ and density $\rhos$ gives
\begin{equation}
\tau_\lambda(p)=2\rhos^2\chi_\lambda(\Ts)\,\sqrt{R^2_\mathrm{out}-p^2} \ .
\end{equation}

Similarly, for $p\leqslant \Rstar$ we have
\begin{equation}
\tau_{\lambda}(p)=\rhos^2\chi_\lambda(\Ts)\,\Big[\sqrt{R^2_\mathrm{out}-p^2}-\sqrt{\Rstar^2-p^2}\Big] \ .
\end{equation}

For $p\geqslant \Rstar$, the specific intensity is given by 

\begin{equation}
I_{\lambda}(p) = B_\lambda(\Ts)(1-e^{-\tau_\lambda(p)}) \ ,
\end{equation}
since $\Ts$ is assumed to be constant along the ray.

For $p\leqslant \Rstar$, both the shell and the stellar photosphere contribute to the specific intensity 
\begin{equation}
I_{\lambda}(p) = B_\lambda(\Ts)\Big(1-e^{-\tau_\lambda(p)}\Big)+ I^\ast_\lambda e^{-\tau_\lambda(p)} \ ,
\end{equation}
where $I^\ast_\lambda$ is the stellar specific intensity.

The observed total emerging flux at a distance $d$ is then computed numerically by quadrature with the following integral
\begin{equation}
F_\lambda=\frac{2\pi}{d^2}\int^{R_\mathrm{out}}_0pI_\lambda(p)dp \ .
\label{eq:totalFlux}
\end{equation}



From the result of the integral in Eq.~(\ref{eq:totalFlux}) we can deduce the magnitude excess defined by
\begin{equation}
\Delta \mathrm{mag}=-2.5\log{\Big(\frac{F_\lambda}{F^\ast_\lambda} \Big)} \ ,
\end{equation}
with $F_{\lambda}^\ast=\pi\left(\frac{\Rstar}{d}\right)^2\,I^\ast_\lambda$.

Note that in the particular case where we have $I^\ast_\lambda = B_\lambda(\Tstar)$ with $\Ts = \Tstar$, Eq.~(\ref{eq:totalFlux}) can be integrated analytically to give
\begin{multline}
F_{\lambda}=\pi\Big(\frac{\Rstar}{d}\Big)^2 B_\lambda(\Ts) \times \\ 
\Big[1 + \Big[\Big(\frac{R_\mathrm{out}}{\Rstar}\Big)^2-1\Big]\,\Big[1+\frac{2}{\tau^{\ast 2}_{\lambda}}\Big[(1+\tau^{\ast}_{\lambda})e^{-\tau^{\ast}_\lambda}-1\Big]\Big]\Big] \ ,
\end{multline}
with $\tau^\ast_\lambda$ defined as $2\,\rhos^2\,\chi_\lambda(\Ts) \,\Rstar \,\sqrt{\left(\frac{R_\mathrm{out}}{\Rstar}\right)^2-1}$.


The corresponding magnitude excess is given by
\begin{equation}
\Delta \mathrm{mag}=2.5\log{ \left[1+\left[\left(\frac{R_\mathrm{out}}{\Rstar}\right)^2-1\right]\\
\left[1+\frac{2}{\tau^{\ast 2}_{\lambda}}\left[(1+\tau^{\ast}_\lambda)e^{-\tau^{\ast}_\lambda}-1\right]\right]\right]}
\end{equation}
\end{appendix} 
\end{document}